\documentclass[useAMS,usenatbib]{mn2e}

\usepackage{epsfig}
\usepackage{amsmath}
\usepackage{longtable,lscape}
\usepackage{rotating}
\usepackage{multirow,multicol}
\usepackage[breaklinks]{hyperref}
\usepackage{afterpage}

\newcommand{\kms}   {km~s$^{-1}$}
\newcommand{\jpb}   {$\rm Jy~beam^{-1}$}    
\newcommand{\lo}    {$L_{\sun}$}
\newcommand{\mo}    {$M_{\sun}$}
\newcommand{\nh}    {NH$_3$}
\newcommand{\nth}   {N$_2$H$^+$}
\newcommand{\et}    {et al.}
\newcommand{\eg}    {e.\,g.,}
\newcommand{\ie}    {i.\,e.,}
\newcommand{\hii}   {H{\small II}}

\newcommand{\supa}  {$^\mathrm{a}$}
\newcommand{\supb}  {$^\mathrm{b}$}
\newcommand{\supc}  {$^\mathrm{c}$}
\newcommand{\supd}  {$^\mathrm{d}$}
\newcommand{\supe}  {$^\mathrm{e}$}
\newcommand{\supf}  {$^\mathrm{f}$}

\newcommand{\phn}   {\phantom{0}}
\newcommand{\phnn}  {\phantom{0}\phantom{0}}
\newcommand{\phu}   {\phantom{$^{\it a}$}}

\title[Properties of dense cores in clustered high-mass star-forming regions]{Properties of dense cores in clustered massive star-forming regions at high angular resolution}
\author[\'A. S\'anchez-Monge et al.]
  {\'Alvaro S\'anchez-Monge$^{1,2}$\thanks{E-mail: asanchez@arcetri.astro.it (ASM)},
  Aina Palau$^{3}$, Francesco Fontani$^{1}$, Gemma Busquet$^{4}$,  
  \newauthor
  Carmen Ju\'arez$^{3,2}$, Robert Estalella$^{2}$, Jonathan C.\ Tan$^{5}$, Inma Sep\'ulveda$^{2}$,
  \newauthor
  Paul T.~P.\ Ho$^{6,7}$, Qizhou Zhang$^{7}$ and Stan Kurtz$^{8}$ \\Ê\\
  $^{1}$Osservatorio Astrofisico di Arcetri, INAF, Largo Enrico Fermi 5, I-50125, Firenze, Italy\\
  $^{2}$Dpt.\ d'Astronomia i Meteorologia (IEEC-UB), ICC, Universitat de Barcelona, Mart\'i i Franqu\`es 1, E-08028, Barcelona, Spain \\
  $^{3}$Institut de Ci\`encies de l'Espai (CSIC-IEEC), Campus UAB, Facultat de Ci\`encies, Torre C-5p, E-08193 Bellaterra, Barcelona, Spain\\
  $^{4}$INAF-Istituto di Astrofisica e Planetologia Spaziali, Via Fosso del Cavaliere 100, I-00133 Roma, Italy\\
  $^{5}$Dpts.\ of Astronomy \& Physics, University of Florida, Gainesville, FL 32611, USAÊ\\
  $^{6}$Institute of Astronomy and Astrophysics, Academia Sinica, P.O.\ Box 23-141, Taipei 106, Taiwan \\
  $^{7}$Harvard-Smithsonian Center for Astrophysics, 60 Garden Street, Cambridge, MA 02138, USA \\
  $^{8}$Centro de Radioastronom\'ia y Astrof'sica, Universidad Nacional Aut\'onoma de M\'exico, P.O.\ Box 3-72, 58090, Morelia, Mich., Mexico}

\begin{document}

\date{Accepted date. Received date; in original form date}

\pagerange{\pageref{firstpage}--\pageref{lastpage}} \pubyear{2012}

\maketitle

\label{firstpage}

\begin{abstract}
We aim at characterising dense cores in the clustered environments associated with intermediate/high-mass star-forming regions. For this, we present an uniform analysis of Very Large Array \nh\,(1,1) and (2,2) observations towards a sample of 15 intermediate/high-mass star-forming regions, where we identify a total of 73 cores, classify them as protostellar, quiescent starless, or perturbed starless, and derive some physical properties. The average sizes and ammonia column densities of the total sample are $\sim 0.06$~pc and $\sim 10^{15}$~cm$^{-2}$, respectively, with no significant differences between the starless and protostellar cores, while the linewidth and rotational temperature of quiescent starless cores are smaller, $\sim 1.0$~\kms\ and 16~K, than linewidths and temperatures of protostellar ($\sim 1.8$~\kms\ and 21~K), and perturbed starless ($\sim 1.4$~\kms\ and 19~K) cores. Such linewidths and temperatures for these quiescent starless cores in the surroundings of intermediate/high-mass stars are still significantly larger than the typical linewidths and rotational temperatures measured in starless cores of low-mass star-forming regions, implying an important non-thermal component. We confirm at high angular resolutions (spatial scales $\sim 0.05$~pc) the correlations previously found with single-dish telescopes (spatial scales $\ga 0.1$~pc) between the linewidth and the rotational temperature of the cores, as well as between the rotational temperature and the linewidth with respect to the bolometric luminosity. In addition, we find a correlation between the temperature of each core and the incident flux from the most massive star in the cluster, suggesting that the large temperatures measured in the starless cores of our sample could be due to heating from the nearby massive star. A simple virial equilibrium analysis seems to suggest a scenario of a self-similar, self-graviting, turbulent, virialised hierarchy of structures from clumps ($\sim 0.1$--10~pc) to cores ($\sim 0.05$~pc). A closer inspection of the dynamical state taking into account external pressure effects, reveal that relatively strong magnetic field support may be needed to stabilise the cores, or that they are unstable and thus on the verge of collapse.
\end{abstract}


\begin{table*}
\caption{Massive star-forming regions observed in \nh\ with the Very Large Array, and few observational and derived parameters}
\label{t:nh3sources}
\centering
\begin{tabular}{l c r c c c c c c c c c c}
\hline\hline

&$d$
&$L_\mathrm{bol}$
&VLA
&\multicolumn{2}{c}{Phase centre\supb}
&
&\multicolumn{2}{c}{Synthesised beam\supc}
&
&\multicolumn{3}{c}{Clumpfind\supe}
\\
\cline{5-6}\cline{8-9}\cline{11-13}
Region\supa
&(kpc)
&(\lo)
&project
&R.A.\
&Dec.\
&
&HPBW
&P.A.
&RMS\supd
&P
&B
&S
\\
\hline
1. 00117+6412	&1.8\phn	&\phn1300	&AB1217\phu	&00:14:27.725	&+64:28:46.17	&&$3.9\times3.5$	&$+46$	&2.2	&\phn66	&35	&10		\\
2. AFGL5142		&1.8\phn	&\phn2200	&AZ120\supf	&05:30:48.024	&+33:47:54.44	&&$3.6\times2.6$	&$-14$	&3.5	&203		&20	&10		\\
3. 05345+3157NE	&1.8\phn	&\phn630		&AF471\phu	&05:37:52.400	&+32:00:06.00	&&$2.3\times2.3$	&$+92$	&1.4	&\phn39	&25	&\phn5	\\
4. 05358+3543NE	&1.8\phn	&\phn3100	&AC733\phu	&05:39:12.600	&+35:45:52.30	&&$5.8\times4.1$	&$+53$	&5.3	&269		&40	&10		\\ 
5. 05373+2349	&1.6\phn	&\phn1100	&AF484\phu	&05:40:24.500	&+23:50:55.00	&&$3.1\times2.2$	&$-73$	&4.0	&116		&30	&10		\\
6. 19035+0641	&2.2\phn	&\phn7700	&AF386\phu	&19:06:01.610	&+06:46:35.80	&&$4.0\times3.7$	&$-13$	&2.6	&378		&28	&10		\\
7. 20081+3122	&2.5\phn	&\phn28200	&AF386\phu	&20:10:09.050	&+31:31:35.20	&&$5.0\times4.4$	&$-56$	&3.3	&602		&25	&\phn5	\\
8. 20126+4104	&1.64	&\phn8900	&AZ113\phu	&20:14:26.053	&+41:13:31.49	&&$3.6\times3.2$	&$+10$	&3.8	&873		&20	&10		\\
9. G75.78+0.34	&3.8\phn	&96000   	&AF386\phu	&20:21:43.970	&+37:26:38.10	&&$3.7\times3.4$	&$-57$	&3.8	&181		&30	&\phn5	\\
10. 20293+3952	&2.0\phn	&\phn8000	&AZ120\phu	&20:31:10.698	&+40:03:10.75	&&$6.9\times3.1$	&$+72$	&3.0	&240		&40	&10		\\
11. 20343+4129	&1.4\phn	&\phn1500	&AS708\phu	&20:36:07.301	&+41:39:57.20	&&$4.2\times3.2$	&$+11$	&3.5	&224		&27	&10		\\
12. 22134+5834	&2.6\phn	&11800   	&AK558\phu	&22:15:08.099	&+58:49:10.00	&&$3.8\times3.1$	&$+88$	&1.5	&\phn31	&40	&10		\\
13. 22172+5549N	&2.4\phn	&\phnn830	&AF484\phu	&22:19:08.600	&+56:05:02.00	&&$3.2\times3.0$	&$+17$	&1.2	&\phn61	&30	&10		\\
14. 22198+6336	&0.76	&\phnn340	&AS926\supf	&22:21:26.764	&+63:51:37.89	&&$3.7\times3.1$	&$-41$	&3.8	&\phn61	&40	&10		\\
15. CepA			&0.7\phn	&\phnn25000	&AF386\phu	&22:56:17.870	&+62:01:48.60	&&$5.4\times4.9$	&$-29$	&3.6	&332		&30	&\phn6	\\
\hline
\end{tabular}
\begin{flushleft}
\supa\ Name of regions starting with numbers (\eg\ 00117+6412) refer to the IRAS name \citep{neugebauer1984}.\\
\supb\ Phase centre coordinates (in Equatorial J2000.0) of the ammonia maps. Right ascension in units h:m:s and Declination in units d:m:s.\\
\supc\ Synthesised beam (HPBW) in arcsec, and position angle (P.A.) in degrees.\\
\supd\ RMS noise level, in units of m\jpb, per channel (of 0.3~\kms\ or 0.6~\kms; see Table~\ref{t:nh3obs}).\\
\supe\ P: intensity peak (in mJy~\kms) of the zero-order moment map. B (bottom) and S (step) parameters, in percentage of the peak, used in the Clumpfind algorithm to extract the position of the cores (see Sect.~\ref{s:cores}).\\
\supf\ AFGL5142 and 22198+6336 includes also data of the project AZ114.
\end{flushleft}
\end{table*}

\section{Introduction}\label{s:intro}

The initial conditions of the star formation process in clusters are still poorly understood. Studies have unveiled the physical and chemical properties of starless {\it isolated low-mass} cores on the verge of gravitational collapse in nearby low-mass star-forming regions, showing that they have dense ($n\sim 10^5$--10$^6$~cm$^{-3}$) and cold ($T\sim 10$~K) nuclei \citep[\eg][]{tafalla2002, tafalla2004, schnee2010}, and that the internal motions are thermally dominated, as demonstrated by their close-to-thermal linewidths, even when observed at low angular resolution \citep[see][for a review]{bergintafalla2007}. In the cold and dense nuclei of these cores, C-bearing molecular species such as CO and CS are strongly depleted \citep[\eg][]{caselli2002, tafalla2002}, while N-bearing species such as \nth\ and \nh\ maintain large abundances in the gas phase \citep[\eg][]{caselli2002, crapsi2005}.

One major issue that has so far been poorly investigated is if, and how, the environment influences the physical and chemical properties of these pre--stellar cores. In clusters containing several forming stars in a small volume (of diametre $\leq 0.1$~pc), turbulence, relative motions, and interaction with nearby forming (proto-)stars can affect the less evolved condensations \citep[\eg][]{wardthompson2007}. Such an interaction is expected to be much more important in clusters containing {\it intermediate-/high-mass protostars} or newly formed massive stars, given the typical energetic feedback associated with the earliest stages of massive star formation (powerful outflows, strong winds, expanding \hii\ regions) and the high pressure of the parental pc-scale clumps. There is also vigorous theoretical debate on how star formation proceeds in clustered regions. Specifically: are the local kinematics of the gas dominated by feedback from protostellar outflows of already-forming, generally low-mass stars or from feedback from high-mass stars \citep{nakamurali2007}? Can a model of star formation starting from quiescent starless cores, which has been successfully developed  to explain observations of regions of isolated low-mass star formation, be applied to clustered regions \citep{mckeetan2003}? To constrain theoretical models, it is crucial to measure the main physical and chemical properties of starless and star-forming cores in clustered environments.

\begin{table*}
\caption{Configurations used, epochs, calibrators and spectral setup of the Very Large Array observational projects}
\label{t:nh3obs}
\centering
\begin{tabular}{l c r l c l l c c c c c c}
\hline\hline

&VLA
&\multicolumn{2}{c}{Epoch}
&Flux
&\multicolumn{2}{c}{Gain calibrators}
&Bandwidth
&\multicolumn{2}{c}{Spectral resolution}
\\
\cline{9-10}
Project
&config.\
&\multicolumn{2}{c}{of observation}
&calibrator
&\multicolumn{2}{c}{(Bootstrapped fluxes, Jy)}
&(MHz)
&(kHz)
&(\kms)
\\
\hline
AB1217	&D	&2007	&May		&3C286,3C48		&J0102+584	&$3.85\pm0.06$	&3.125	&48.8	&0.6		\\	
AC733	&DnC&2004	&Jun		&3C84			&J0530+135	&$3.33\pm0.02$	&3.125	&48.8	&0.6		\\	
AF386	&D	&2001	&Jan		&3C286			&J2025+337	&$2.50\pm0.04$	&3.125	&48.8	&0.6		\\	
		&	&		&		&				&J1849+005	&$0.81\pm0.01$	&		&		&		\\	
		&	&		&		&				&J2322+509	&$0.72\pm0.01$	&		&		&		\\	
AF471	&D	&2009	&Oct/Nov	&3C286,3C48		&J0555+398	&$3.15\pm0.01$	&3.125	&24.4	&0.3		\\	
AF484	&D	&2009	&Oct/Nov	&3C48			&J0559+238	&$0.80\pm0.01$	&3.125	&24.4	&0.3		\\	
		&	&		&		&				&J2148+611	&$0.63\pm0.01$	&		&		&		\\	
AK558	&D	&2003	&May		&3C286			&J2148+611	&$0.60\pm0.01$	&3.125	&48.8	&0.6		\\	
AS708	&C	&2001	&Jul		&3C286			&J2015+371	&$2.34\pm0.04$	&3.125	&48.8	&0.6		\\	
AS926	&C	&2008	&Apr		&3C286,3C48		&J2146+611	&$0.72\pm0.02$	&3.125	&48.8	&0.6		\\	
AZ113	&D	&1999	&May		&3C286,3C48		&J2013+370	&$2.27\pm0.21$	&3.125	&24.4	&0.3		\\	
AZ114	&D	&1999	&Mar		&3C48			&J2230+697	&$0.43\pm0.01$	&3.125	&48.8	&0.6		\\	
		&	&		&		&				&J0552+398	&$3.46\pm0.05$	&		&		&		\\	
AZ120	&C	&2000	&Apr		&3C48			&J0552+398	&$4.72\pm0.15$	&3.125	&48.8	&0.6		\\	
		&D	&2000	&Sep		&3C48			&J0552+398	&$3.38\pm0.04$	&		&		&		\\	
		&	&		&		&				&J2015+371	&$3.84\pm0.02$	&		&		&		\\	
\hline
\end{tabular}
\end{table*}

An observational effort in this direction has started, but it is mostly concentrated on nearby low-mass star-forming regions like Ophiuchus \citep[\eg][]{andre2007, friesen2009} and Perseus \citep[\eg][]{foster2009}, due to the fact that the typical large distances ($\geq 1$~kpc) of high-mass star-forming regions make the study of clustered environments very challenging. \citet{foster2009} and \citet{friesen2009} show that starless cores (studied at spatial scales $\sim 0.01$--0.05~pc) within low-mass star-forming clusters have typically higher kinetic temperatures ($\sim 13$--14~K) than low-mass isolated cores ($\sim 10$~K). On the other hand, the kinematics of these low-mass protoclusters seem to be dominated by thermal motions like in more isolated cores, even though the external environment is turbulent \citep{andre2007}. On the contrary, in the very few published studies of (proto-)clusters containing an intermediate-/high-mass forming star, the internal motions of starless cores are dominated by turbulence (\eg\ \citealt{fontani2008, fontani2009} for IRAS~05345$+$3157; \citealt{wang2008} for G28.34$+$0.06; \citealt{pillai2011} for G29.96$-$0.02 and G35.20$-$1.74; \citealt{fontani2012b} for IRAS~20343$+$4129).

By using ammonia, \citet{palau2007b} measured the temperature of the starless cores in the high-mass proto-cluster IRAS~20293$+$3952 at spatial scales of $\sim 0.05$~pc, through Very Large Array observations of the \nh\,(1,1) and (2,2) inversion transitions. In fact, \nh\ is the ideal thermometer for cold and dense gas because it is produced by the volatile molecular nitrogen, which is not expected to freeze out even in very cold and dense gas, and the \nh\,(2,2) to (1,1) line ratio is sensitive to temperature \citep[\eg][]{hotownes1983}. Moreover, because other volatile species (like \nth) cannot be used to derive the temperature, {\it ammonia is the unique tracer for this purpose}. \citet{palau2007b} found temperatures of 16~K for starless cores in IRAS~20293$+$3952, \ie\ higher than those measured in isolated low-mass starless cores (or infrared dark clouds with no active star formation; \eg\ \citealt{ragan2011}), but similar to the values measured in active clustered cores in Perseus and Ophiuchus.

In this paper, we study a sample of 15 massive star-forming regions (see Table~\ref{t:nh3sources}) observed with the Very Large Array (providing an average spatial resolution of $\sim0.05$~pc) and analysed them in an uniform way. The main goal is to understand whether the results obtained in the few examples shown above are also obtained when analysing a statistically significant sample. The regions were selected according to the following criteria: i) regions must be at distances $\la 3.5$~kpc, to obtain spatial resolution $\la 0.05$~pc, similar to the scale of the cores (0.03--0.2~pc: \eg\ \citealt{bergintafalla2007}); ii) regions must have bolometric luminosities larger than $\sim 300$~\lo, which are the luminosities where clustering seems to be important \citep[\eg][]{testi1999, palau2013}; and iii) regions must be still associated with important amounts of gas and dust judging from single-dish observations of dust and molecular gas studies \citep[\eg][]{molinari1996, beuther2002b}.

\begin{figure*}
\centering
\begin{tabular}[b]{c c c}
\vspace{0.5cm}
  \epsfig{file=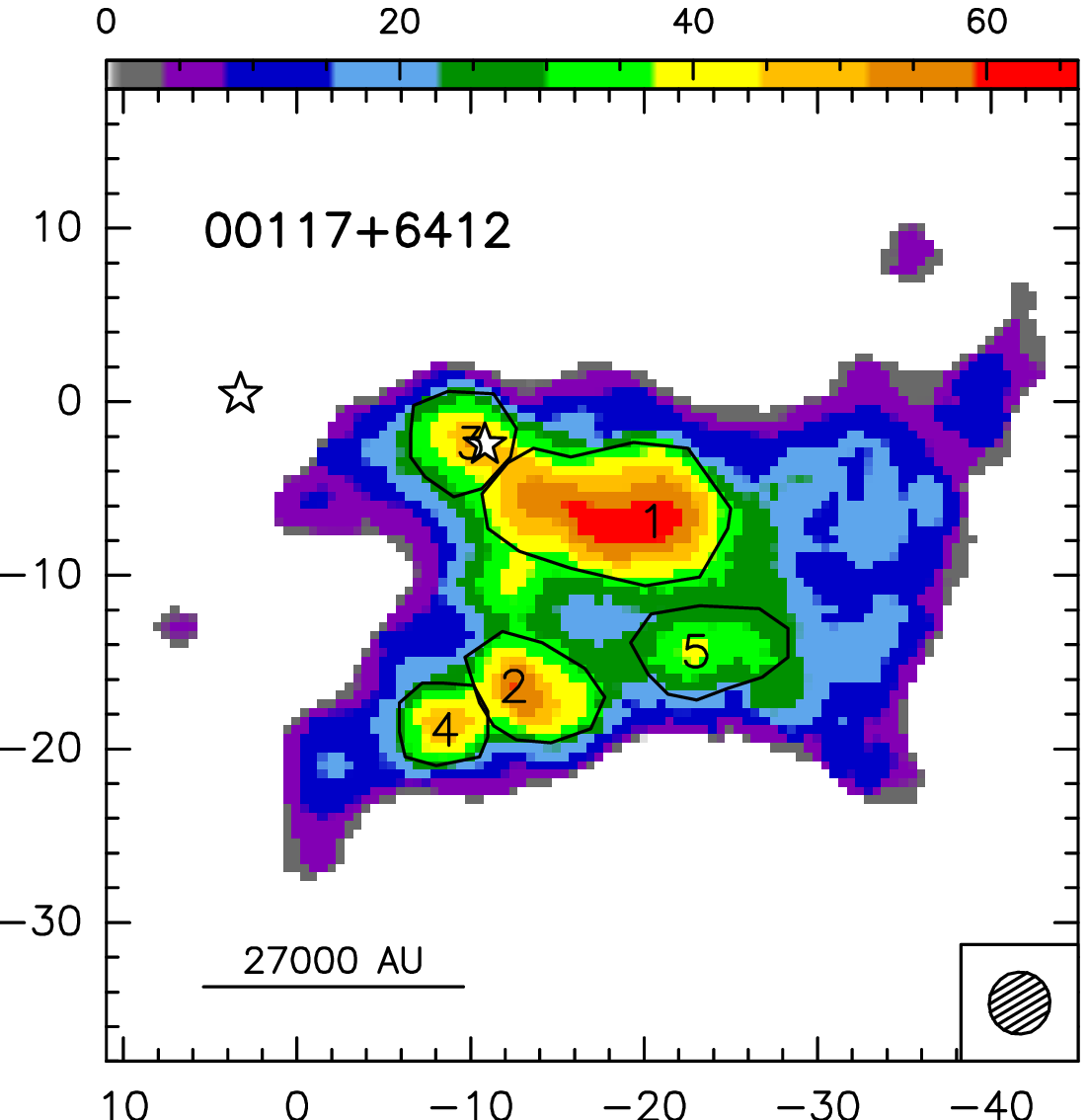, scale=0.48} &
  \epsfig{file=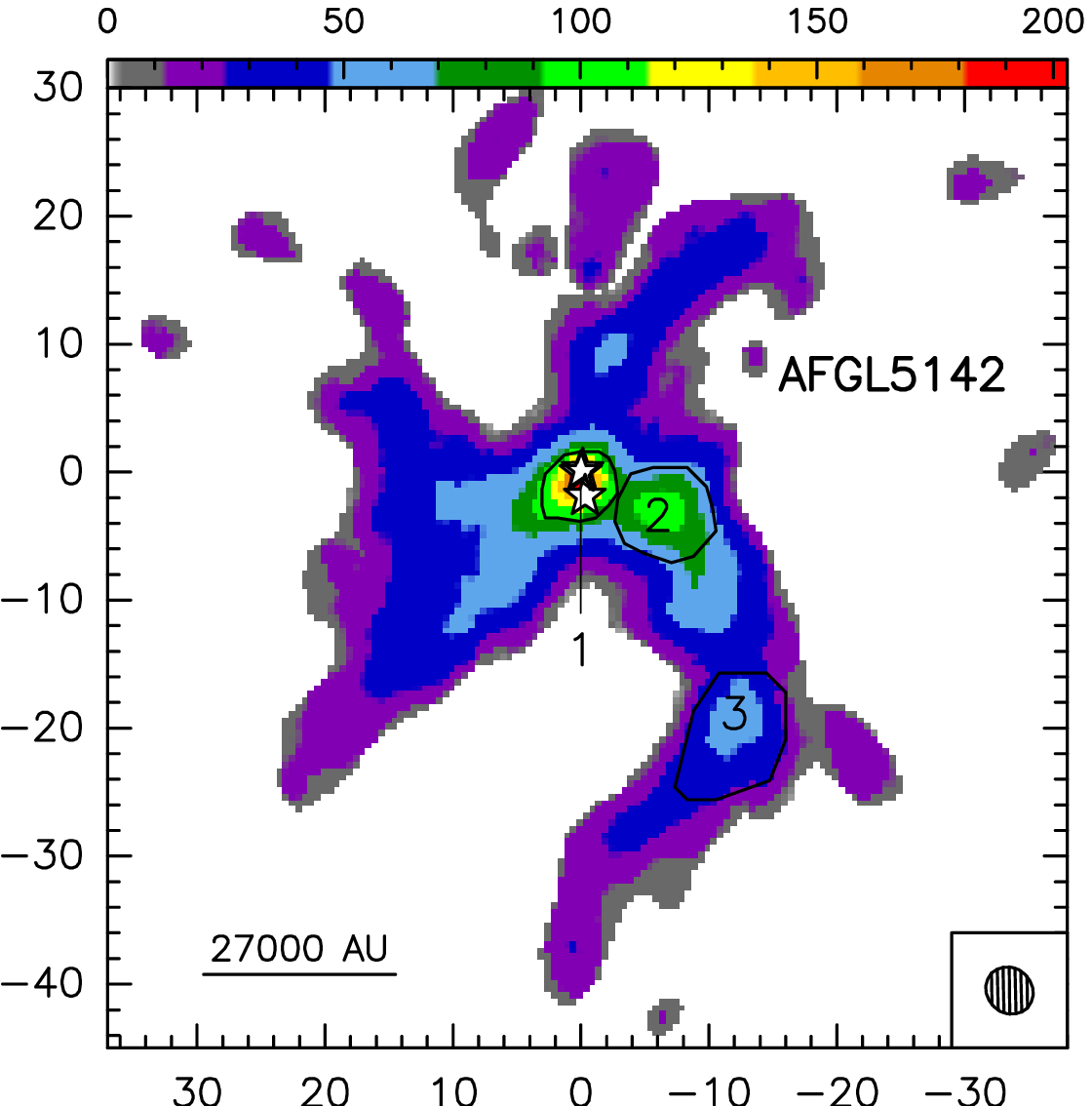, scale=0.48} &
  \epsfig{file=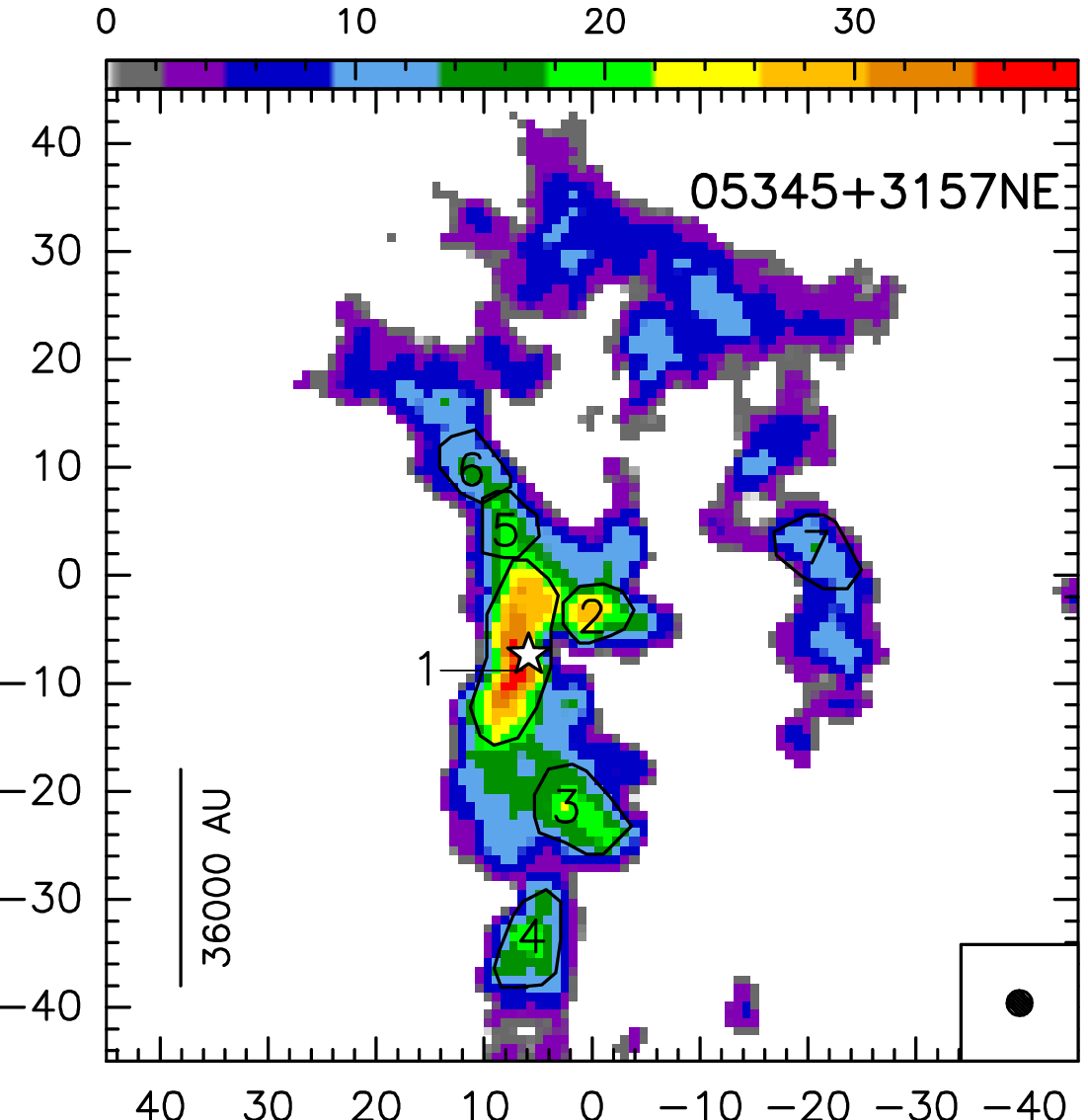, scale=0.48} \\
\vspace{0.5cm}
  \epsfig{file=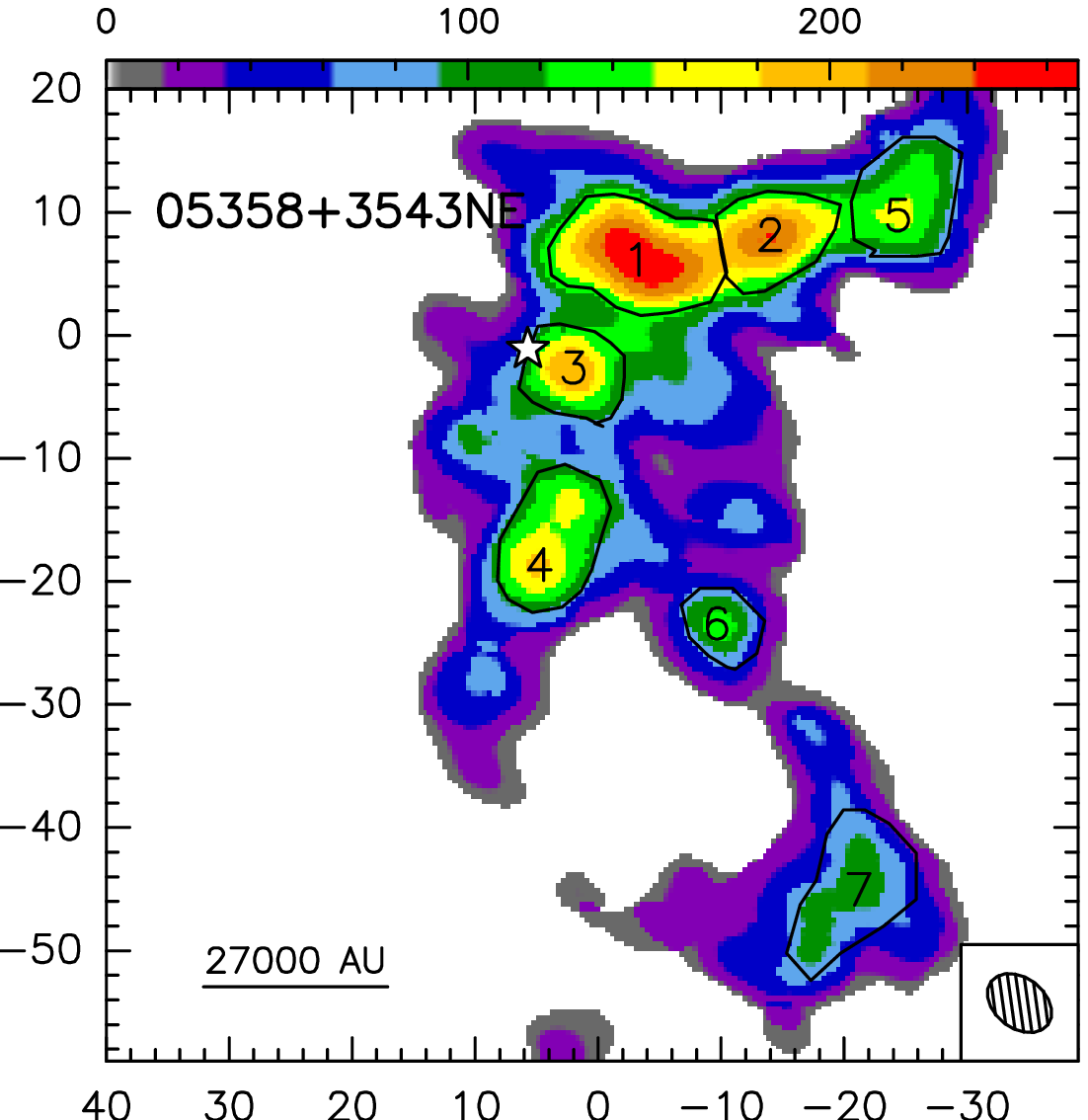, scale=0.48} &
  \epsfig{file=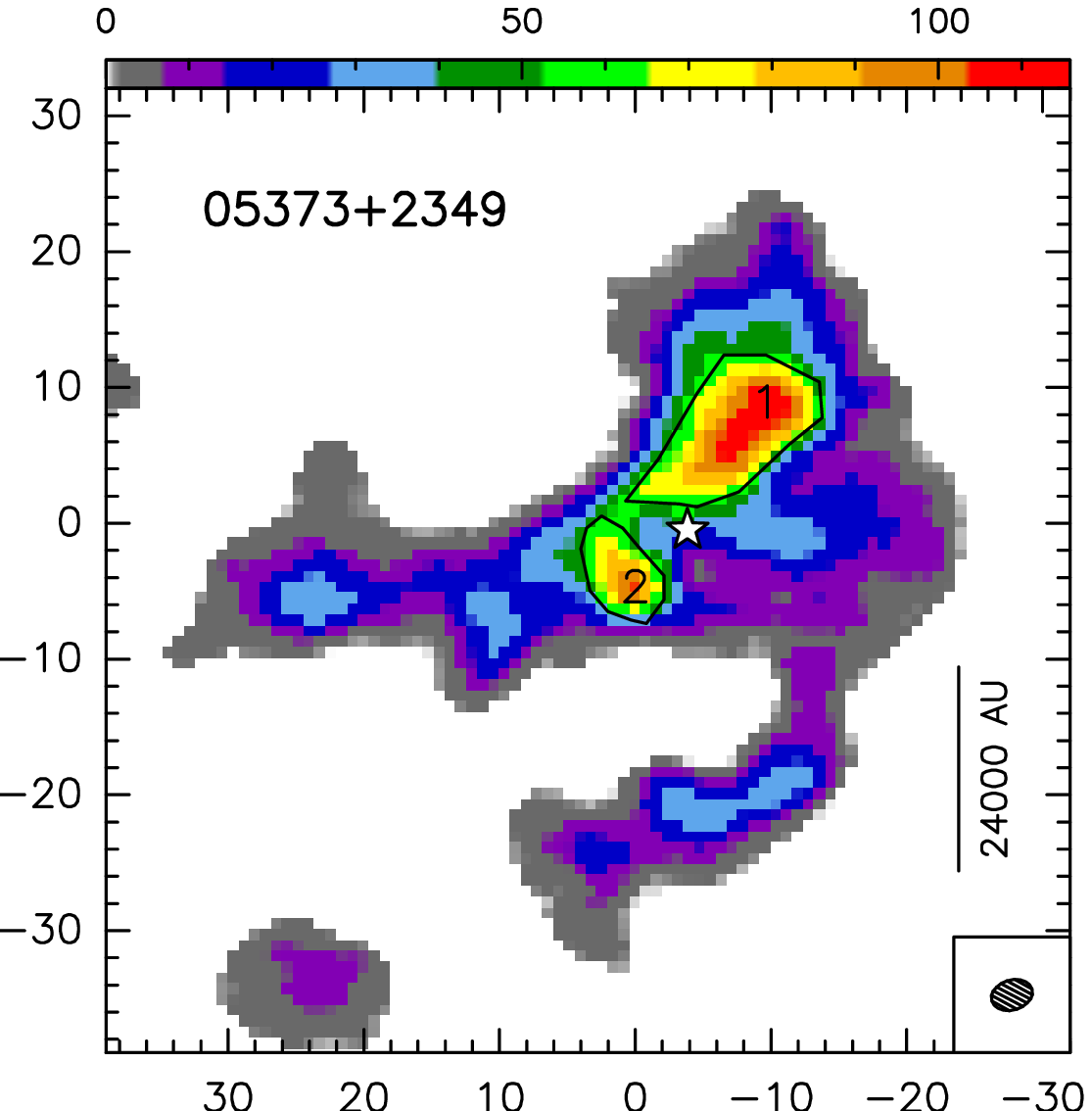, scale=0.48} &
  \epsfig{file=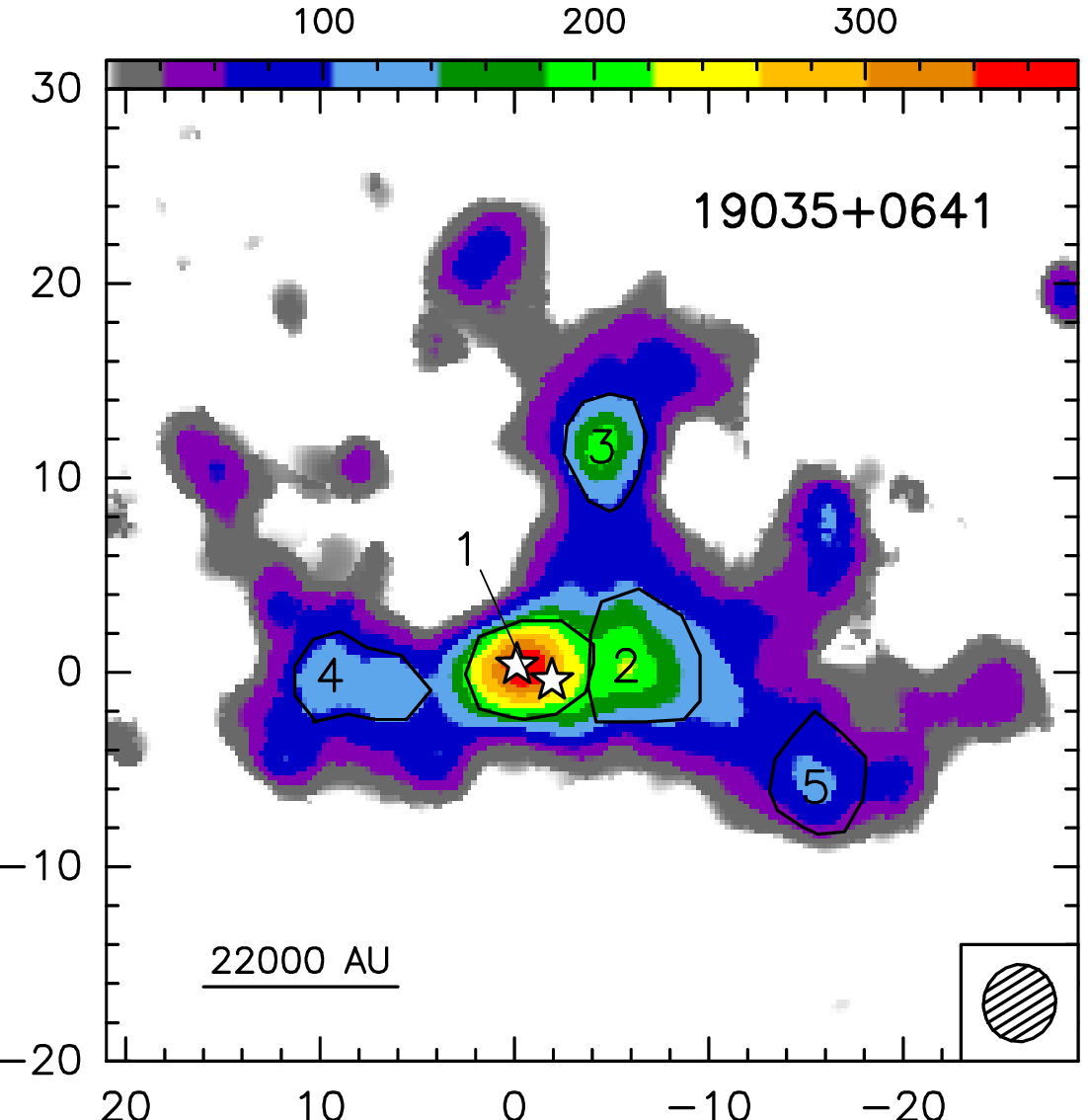, scale=0.48} \\
  \epsfig{file=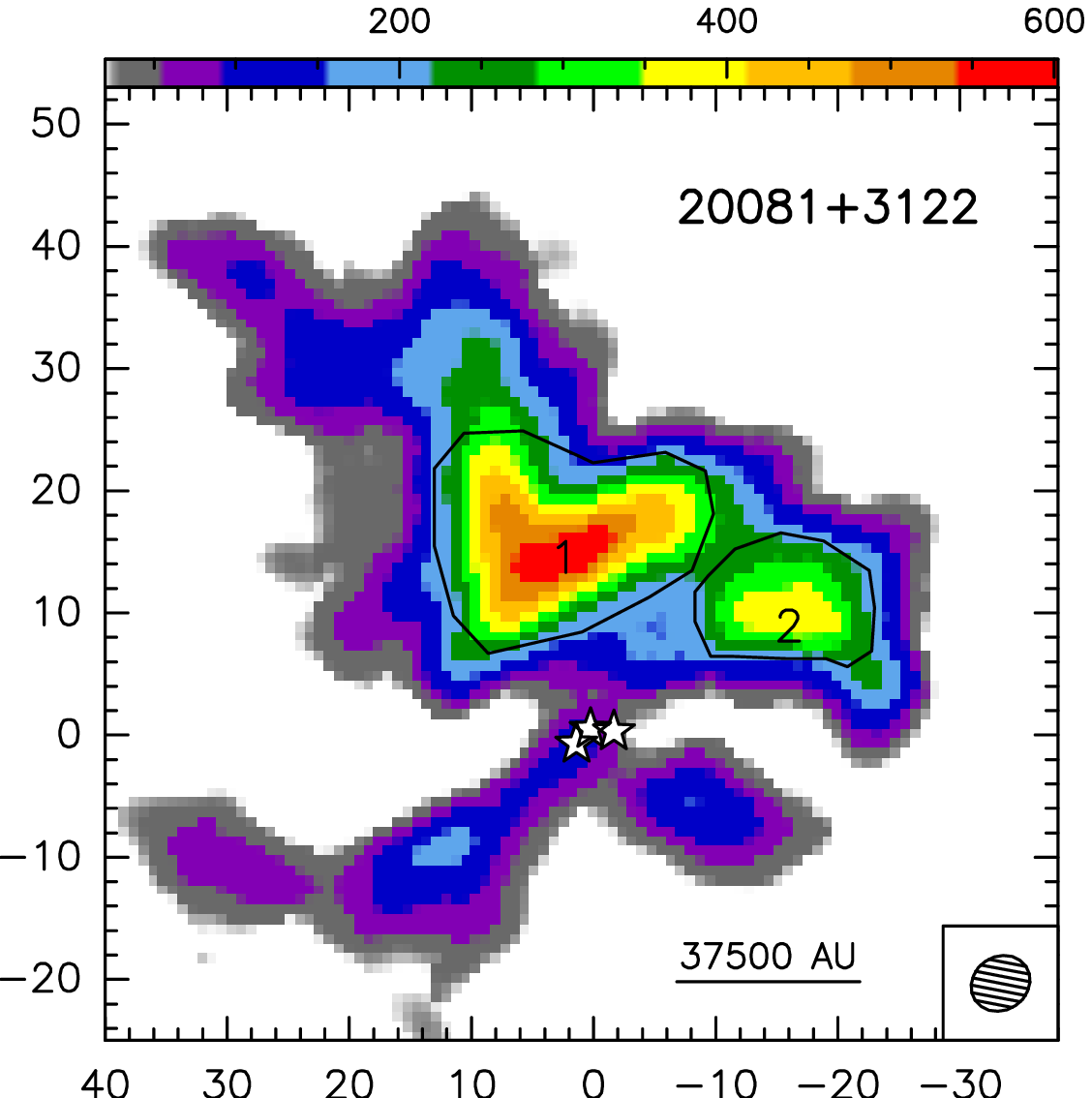, scale=0.48} &
  \epsfig{file=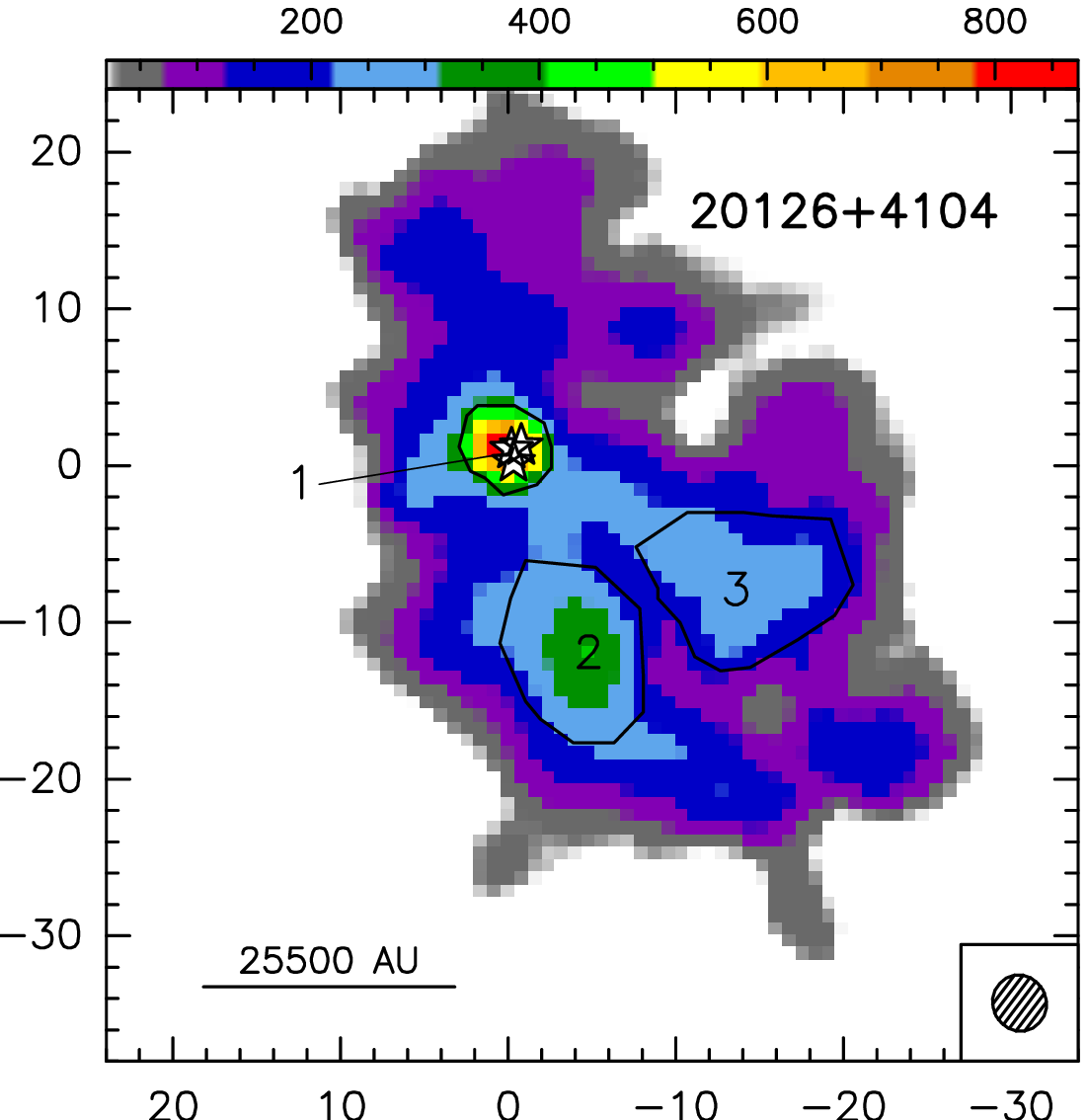, scale=0.48} &
  \epsfig{file=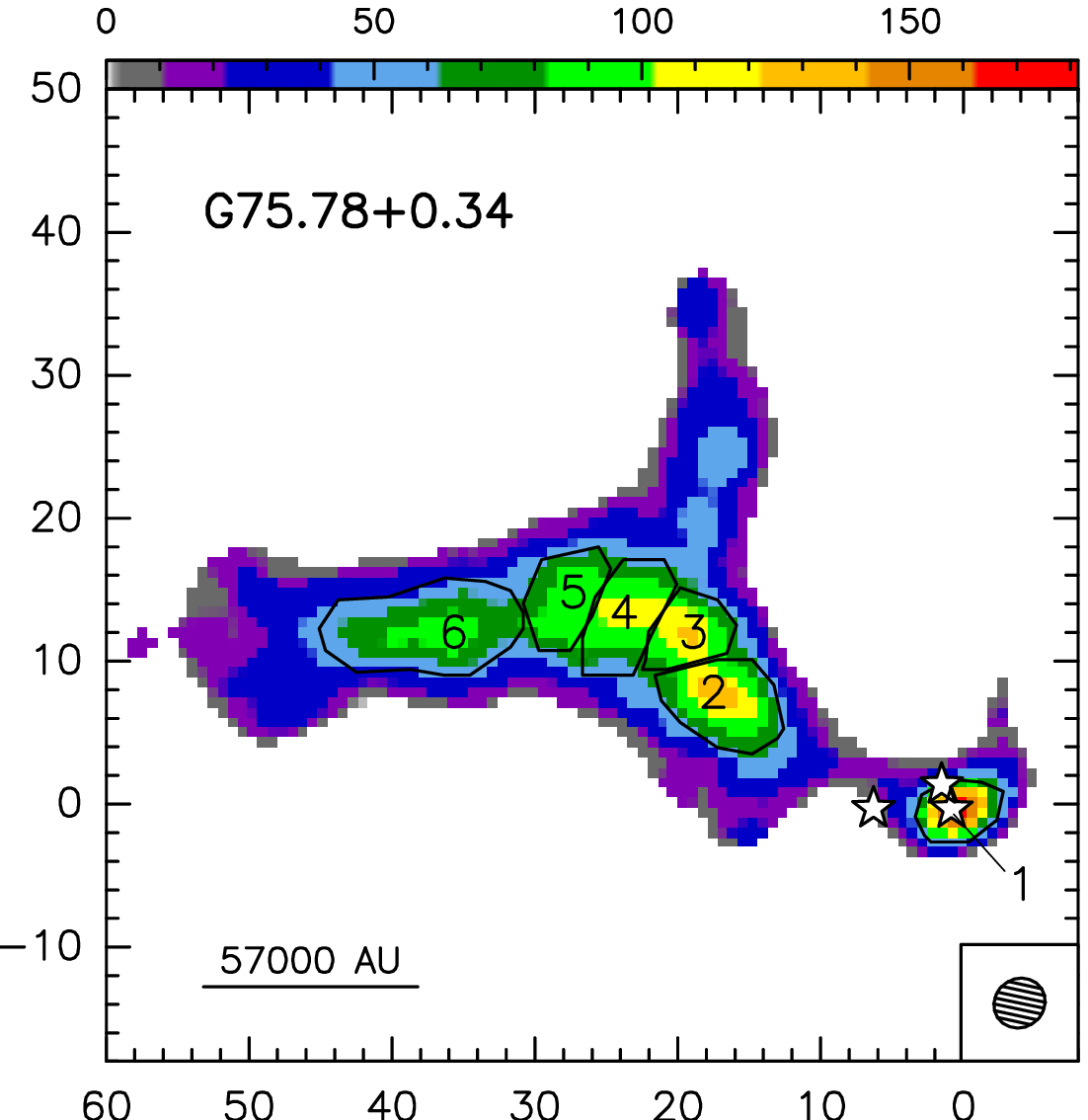, scale=0.48} \\
\end{tabular}
\caption{Colour image: \nh\,(1,1) zero-order moment (integrated intensity) maps in units of mJy~\kms\ (horizontal colour-bar on the top of each panel). Coordinates in the x and y axis are right ascension and declination offsets in arcsec, with the (0,0) corresponding to the phase centre of the map (see Table~\ref{t:nh3sources}). Black thick lines show the polygons for each core listed in Table~\ref{t:nh3cores}, as obtained from the CLUMPFIND analysis (see Sect.~\ref{s:results}). The numbers inside the polygons identify the core (see Table~\ref{t:nh3cores}). The synthesised beams (listed in Table~\ref{t:nh3sources}) are shown in the bottom-right corner of each panel. Star symbols in all panels correspond to centimetre continuum sources.}
\label{f:cores}
\end{figure*}
\begin{figure*}
\centering
\begin{tabular}[b]{c c c}
\vspace{0.5cm}
  \epsfig{file=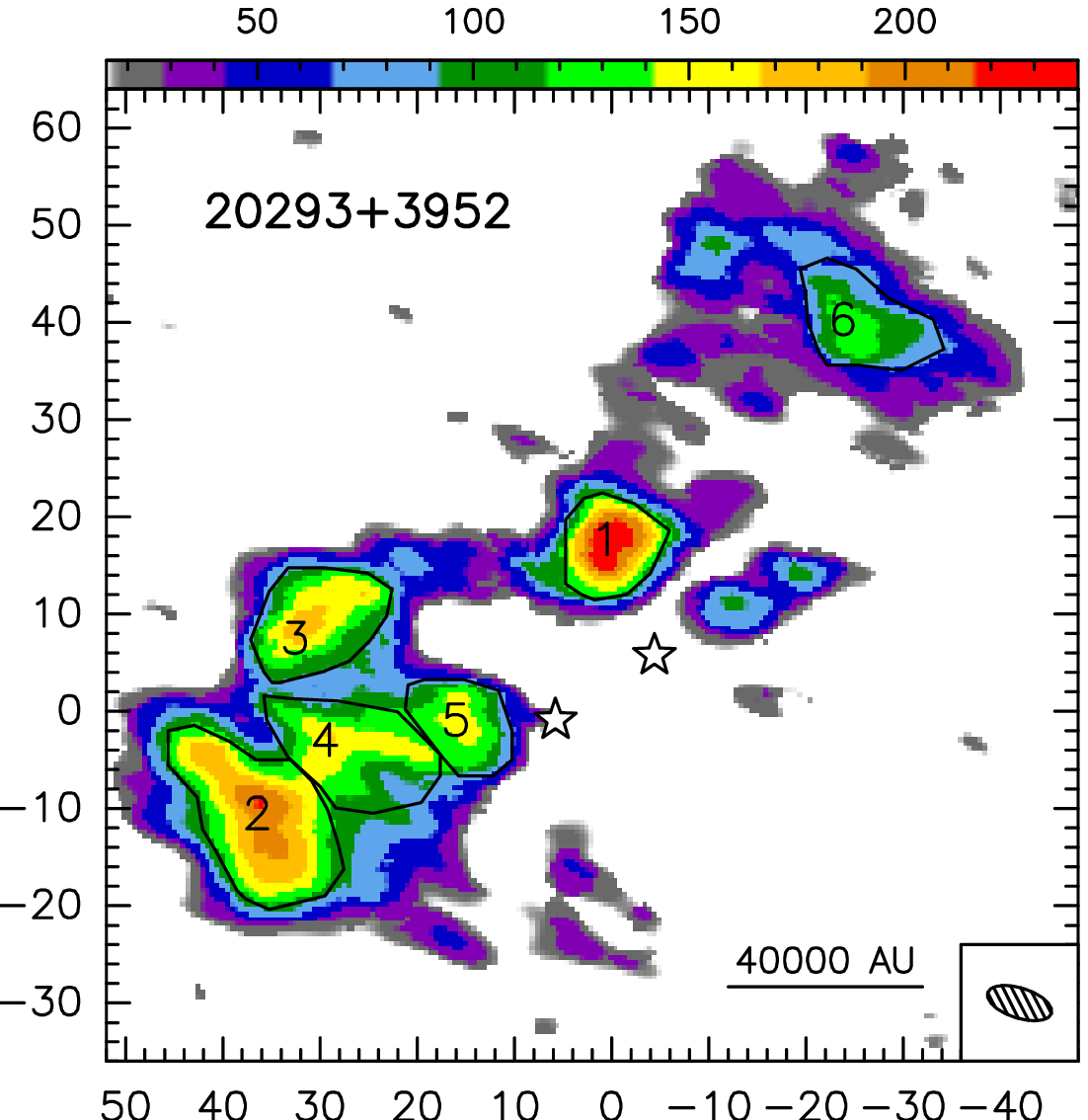, scale=0.48} &
  \epsfig{file=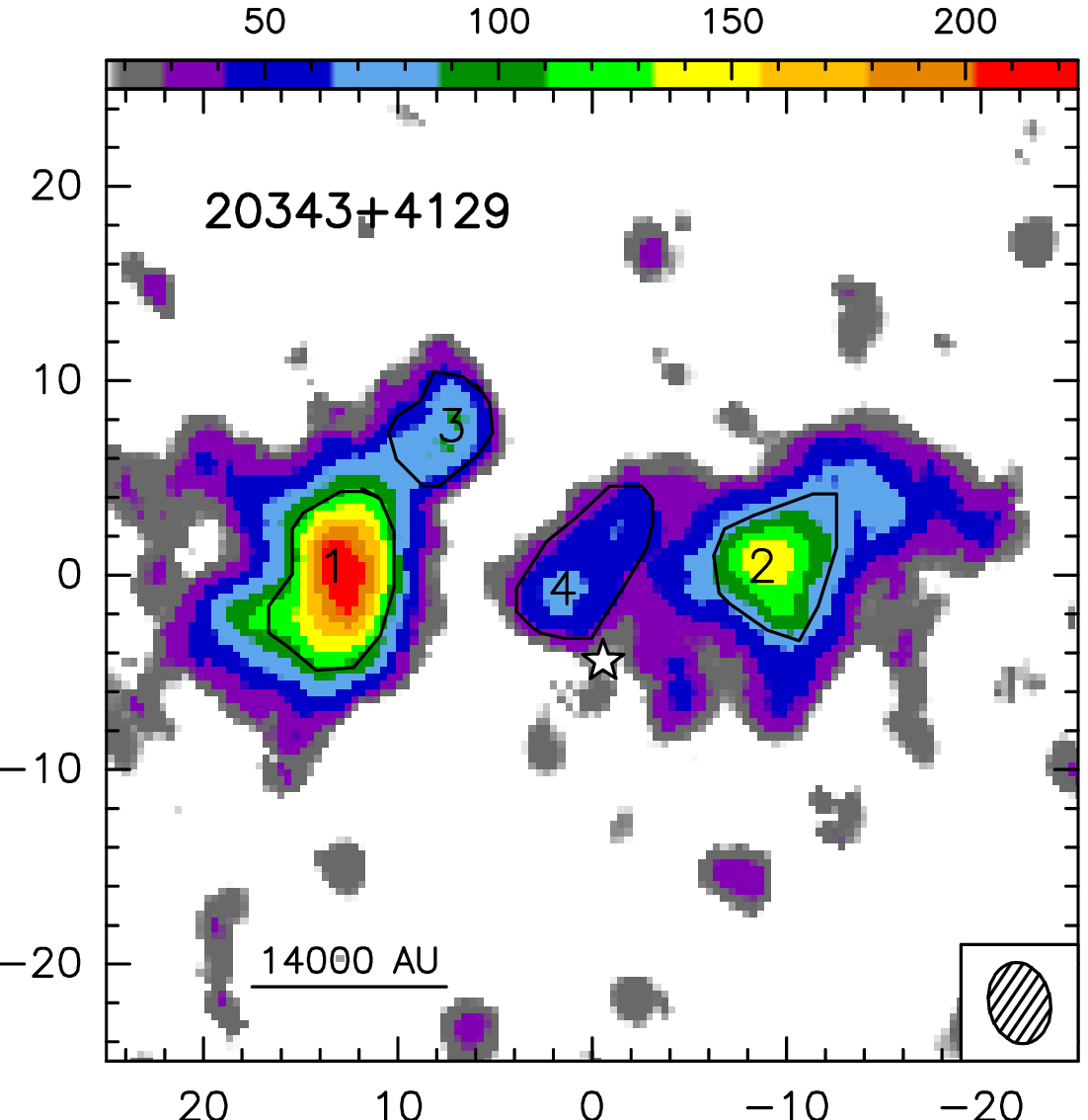, scale=0.48} &
  \epsfig{file=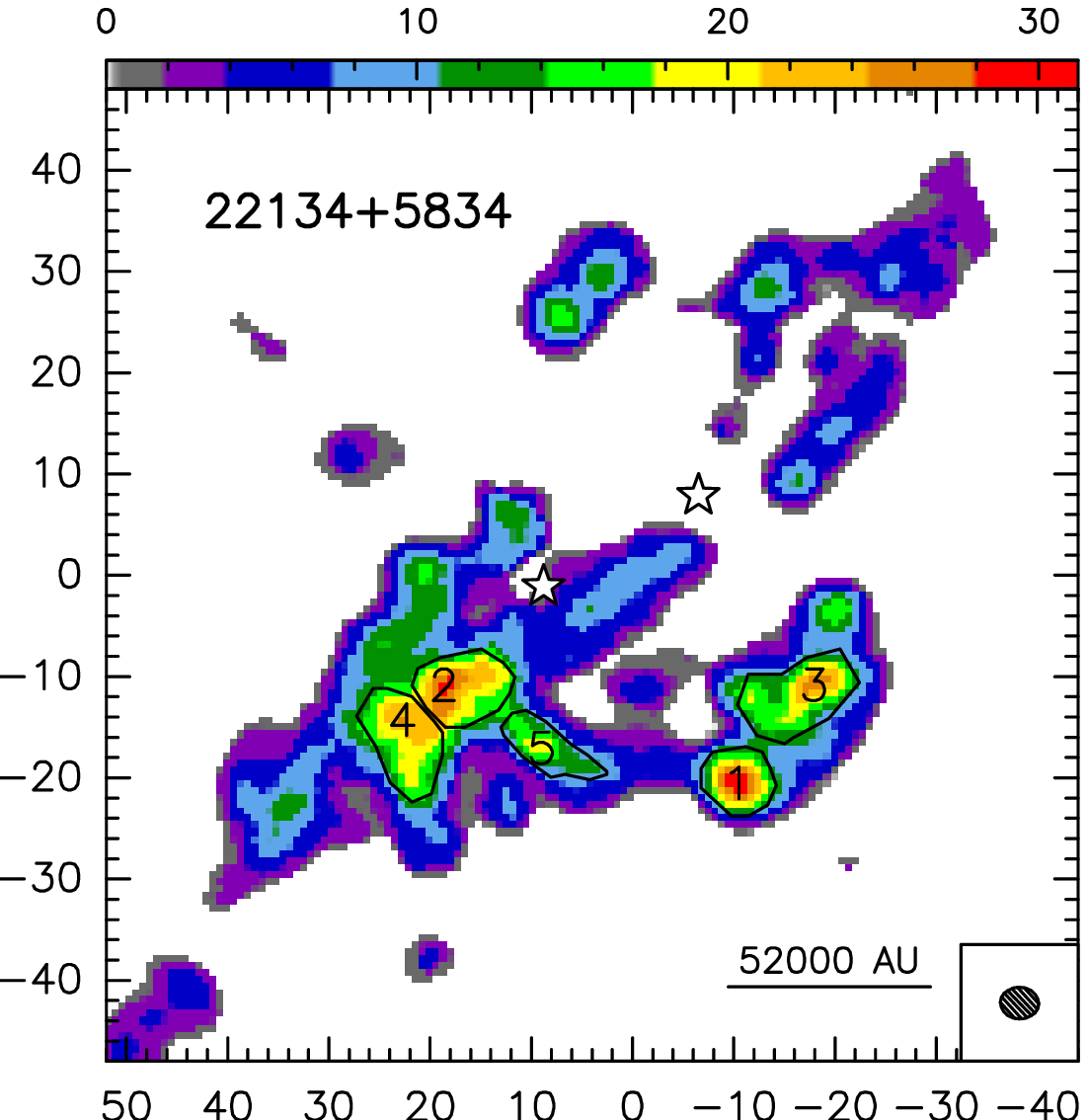, scale=0.48} \\
  \epsfig{file=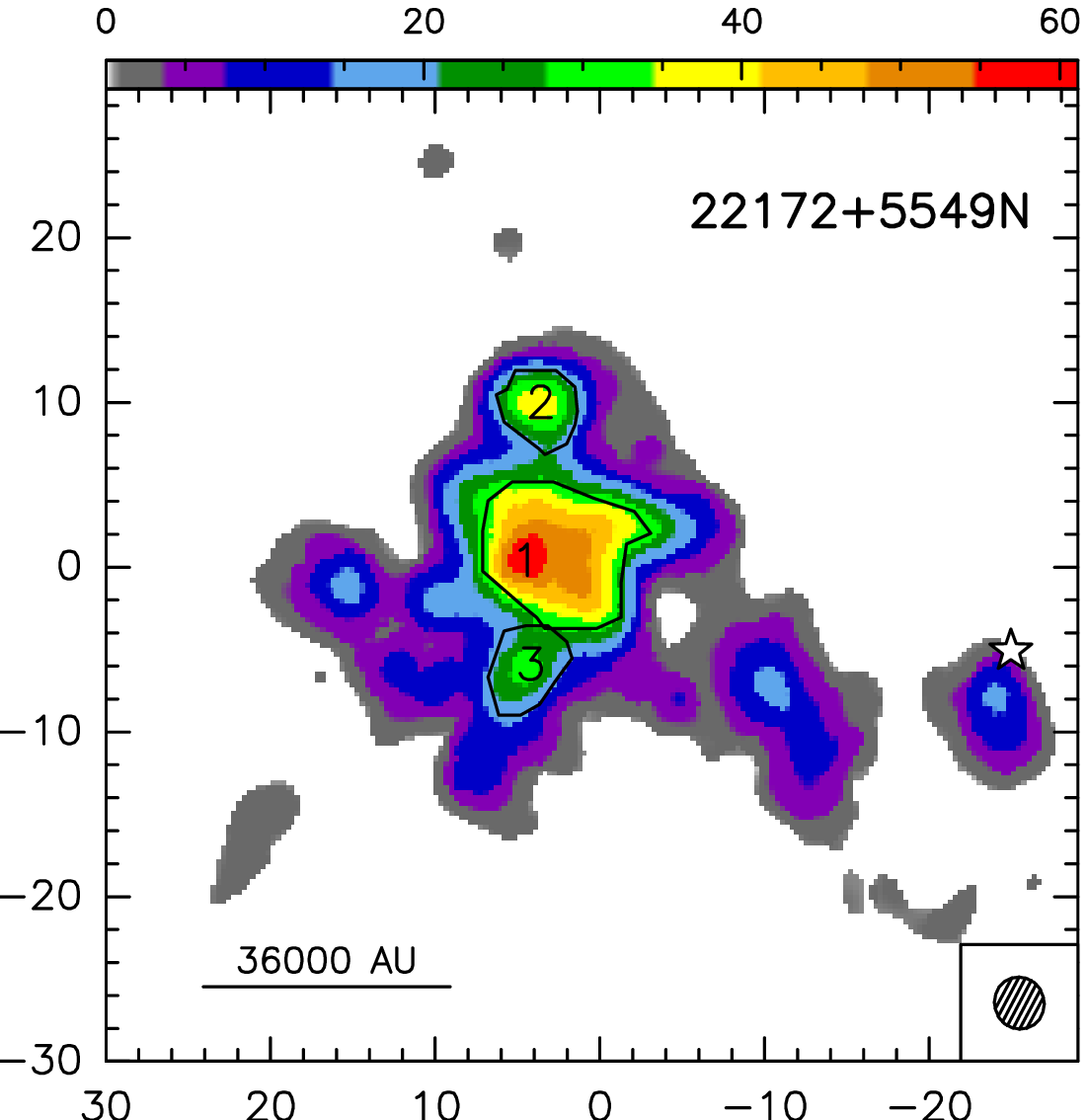, scale=0.48} &
  \epsfig{file=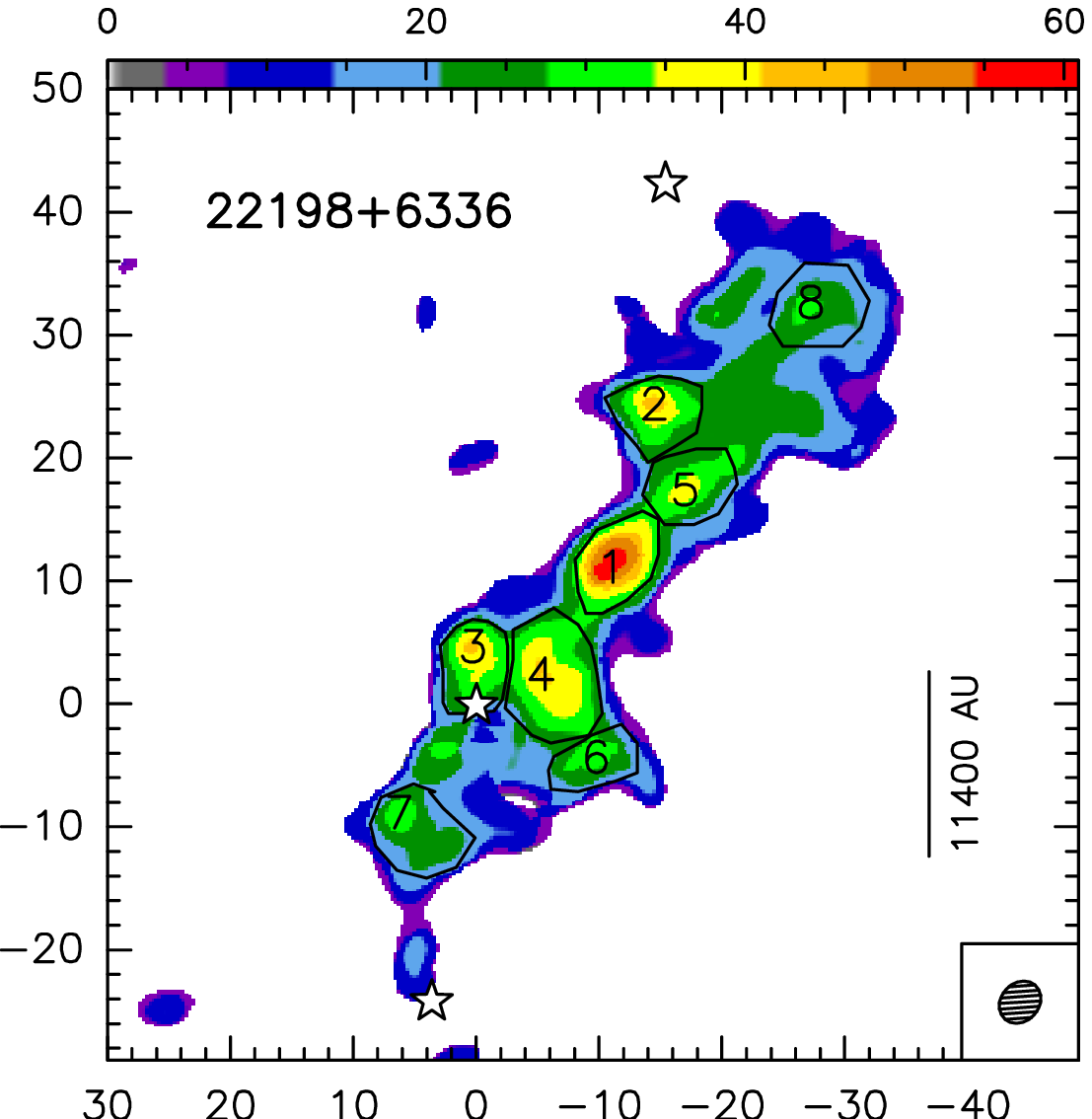, scale=0.48} &
  \epsfig{file=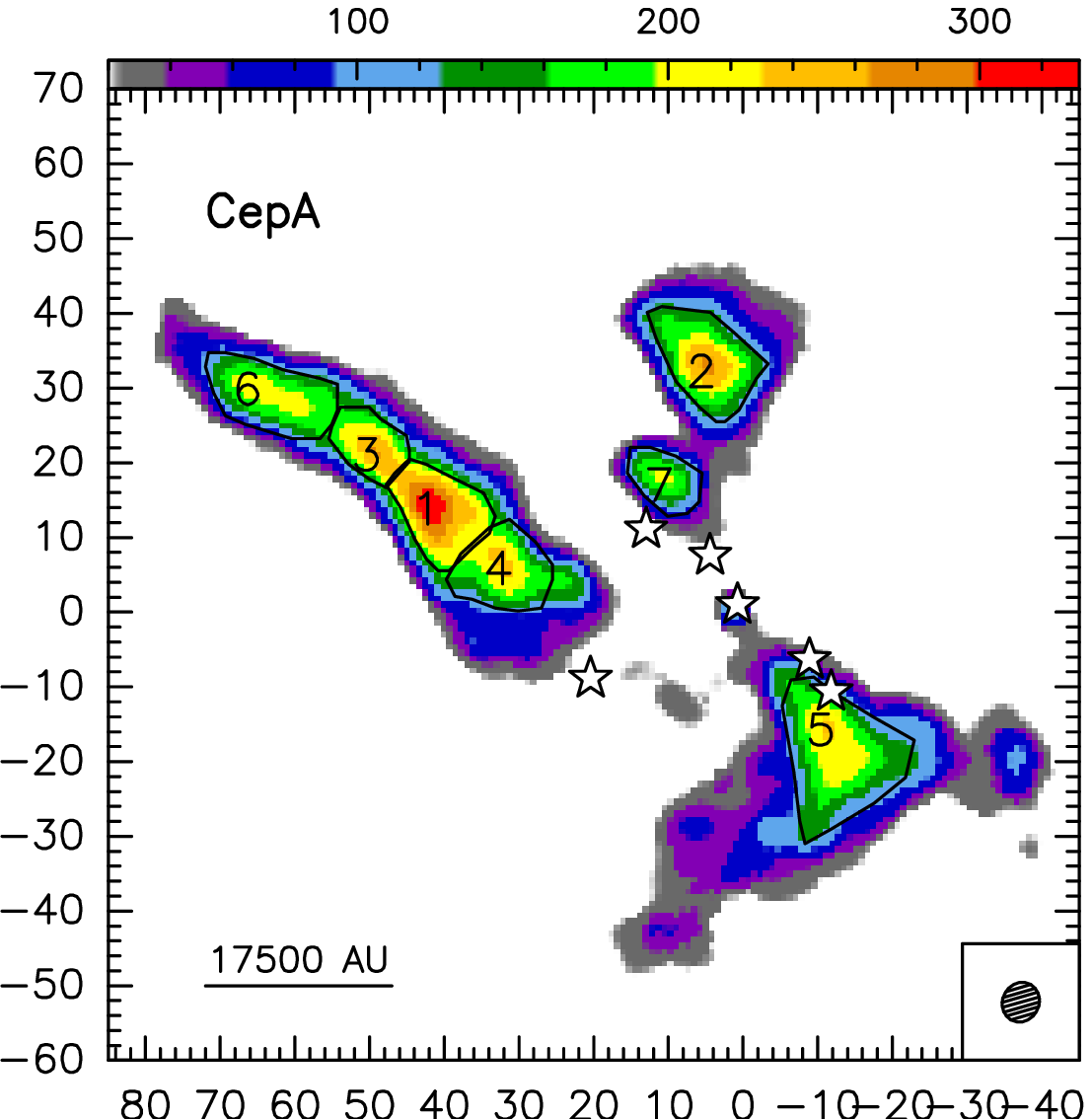, scale=0.48} \\
\end{tabular}
\contcaption{}
\end{figure*}

\section{Observations}\label{s:obs}

The Very Large Array (VLA\footnote{The Very Large Array is operated by the National Radio Astronomy Observatory (NRAO). The NRAO is a facility of the National Science Foundation operated under cooperative agreement by Associated Universities, Inc.}) was used to observe the ammonia ($J$,$K$) = (1,1) and (2,2) inversion transitions toward 15 intermediate/high-mass star-forming regions. These observations are part of different observational projects carried out on different epochs (from 1999 to 2009), and with the array in compact configurations (C, DnC or D). In Table~\ref{t:nh3sources} we list the 15 massive star-forming regions, indicating the name of the VLA project and the coordinates of the phase centre for each region. More technical details of each observational project, such as calibrators and epochs, are provided in Table~\ref{t:nh3obs}. The full-width at half-maximum of the primary beam at the frequency of the observations ($\sim 23.7$~GHz) is approximately 2~arcmin. The spectral setup configuration used in each project (\ie\ bandwidth and spectral resolution) are indicated in Table~\ref{t:nh3obs}. The absolute flux scale was set by observing the quasars 1331+305 (3C286), 0137+331 (3C48) or 0319+415 (3C84), for which we adopted a flux of 2.41~Jy, 1.05~Jy and 16~Jy, respectively. Gain-calibration was done against nearby quasars, which were typically observed at regular time intervals of $\sim 5$--10~minutes before and after each target observation. Flux calibration errors could be of up to 20--30\%. The data reduction followed the VLA standard guidelines for calibration of high-frequency data, using the NRAO package AIPS. Final images were produced with the task IMAGR of AIPS, with the robust parameter of \citet{briggsphd1995} typically set to 5, corresponding to natural weighting. Two sources (AFGL\,5142 and 22198+6336) were observed in different projects with different angular resolutions. For these sources we combined the \emph{uv}-data to improve the sensitivity and \emph{uv}-coverage of the final images. The resulting synthesised beams and rms noise levels (per channel) are indicated in Table~\ref{t:nh3sources}.

\section{Results and Analysis}\label{s:results}

\subsection{Core identification and classification}\label{s:cores}

In Fig.~\ref{f:cores}, we present the zero-order moment (velocity-integrated intensity) maps of the \nh\,(1,1) emission for the 15 massive star-forming regions studied in this work, while in Fig.~\ref{f:mom12}, we present the first-order (intensity weighted mean $v_\mathrm{LSR}$) and second-order (intensity weighted mean velocity linewidth) moment maps. The zero-order moment maps were constructed by integrating the emission of all the hyperfine components, while for the first and second-order moment maps we considered only the main hyperfine component, except for 20126+4104 for which the inner satellite hyperfine component was used (in the zeroth, first and second order moments) to avoid the high opacities in the main line. The integrated intensity maps reveal ammonia emission in all the regions, coming from compact condensations surrounded by faint and more extended emission. In the first-order moment maps (see Fig.~\ref{f:mom12}), we identify large-scale velocity gradients (see \eg\ AFGL\,5142, 20081$+$3122, G75.78$+$0.34, 22172$+$5549N), not only associated with the compact condensations or cores but with the whole ammonia emission, probably revealing the large scale kinematics of the cloud. To extract and identify the cores within each region we ran the two-dimensional CLUMPFIND algorithm \citep{williams1994} on the \nh\,(1,1) zero-order moment map. This algorithm requires two input parameters: the `bottom' level, corresponding to the lowest intensity to be included in the core, and the `step', which determines the intensity increment required to differentiate one core from another core (see \eg\ \citealt{miettinen2012} for further details). In our case, we used `bottom' values of around 20--40\% of the peak intensity of the zero-order moment, and `steps' in the range 5--10\% of the peak intensity (see last columns in Table~\ref{t:nh3sources}). The output parameters defining each core returned by CLUMPFIND are the peak position, peak intensity and the flux density of the core. For each identified core we defined a polygon at the level of the full width at half maximum (FWHM; see black thick lines in Fig.~\ref{f:cores}). In Table~\ref{t:nh3cores} we list the offset position of each core (relative to the phase centre; listed in Table~\ref{t:nh3sources}) and the deconvolved FWHM size. A total of 73 cores were identified.

The studied field of view (VLA primary beam at 1.3~cm) is $\sim 2$~arcmin, or 1.5~pc at a distance of 2.5~kpc. On average, we found of the order of 5--6 cores in each region, with diametres $\sim 0.06$~pc. We classify the identified cores as starless or protostellar (see Table~\ref{t:nh3extra}), based on their infrared properties (from the \emph{Spitzer}/IRAC, \emph{Spitzer}/MIPS or WISE\footnote{The Wide-filed Infrared Survey Explorer (WISE) is described in \citep{wright2010}.} images) and other observational evidence reported in the literature (\eg\ molecular outflows, masers, \hii\ regions, radiojets; see references in Table~\ref{t:nh3types}), summarized in Table~\ref{t:nh3types}. Within the starless cores we consider two groups: `quiescent starless' and `perturbed starless' cores. The former group includes cores appearing dark in the IRAC images, away ($\ga$10\arcsec) from strong IR sources or outflows, with or without millimetre continuum. On the other hand, perturbed starless cores are neither associated with a point infrared source but lie close ($\la$10\arcsec) to a bright IR source and/or along the axis of a molecular outflow, suggesting that these cores might be perturbed\footnote{The differentiation between quiescent and perturbed starless cores was already introduced in \citet{fontani2011}.}. A core is classified as `protostellar' if it is associated with an infrared point source, and with other star formation signposts (\eg\ maser, outflow, centimetre continuum). Some cores were tentatively classified as `protostellar?' if the only hint of star formation is the association with an infrared point source, which could be a background or foreground star not associated with the dense core. These tentatively classified cores are not included in our statistical analysis when comparing the different types of cores, to avoid contamination by misclassified cores. In Fig.~\ref{f:cores}, the stars indicate the position of centimetre continuum sources, likely tracing \hii\ regions or thermal radio jets. In many cases, the most massive source of the region, usually associated with a strong centimetre source, tracing an (ultracompact) \hii\ region is located offset of the core peaks (\eg\ 20081+3122, 20293+3952, 20343+4129).

\begin{landscape}
\begin{table}
\caption{Physical parameters of the ammonia cores identified toward the 15 massive star-forming regions}
\label{t:nh3cores}
\begin{center}
\begin{tabular}{ l c c c c c c c c c c c c c c}
\hline\hline

&
&\multicolumn{2}{c}{Offset\supa}
&Size\supb
&\multicolumn{4}{c}{\nh\,(1,1)\supc}
&
&\multicolumn{3}{c}{\nh\,(2,2)\supc}
&
&
\\
\cline{3-4}\cline{6-9}\cline{11-13}
\texttt{ID}~Region
&\#
&$\Delta x$
&$\Delta y$
&(pc)
&$A\times\tau$
&$v$
&$\Delta v$
&$\tau$\supd
&
&$T_\mathrm{mb}$
&$v$
&$\Delta v$
&$T_\mathrm{rot}$\supe
&$N_\mathrm{NH_\mathrm{3}}$\supe
\\
\hline
01~00117+6412	&1	&$-20.5$		&\phn$-7.0$	&0.085	 &\phn$8.2\pm1.0$  &$-36.2\pm0.1$     &$0.8\pm0.1$  &$1.80\pm0.29$ & &\phn$1.4\pm0.2$  &$-36.1\pm0.1$	  &$1.0\pm0.2$   &$14.6\pm0.9$     &$\phn8.4\pm0.9$\\
02				&2	&$-12.5$		&$-16.5$		&0.046	 &\phn$4.5\pm1.3$  &$-35.5\pm0.1$     &$1.2\pm0.2$  &$1.40\pm0.62$ & &\phn$0.5\pm0.2$  &$-35.6\pm0.6$	  &$2.4\pm1.3$   &$12.3\pm1.5$     &$10.1\pm2.3$\\
03				&3	&$-10.0$		&\phn$-2.5$	&0.039	 &\phn$2.5\pm0.3$  &$-36.4\pm0.1$     &$1.6\pm0.3$  &$<0.26$       & &\phn$0.6\pm0.2$  &$-36.0\pm0.5$	  &$2.2\pm1.0$   &$15.8\pm0.8$     &$\phn2.9\pm0.6$\\
04				&4	&\phn$-8.5$	&$-19.0$		&0.026	 &\phn$3.5\pm1.4$  &$-35.1\pm0.2$     &$1.4\pm0.3$  &$0.77\pm0.82$ & &\phn$0.8\pm0.3$  &$-35.0\pm0.3$	  &$1.0\pm0.6$   &$15.4\pm2.9$     &$\phn6.4\pm2.0$\\
05				&5	&$-23.0$		&$-14.5$		&0.048	 &\phn$8.1\pm2.3$  &$-36.3\pm0.1$     &$0.6\pm0.1$  &$3.45\pm0.97$ & &\phn$0.6\pm0.3$  &$-36.3\pm0.3$	  &$0.8\pm0.6$   &$11.3\pm1.8$     &$13.2\pm3.4$\\
06~AFGL5142		&1	&\phn$+0.0$	&\phn$-1.0$	&0.040	 &$12.3\pm1.7$     &\phn$-2.2\pm0.2$  &$3.8\pm0.2$  &$1.35\pm0.30$ & &\phn$7.1\pm0.6$  &\phn$-4.0\pm0.2$  &$4.6\pm0.5$   &$36.0\pm3.9$     &$41.1\pm6.0$\\
07				&2	&\phn$-6.0$	&\phn$-3.5$	&0.059	 &$12.9\pm1.5$     &\phn$-3.2\pm0.1$  &$1.7\pm0.1$  &$0.69\pm0.25$ & &\phn$9.2\pm0.5$  &\phn$-3.8\pm0.1$  &$1.4\pm0.1$   &$32.5\pm3.1$     &$16.5\pm2.3$\\
08				&3	&$-12.0$		&$-19.0$		&0.073	 &\phn$9.2\pm2.2$  &\phn$-5.0\pm0.1$  &$1.4\pm0.2$  &$1.19\pm0.53$ & &\phn$3.5\pm0.2$  &\phn$-4.5\pm0.1$  &$1.6\pm0.1$   &$21.2\pm3.5$     &$12.4\pm2.7$\\
09~05345+3157NE	&1	&\phn$+6.4$	&\phn$-8.8$	&0.090	 &$11.8\pm0.6$     &$-17.7\pm0.1$     &$1.3\pm0.1$  &$1.64\pm0.13$ & &\phn$2.3\pm0.2$  &$-17.7\pm0.1$	  &$1.7\pm0.1$   &$15.3\pm0.4$     &$16.4\pm0.9$\\
10				&2	&\phn$+0.0$	&\phn$-4.0$	&0.046	 &$10.8\pm0.9$     &$-18.3\pm0.1$     &$1.0\pm0.1$  &$1.06\pm0.17$ & &\phn$3.2\pm0.3$  &$-18.4\pm0.1$	  &$0.9\pm0.1$   &$18.1\pm0.8$     &$\phn9.0\pm0.8$\\
11				&3	&\phn$+2.4$	&$-21.6$		&0.065	 &$17.7\pm1.1$     &$-18.1\pm0.1$     &$0.5\pm0.1$  &$1.90\pm0.16$ & &\phn$2.9\pm0.3$  &$-18.0\pm0.1$	  &$0.5\pm0.1$   &$14.5\pm0.5$     &$10.2\pm0.7$\\
12				&4	&\phn$+5.6$	&$-33.6$		&0.059	 &\phn$8.8\pm0.8$  &$-17.1\pm0.1$     &$1.0\pm0.1$  &$1.27\pm0.21$ & &\phn$2.0\pm0.2$  &$-17.2\pm0.1$	  &$1.1\pm0.2$   &$16.1\pm0.8$     &$10.8\pm0.8$\\
13				&5	&\phn$+8.0$	&\phn$+4.0$	&0.045	 &$17.3\pm1.4$     &$-17.6\pm0.1$     &$0.5\pm0.1$  &$1.95\pm0.19$ & &\phn$2.3\pm0.3$  &$-17.6\pm0.1$	  &$0.8\pm0.1$   &$13.2\pm0.6$     &$10.2\pm0.9$\\
14				&6	&$+11.2$		&\phn$+9.6$	&0.047	 &$17.4\pm1.3$     &$-17.9\pm0.1$     &$0.4\pm0.1$  &$2.53\pm0.23$ & &\phn$1.1\pm0.4$  &$-17.8\pm0.1$	  &$0.6\pm0.2$   &$10.7\pm0.4$     &$12.0\pm1.0$\\
15				&7	&$-20.8$		&\phn$+2.4$	&0.056	 &$10.9\pm1.1$     &$-17.9\pm0.1$     &$0.4\pm0.1$  &$1.48\pm0.23$ & &\phn$2.5\pm0.4$  &$-17.9\pm0.1$	  &$0.4\pm0.1$   &$16.3\pm1.0$     &$\phn5.1\pm0.5$\\
16~05358+3543NE	&1	&\phn$-3.2$	&\phn$+6.0$	&0.089	 &\phn$9.5\pm0.7$  &$-34.6\pm0.1$     &$1.2\pm0.1$  &$<0.14$       & &\phn$5.7\pm0.3$  &$-34.5\pm0.1$	  &$1.4\pm0.1$   &$23.4\pm0.9$     &$\phn7.1\pm0.8$\\
17				&2	&$-14.0$		&\phn$+8.0$	&0.061	 &$14.5\pm1.5$     &$-35.4\pm0.1$     &$0.8\pm0.1$  &$1.14\pm0.22$ & &\phn$4.3\pm0.4$  &$-35.4\pm0.1$	  &$1.3\pm0.1$   &$18.2\pm1.1$     &$\phn9.9\pm1.1$\\
18				&3	&\phn$+2.0$	&\phn$-2.8$	&0.055	 &\phn$5.9\pm0.3$  &$-35.0\pm0.1$     &$1.7\pm0.1$  &$<0.07$       & &\phn$4.4\pm0.4$  &$-35.1\pm0.1$	  &$1.8\pm0.2$   &$26.6\pm1.0$     &$\phn6.2\pm0.6$\\
19				&4	&\phn$+4.8$	&$-18.8$		&0.074	 &\phn$6.7\pm1.2$  &$-35.8\pm0.1$     &$1.3\pm0.1$  &$0.30\pm0.36$ & &\phn$3.8\pm0.3$  &$-35.7\pm0.1$	  &$1.2\pm0.1$   &$23.9\pm2.5$     &$\phn6.5\pm1.2$\\
20				&5	&$-24.4$		&\phn$+9.6$	&0.069	 &$16.5\pm3.0$     &$-36.1\pm0.1$     &$0.5\pm0.1$  &$2.26\pm0.50$ & &\phn$3.6\pm0.7$  &$-36.1\pm0.1$	  &$0.7\pm0.2$   &$16.9\pm2.2$     &$\phn9.5\pm1.7$\\
21				&6	&\phn$-9.6$	&$-23.6$		&0.031	 &$15.6\pm2.6$     &$-36.7\pm0.1$     &$0.5\pm0.1$  &$1.69\pm0.38$ & &\phn$2.1\pm0.4$  &$-36.8\pm0.2$	  &$1.3\pm0.3$   &$13.2\pm1.1$     &$\phn8.7\pm1.4$\\
22				&7	&$-21.2$		&$-45.2$		&0.078	 &\phn$7.6\pm1.3$  &$-37.3\pm0.1$     &$0.9\pm0.1$  &$1.10\pm0.36$ & &\phn$1.3\pm0.4$  &$-37.3\pm0.2$	  &$1.1\pm0.4$   &$14.2\pm1.1$     &$12.9\pm1.2$\\
23~05373+2349	&1	&\phn$-9.6$	&\phn$+8.8$	&0.081	 &$12.3\pm0.5$     &\phn$+2.0\pm0.1$  &$1.0\pm0.1$  &$1.23\pm0.13$ & &\phn$3.4\pm0.1$  &\phn$+2.1\pm0.1$  &$1.4\pm0.1$   &$17.8\pm0.5$     &$10.4\pm0.6$\\
24				&	&			&			&		 &$18.3\pm1.6$     &\phn$+3.2\pm0.1$  &$0.6\pm0.1$  &$3.15\pm0.28$ & &\phn$2.9\pm0.1$  &\phn$+3.3\pm0.1$  &$0.7\pm0.1$   &$14.9\pm1.0$     &$13.2\pm1.3$\\
25				&2	&\phn$+0.0$	&\phn$-4.8$	&0.044	 &$11.1\pm0.6$     &\phn$+1.5\pm0.1$  &$1.2\pm0.1$  &$1.16\pm0.16$ & &\phn$3.4\pm0.1$  &\phn$+2.1\pm0.1$  &$1.4\pm0.1$   &$18.7\pm0.6$     &$11.2\pm0.6$\\
26				&	&			&			&		 &\phn$9.7\pm1.5$  &\phn$+2.6\pm0.1$  &$0.5\pm0.1$  &$0.57\pm0.30$ & &\phn$4.2\pm0.2$  &\phn$+2.5\pm0.1$  &$0.5\pm0.1$   &$21.1\pm1.9$     &$\phn3.4\pm0.6$\\
27~19035+0641	&1	&\phn$-0.5$	&\phn$+0.3$	&0.044	 &\phn$5.9\pm1.1$  &$+32.4\pm0.2$     &$4.4\pm0.3$  &$0.61\pm0.36$ & &\phn$3.3\pm0.2$  &$+33.0\pm0.2$	  &$5.0\pm0.4$   &$25.0\pm3.1$     &$21.0\pm3.8$\\
28				&2	&\phn$-5.8$	&\phn$+0.3$	&0.050	 &\phn$4.7\pm0.3$  &$+32.9\pm0.1$     &$3.1\pm0.2$  &$<0.15$       & &\phn$3.4\pm0.2$  &$+33.1\pm0.1$	  &$3.0\pm0.2$   &$26.0\pm1.0$     &$\phn9.2\pm0.9$\\
29				&3	&\phn$-4.5$	&$+11.5$		&0.028	 &\phn$5.1\pm0.9$  &$+34.2\pm0.1$     &$2.7\pm0.2$  &$0.30\pm0.33$ & &\phn$1.4\pm0.3$  &$+34.2\pm0.3$	  &$2.3\pm0.4$   &$16.4\pm1.2$     &$11.7\pm2.0$\\
30				&4	&\phn$+9.5$	&\phn$-0.3$	&0.036	 &\phn$4.0\pm0.8$  &$+32.2\pm0.2$     &$2.8\pm0.3$  &$0.33\pm0.43$ & &\phn$1.3\pm0.2$  &$+32.3\pm0.3$	  &$2.8\pm0.5$   &$17.9\pm1.7$     &$\phn9.1\pm1.8$\\
31				&5	&$-15.5$		&\phn$-6.0$	&0.034	 &\phn$6.4\pm1.3$  &$+31.9\pm0.1$     &$1.6\pm0.2$  &$1.52\pm0.46$ & &\phn$1.2\pm0.3$  &$+31.7\pm0.2$	  &$1.3\pm0.4$   &$15.0\pm1.7$     &$13.2\pm2.3$\\
32~20081+3122	&1	&\phn$+1.6$	&$+15.2$		&0.232	 &$15.6\pm1.7$     &$+10.5\pm0.1$     &$2.1\pm0.1$  &$2.10\pm0.28$ & &\phn$4.0\pm0.3$  &$+10.4\pm0.1$	  &$2.7\pm0.2$   &$18.2\pm1.5$     &$31.1\pm3.4$\\
33				&2	&$-16.8$		&\phn$+9.6$	&0.140	 &\phn$7.6\pm0.9$  &$+11.9\pm0.1$     &$1.2\pm0.1$  &$0.89\pm0.36$ & &\phn$2.4\pm0.2$  &$+11.9\pm0.1$	  &$1.4\pm0.2$   &$18.4\pm1.2$     &$\phn8.7\pm1.0$\\
34				&	&			&			&		 &\phn$6.4\pm0.9$  &$+13.6\pm0.1$     &$1.3\pm0.1$  &$0.23\pm0.30$ & &\phn$3.0\pm0.2$  &$+13.5\pm0.1$	  &$1.6\pm0.2$   &$20.9\pm1.6$     &$\phn6.5\pm1.0$\\
35~20126+4104	&1	&\phn$+0.0$	&\phn$+0.8$	&0.033	 &$21.9\pm2.4$     &\phn$-4.3\pm0.2$  &$4.7\pm0.2$  &$2.09\pm0.28$ & &\phn$8.4\pm0.6$  &\phn$-4.1\pm0.2$  &$5.5\pm0.5$   &$27.7\pm2.5$     &$81.5\pm9.6$\\
36				&2	&\phn$-4.8$	&$-12.0$		&0.076	 &$85.7\pm5.0$     &\phn$-4.5\pm0.1$  &$0.8\pm0.1$  &$5.40\pm0.34$ & &$11.8\pm0.4$     &\phn$-4.5\pm0.1$  &$1.2\pm0.1$   &$15.7\pm1.0$     &$58.8\pm5.0$\\
37				&3	&$-13.6$		&\phn$-8.0$	&0.084	 &$52.0\pm4.0$     &\phn$-4.5\pm0.1$  &$0.9\pm0.1$  &$3.89\pm0.26$ & &\phn$9.3\pm0.3$  &\phn$-4.5\pm0.1$  &$1.2\pm0.1$   &$17.0\pm1.3$     &$39.3\pm3.8$\\
38~G75.78+0.34	&1	&\phn$+0.7$	&\phn$-0.7$	&0.062	 &\phn$5.1\pm0.6$  &\phn$+0.2\pm0.3$  &$4.6\pm0.6$  &$<0.22$       & &\phn$5.6\pm0.6$  &\phn$+0.4\pm0.3$  &$4.5\pm0.6$   &$35.2\pm3.3$     &$15.7\pm2.8$\\
39				&2	&$+17.5$		&\phn$+7.7$	&0.113	 &\phn$5.2\pm0.4$  &\phn$+0.4\pm0.2$  &$3.0\pm0.3$  &$<0.11$       & &\phn$3.6\pm0.4$  &\phn$+0.4\pm0.2$  &$2.8\pm0.3$   &$25.4\pm1.2$     &$10.5\pm1.2$\\
40				&3	&$+18.9$		&$+11.9$		&0.077	 &\phn$6.3\pm0.8$  &\phn$+1.1\pm0.2$  &$2.6\pm0.4$  &$<0.27$       & &\phn$4.2\pm0.4$  &\phn$+0.9\pm0.2$  &$2.8\pm0.3$   &$25.1\pm1.9$     &$11.1\pm1.9$\\
41				&4	&$+23.8$		&$+13.3$		&0.102	 &\phn$5.6\pm0.3$  &\phn$+1.5\pm0.1$  &$1.3\pm0.2$  &$0.22\pm0.03$ & &\phn$3.3\pm0.3$  &\phn$+1.5\pm0.2$  &$1.7\pm0.3$   &$24.1\pm0.7$     &$\phn5.8\pm0.6$\\
42				&	&			&			&		 &\phn$5.7\pm0.7$  &\phn$-0.2\pm0.1$  &$1.3\pm0.2$  &$<0.20$       & &\phn$2.9\pm0.3$  &\phn$-0.5\pm0.2$  &$1.8\pm0.3$   &$21.5\pm1.3$     &$\phn5.4\pm0.9$\\
43				&5	&$+27.3$		&$+14.7$		&0.080	 &\phn$6.3\pm0.9$  &\phn$+0.3\pm0.1$  &$2.4\pm0.2$  &$0.14\pm0.37$ & &\phn$3.2\pm0.4$  &\phn$+0.3\pm0.2$  &$3.0\pm0.4$   &$21.8\pm1.6$     &$12.4\pm1.7$\\
44				&6	&$+35.7$		&$+11.9$		&0.162	 &\phn$8.6\pm1.2$  &\phn$-0.2\pm0.1$  &$1.4\pm0.1$  &$0.17\pm0.36$ & &\phn$4.5\pm0.4$  &\phn$-0.2\pm0.1$  &$1.3\pm0.1$   &$22.3\pm1.8$     &$10.8\pm1.4$\\
\hline
\end{tabular}
\end{center}
\end{table}
\end{landscape}
\begin{landscape}
\begin{table}
\contcaption{}
\begin{center}
\begin{tabular}{l c c c c c c c c c c c c c c}
\hline\hline
&
&\multicolumn{2}{c}{Offset\supa}
&Size\supb
&\multicolumn{4}{c}{\nh\,(1,1)\supc}
&
&\multicolumn{3}{c}{\nh\,(2,2)\supc}
&
&
\\
\cline{3-4}\cline{6-9}\cline{11-13}
\texttt{ID}~Region
&\#
&$\Delta x$
&$\Delta y$
&(pc)
&$A\times\tau$
&$v$
&$\Delta v$
&$\tau$\supd
&
&$T_\mathrm{mb}$
&$v$
&$\Delta v$
&$T_\mathrm{rot}$\supe
&$N_\mathrm{NH_\mathrm{3}}$\supe
\\
\hline
45~20293+3952	&1	&\phn$+0.0$	&$+18.0$		&0.089	 &$16.7\pm4.4$     &\phn$+7.3\pm0.1$  &$1.1\pm0.1$  &$3.33\pm0.88$ & &\phn$3.0\pm0.6$  &\phn$+7.4\pm0.2$  &$1.3\pm0.3$   &$16.3\pm3.6$     &$20.4\pm5.5$\\
46				&2	&$+36.0$		&$-10.0$		&0.148	 &$23.1\pm3.9$     &\phn$+6.0\pm0.1$  &$0.9\pm0.1$  &$4.84\pm0.72$ & &\phn$2.6\pm0.3$  &\phn$+6.1\pm0.1$  &$1.6\pm0.2$   &$13.4\pm2.0$     &$35.4\pm5.8$\\
47				&3	&$+32.0$		&\phn$+8.0$	&0.112	 &\phn$6.4\pm1.0$  &\phn$+5.9\pm0.1$  &$2.1\pm0.2$  &$1.63\pm0.37$ & &\phn$2.0\pm0.2$  &\phn$+6.0\pm0.2$  &$2.9\pm0.4$   &$20.1\pm2.2$     &$18.7\pm2.1$\\
48				&4	&$+29.0$		&\phn$-2.5$	&0.125	 &$13.3\pm2.3$     &\phn$+5.5\pm0.1$  &$1.0\pm0.1$  &$2.87\pm0.55$ & &\phn$2.4\pm0.3$  &\phn$+5.6\pm0.1$  &$1.8\pm0.2$   &$15.6\pm2.0$     &$18.5\pm2.8$\\
49				&5	&$+15.5$		&\phn$-0.5$	&0.083	 &$12.8\pm2.8$     &\phn$+5.3\pm0.1$  &$1.0\pm0.1$  &$2.80\pm0.63$ & &\phn$2.7\pm0.3$  &\phn$+5.4\pm0.1$  &$1.6\pm0.2$   &$17.5\pm3.2$     &$15.0\pm3.2$\\
50				&6	&$-24.5$		&$+40.5$		&0.103	 &\phn$8.5\pm1.8$  &\phn$+4.9\pm0.1$  &$1.3\pm0.1$  &$2.18\pm0.54$ & &\phn$1.4\pm0.2$  &\phn$+4.9\pm0.2$  &$1.9\pm0.3$   &$14.8\pm1.9$     &$23.9\pm2.8$\\
51~20343+4129	&1	&$+12.9$		&\phn$+0.8$	&0.044	 &$11.6\pm2.3$     &$+11.8\pm0.1$     &$1.9\pm0.2$  &$2.04\pm0.52$ & &\phn$2.6\pm0.6$  &$+11.9\pm0.3$	  &$2.5\pm0.6$   &$16.7\pm2.3$     &$23.9\pm4.6$\\
52				&2	&\phn$-9.1$	&\phn$-0.8$	&0.035	 &\phn$6.7\pm0.9$  &$+10.9\pm0.1$     &$1.6\pm0.3$  &$<0.24$       & &\phn$2.5\pm0.7$  &$+10.9\pm0.2$	  &$0.9\pm0.3$   &$18.5\pm1.2$     &$\phn7.1\pm1.6$\\
53				&3	&\phn$+6.8$	&\phn$+8.0$	&0.023	 &\phn$4.6\pm1.1$  &$+13.0\pm0.2$     &$1.3\pm0.4$  &$<0.25$       & &\phn$2.2\pm1.7$  &$+12.4\pm0.4$	  &$0.4\pm0.3$   &$20.8\pm2.5$     &$\phn3.8\pm1.5$\\
54				&4	&\phn$+1.1$	&\phn$-0.4$	&0.036	 &\phn$6.1\pm2.0$  &$+10.2\pm0.1$     &$1.1\pm0.2$  &$0.60\pm0.67$ & &\phn$2.1\pm0.5$  &$+10.2\pm0.2$	  &$1.0\pm0.3$   &$18.8\pm3.3$     &$\phn5.3\pm1.8$\\
55~22134+5834	&1	&$-10.4$		&$-20.7$		&0.071	 &\phn$9.7\pm1.8$  &$-19.2\pm0.1$     &$0.7\pm0.1$  &$3.15\pm0.57$ & &\phn$1.2\pm0.2$  &$-19.2\pm0.1$	  &$0.8\pm0.2$   &$13.1\pm1.6$     &$12.2\pm2.1$\\
56				&2	&$+18.6$		&$-11.0$		&0.094	 &\phn$5.1\pm0.5$  &$-18.7\pm0.1$     &$1.0\pm0.1$  &$1.21\pm0.22$ & &\phn$2.1\pm0.2$  &$-18.7\pm0.1$	  &$1.0\pm0.1$   &$22.7\pm1.7$     &$\phn5.7\pm0.5$\\
57				&3	&$-17.9$		&$-11.0$		&0.107	 &\phn$7.4\pm0.6$  &$-18.7\pm0.1$     &$0.6\pm0.1$  &$1.94\pm0.22$ & &\phn$1.1\pm0.2$  &$-18.7\pm0.1$	  &$0.8\pm0.2$   &$13.9\pm0.7$     &$\phn7.1\pm0.6$\\
58				&4	&$+22.8$		&$-14.5$		&0.100	 &\phn$3.2\pm0.6$  &$-18.4\pm0.1$     &$1.2\pm0.1$  &$0.49\pm0.35$ & &\phn$1.0\pm0.1$  &$-18.4\pm0.1$	  &$1.2\pm0.2$   &$17.3\pm1.6$     &$\phn4.1\pm0.6$\\
59				&5	&\phn$+9.0$	&$-17.3$		&0.068	 &\phn$3.5\pm0.7$  &$-18.7\pm0.1$     &$0.9\pm0.1$  &$0.92\pm0.41$ & &\phn$1.2\pm0.2$  &$-18.9\pm0.1$	  &$1.0\pm0.2$   &$19.7\pm2.3$     &$\phn3.6\pm0.5$\\
60~22172+5549N	&1	&\phn$+4.5$	&\phn$+0.3$	&0.099	 &\phn$4.2\pm0.6$  &$-43.2\pm0.1$     &$1.4\pm0.1$  &$0.25\pm0.26$ & &\phn$1.8\pm0.2$  &$-43.3\pm0.1$	  &$1.5\pm0.2$   &$20.3\pm1.4$     &$\phn4.5\pm0.6$\\
61				&2	&\phn$+3.6$	&\phn$+9.9$	&0.040	 &\phn$5.6\pm1.5$  &$-43.9\pm0.1$     &$0.9\pm0.1$  &$1.06\pm0.57$ & &\phn$1.3\pm1.7$  &$-44.0\pm2.0$	  &$0.8\pm3.0$   &$16.0\pm2.2$     &$\phn5.3\pm1.3$\\
62				&3	&\phn$+4.2$	&\phn$-6.0$	&0.039	 &\phn$3.2\pm0.8$  &$-43.6\pm0.1$     &$1.8\pm0.2$  &$1.31\pm0.54$ & &\phn$0.9\pm0.6$  &$-43.5\pm1.0$	  &$2.3\pm1.0$   &$17.5\pm2.6$     &$\phn8.2\pm1.5$\\
63~22198+6336	&1	&$-11.0$		&$+11.0$		&0.023	 &$11.2\pm2.2$     &$-10.6\pm0.1$     &$1.0\pm0.1$  &$1.38\pm0.43$ & &\phn$1.5\pm0.6$  &$-10.7\pm0.2$	  &$1.0\pm0.5$   &$13.2\pm1.2$     &$12.9\pm2.5$\\
64				&2	&$-14.5$		&$+24.3$		&0.021	 &\phn$7.2\pm1.9$  &$-10.8\pm0.1$     &$0.9\pm0.1$  &$0.83\pm0.50$ & &\phn$1.9\pm0.9$  &$-10.8\pm0.1$	  &$0.5\pm0.2$   &$16.6\pm2.1$     &$\phn7.3\pm1.6$\\
65				&3	&\phn$+0.3$	&\phn$+4.5$	&0.021	 &\phn$6.4\pm1.4$  &$-10.5\pm0.1$     &$1.1\pm0.1$  &$0.60\pm0.40$ & &\phn$2.4\pm0.4$  &$-10.7\pm0.2$	  &$1.4\pm0.3$   &$19.4\pm2.2$     &$\phn5.9\pm1.2$\\
66				&4	&\phn$-5.3$	&\phn$+2.3$	&0.031	 &\phn$9.2\pm1.4$  &$-10.6\pm0.1$     &$0.9\pm0.1$  &$1.56\pm0.33$ & &\phn$1.5\pm0.3$  &$-10.7\pm0.2$	  &$1.9\pm0.5$   &$14.1\pm1.1$     &$10.2\pm1.4$\\
67				&5	&$-17.0$		&$+17.3$		&0.020	 &$16.1\pm3.6$     &$-10.8\pm0.1$     &$0.5\pm0.1$  &$2.33\pm0.62$ & &\phn$2.4\pm1.2$  &$-10.7\pm0.2$	  &$0.4\pm0.2$   &$14.1\pm1.9$     &$\phn9.0\pm2.1$\\
68				&6	&\phn$-9.8$	&\phn$-4.5$	&0.017	 &\phn$3.8\pm0.5$  &$-10.9\pm0.1$     &$1.2\pm0.2$  &$<0.29$       & &\phn$1.6\pm0.6$  &$-11.1\pm0.2$	  &$0.9\pm0.5$   &$19.4\pm1.2$     &$\phn3.2\pm0.7$\\
69				&7	&\phn$+6.3$	&\phn$-9.0$	&0.024	 &$11.5\pm2.9$     &$-11.5\pm0.1$     &$0.6\pm0.1$  &$2.52\pm0.70$ & &\phn$2.5\pm0.6$  &$-11.4\pm0.1$	  &$0.8\pm0.3$   &$17.1\pm3.3$     &$\phn8.2\pm2.1$\\
70				&8	&$-27.3$		&$+32.5$		&0.023	 &$16.3\pm3.9$     &$-11.5\pm0.1$     &$0.3\pm0.1$  &$2.93\pm0.73$ & &\ldots	       &\ldots  	  &\ldots        &       	   & \\
71~CepA			&1	&$+41.6$		&$+14.4$		&0.039	 &\phn$8.9\pm1.0$  &$-11.7\pm0.1$     &$1.3\pm0.1$  &$0.54\pm0.24$ & &\phn$4.1\pm0.3$  &$-11.6\pm0.1$	  &$1.5\pm0.1$   &$21.9\pm1.5$     &$13.4\pm1.1$\\
72				&2	&\phn$+4.8$	&$+32.8$		&0.041	 &$22.4\pm2.1$     &$-10.2\pm0.1$     &$0.5\pm0.1$  &$2.13\pm0.26$ & &\phn$5.7\pm0.5$  &$-10.2\pm0.1$	  &$0.8\pm0.1$   &$18.3\pm1.3$     &$11.5\pm1.0$\\
73				&3	&$+49.6$		&$+21.6$		&0.027	 &\phn$7.4\pm1.4$  &$-11.9\pm0.1$     &$1.5\pm0.1$  &$0.39\pm0.32$ & &\phn$3.1\pm0.3$  &$-12.0\pm0.1$	  &$1.9\pm0.2$   &$20.2\pm2.0$     &$15.3\pm1.6$\\
74				&4	&$+32.0$		&\phn$+6.4$	&0.036	 &\phn$9.7\pm0.8$  &$-12.0\pm0.1$     &$0.9\pm0.1$  &$0.48\pm0.18$ & &\phn$5.2\pm0.3$  &$-12.0\pm0.1$	  &$1.0\pm0.1$   &$23.8\pm1.3$     &$\phn7.7\pm0.6$\\
75				&5	&$-11.2$		&$-15.2$		&0.056	 &$10.4\pm1.6$     &\phn$-9.5\pm0.1$  &$0.8\pm0.1$  &$0.49\pm0.27$ & &\phn$6.3\pm0.3$  &\phn$-9.5\pm0.1$  &$0.9\pm0.1$   &$26.0\pm2.8$     &$\phn6.1\pm1.0$\\
76				&6	&$+65.6$		&$+30.4$		&0.041	 &$13.7\pm1.5$     &$-12.1\pm0.1$     &$0.7\pm0.1$  &$2.17\pm0.30$ & &\phn$3.2\pm0.2$  &$-12.1\pm0.1$	  &$1.0\pm0.1$   &$17.5\pm1.4$     &$31.1\pm1.2$\\
77				&7	&$+10.4$		&$+17.6$		&0.023	 &$17.8\pm2.4$     &$-10.8\pm0.1$     &$0.5\pm0.1$  &$2.33\pm0.34$ & &\phn$5.1\pm0.6$  &$-10.8\pm0.1$	  &$0.7\pm0.1$   &$20.6\pm2.3$     &$\phn8.0\pm1.1$\\
\hline
\end{tabular}
\end{center}
\begin{flushleft}
\supa\ Offset (in arcsec) between the position of the core and the phase centre (see Table~\ref{t:nh3obs}). $x$ and $y$ correspond to right ascension and declination, respectively.\\
\supb\ Linear diametre of the clump computed from the deconvolved angular diametre, $\theta_\mathrm{S}$, which has been calculated assuming the cores are Gaussians from the expression $\theta_\mathrm{S}^2=\mathrm{FWHM}^2-\mathrm{HPBW}^2$, with $\mathrm{FWHM}=2\sqrt{A/\pi}$ and $A$ being the area inside the polygons (shown in Fig.~\ref{f:cores}) and with HPBW corresponding to the geometric mean of the minor and major axes of the synthesised beams (see Table~\ref{t:nh3obs}).\\
\supc\ Parameters obtained from the spectral fit to the \nh\,(1,1) and (2,2) lines: $v$ and $\Delta v$ are the velocity of the local standard of rest and the linewidth in \kms, and $T_\mathrm{mb}$ is the temperature of the line in K. The hyperfine structure of the (1,1) transition allows us to measure the opacity $\tau$ of the main line. The parameter $A\times\tau$ is defined as $A = f\,[J_\nu(T_\mathrm{ex})-J_\nu(T_\mathrm{bg})]$ with $f = 1$ (see Appendix~\ref{a:hfs}).\\
\supd\ For a few cases, corresponding to the cores with optically thin emission, we provide an upper limit in the opacity. Even if $\tau$ cannot be determined with high precision, the derived quantities, $T_\mathrm{rot}$ and $N_\mathrm{NH_\mathrm{3}}$, in the optically thin limit, depend on the line intensity, $A\times\tau$, which is very well constrained by the data (see Appendix~\ref{a:hfs}).\\
\supe\ Rotational temperature (in K) and column density (in 10$^{14}$~cm$^{-2}$).
\end{flushleft}
\end{table}
\end{landscape}

\subsection{Averaged physical parameters for ammonia dense cores}\label{s:dense}

\begin{figure*}
\centering
\begin{tabular}[b]{c c c}
\vspace{0.5cm}
  \epsfig{file=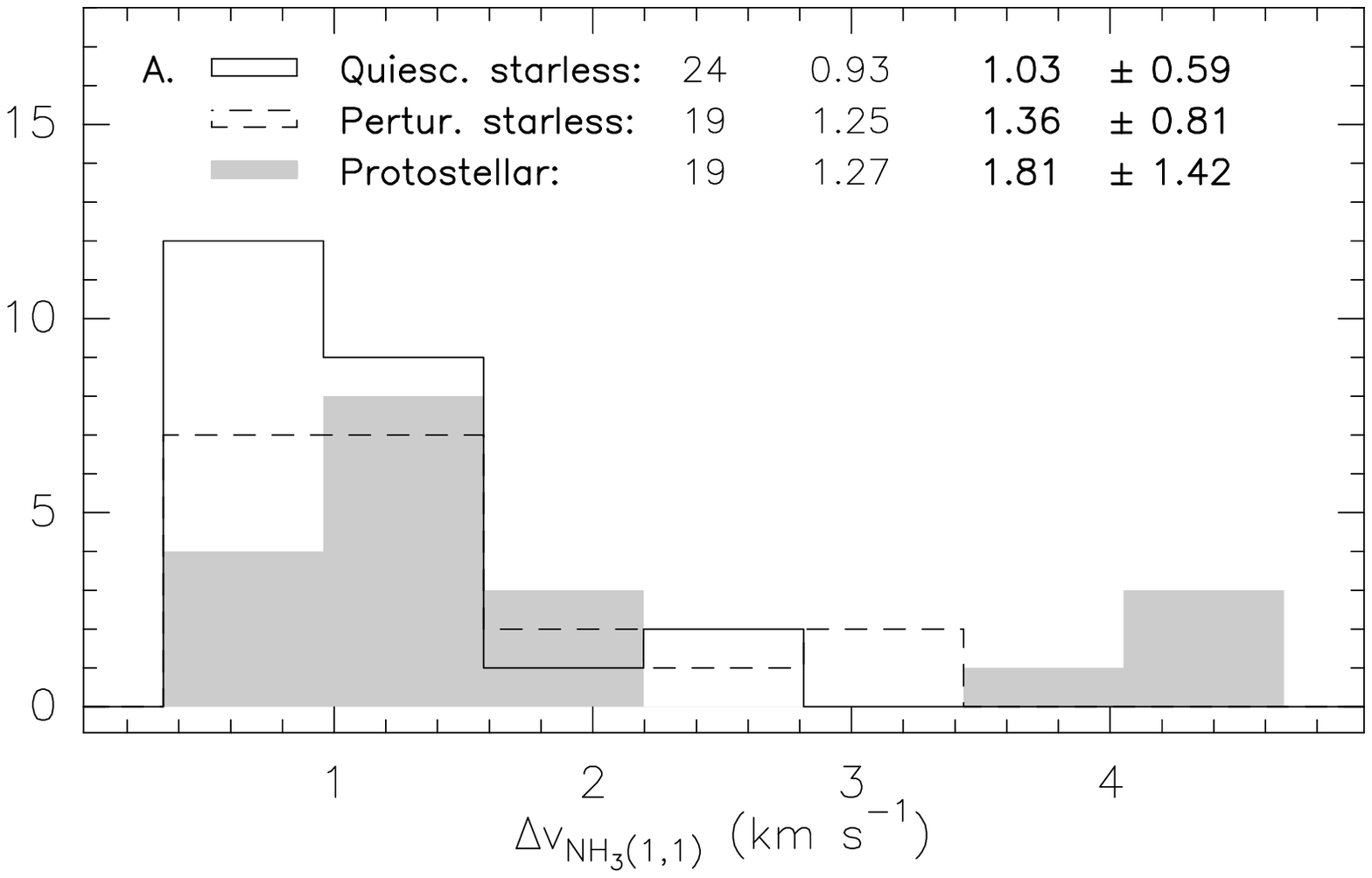, scale=0.34} &
  \epsfig{file=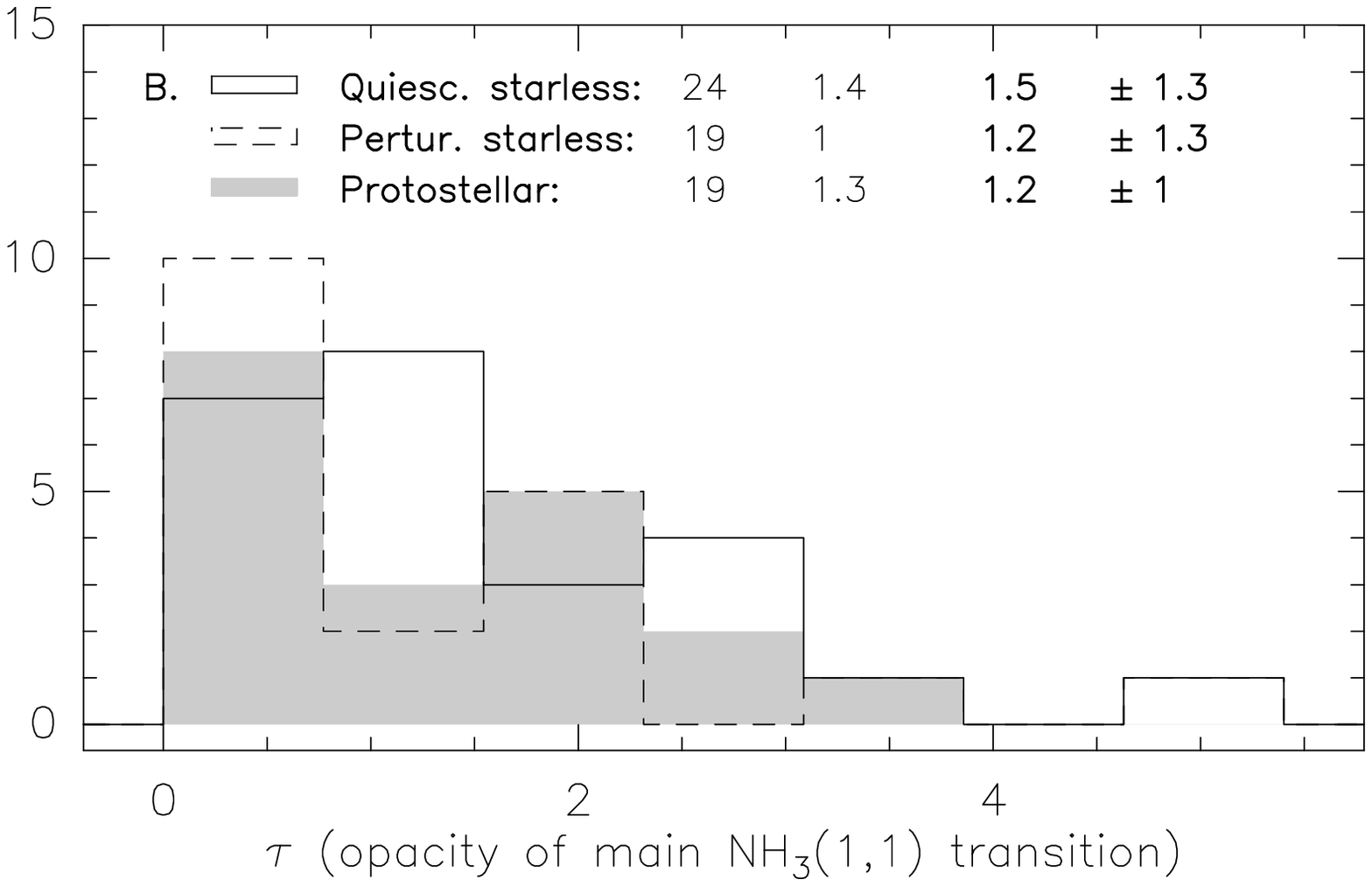, scale=0.34} &
  \epsfig{file=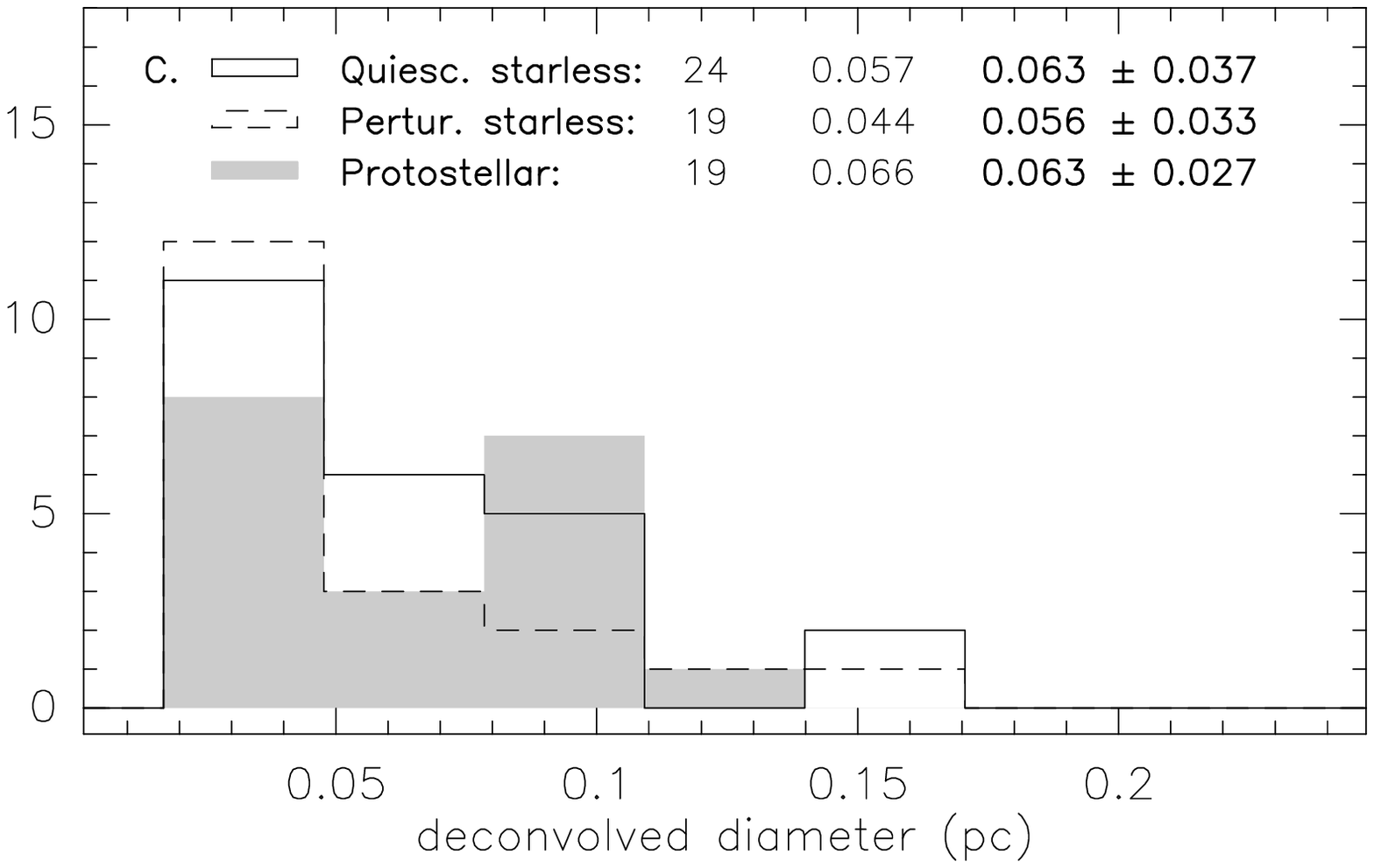, scale=0.34} \\
\vspace{0.5cm}
  \epsfig{file=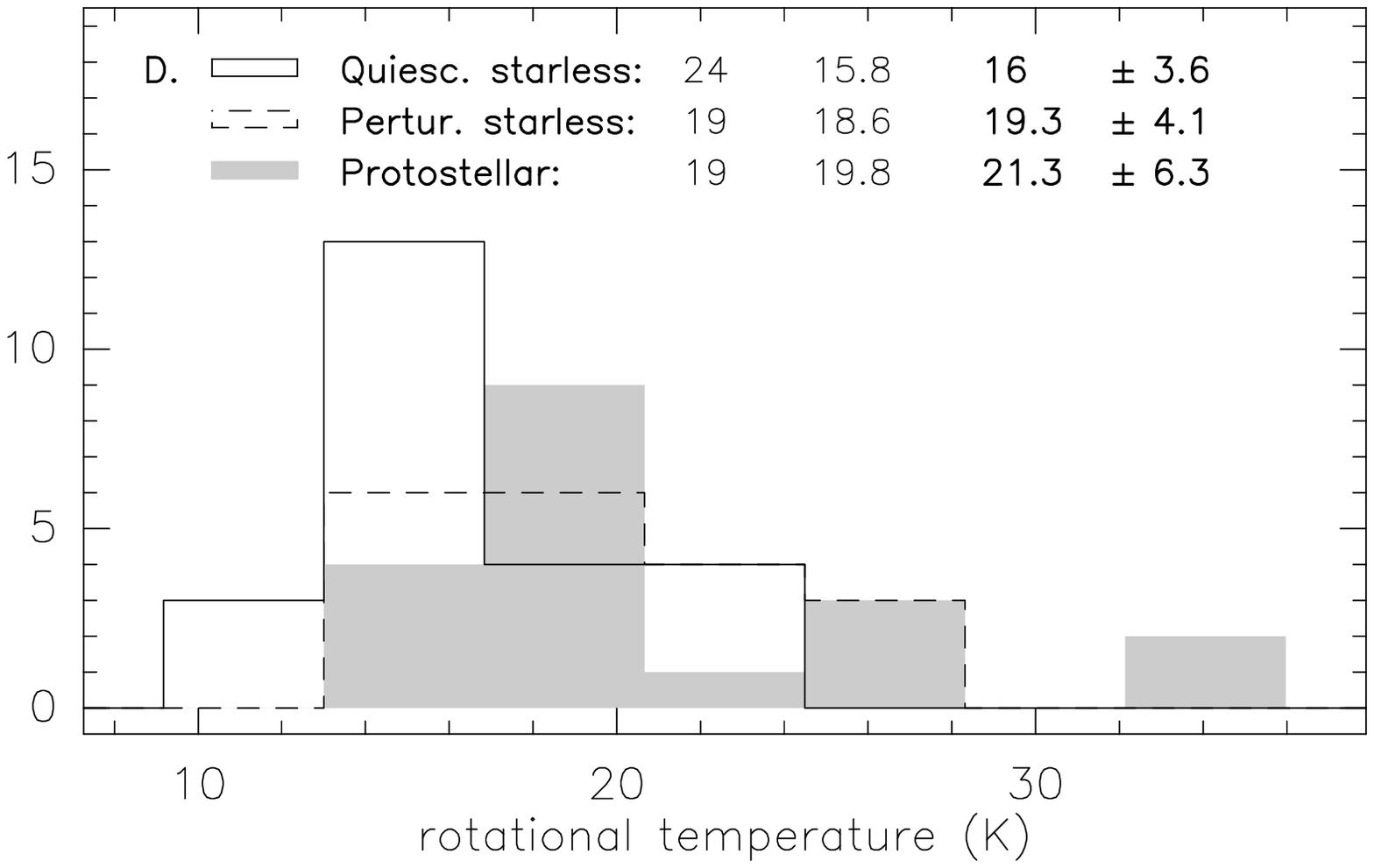, scale=0.34} &
  \epsfig{file=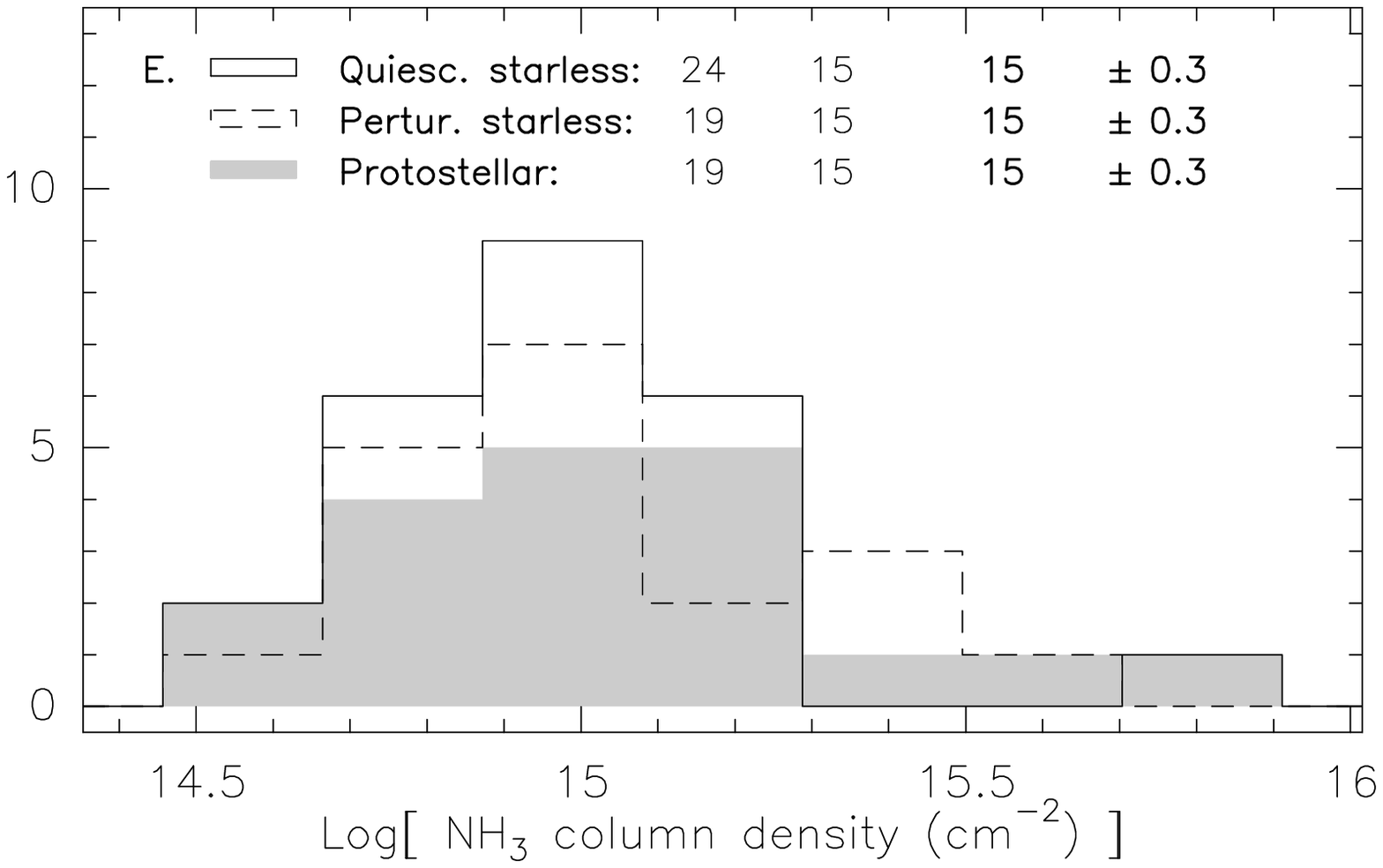, scale=0.34} &
  \epsfig{file=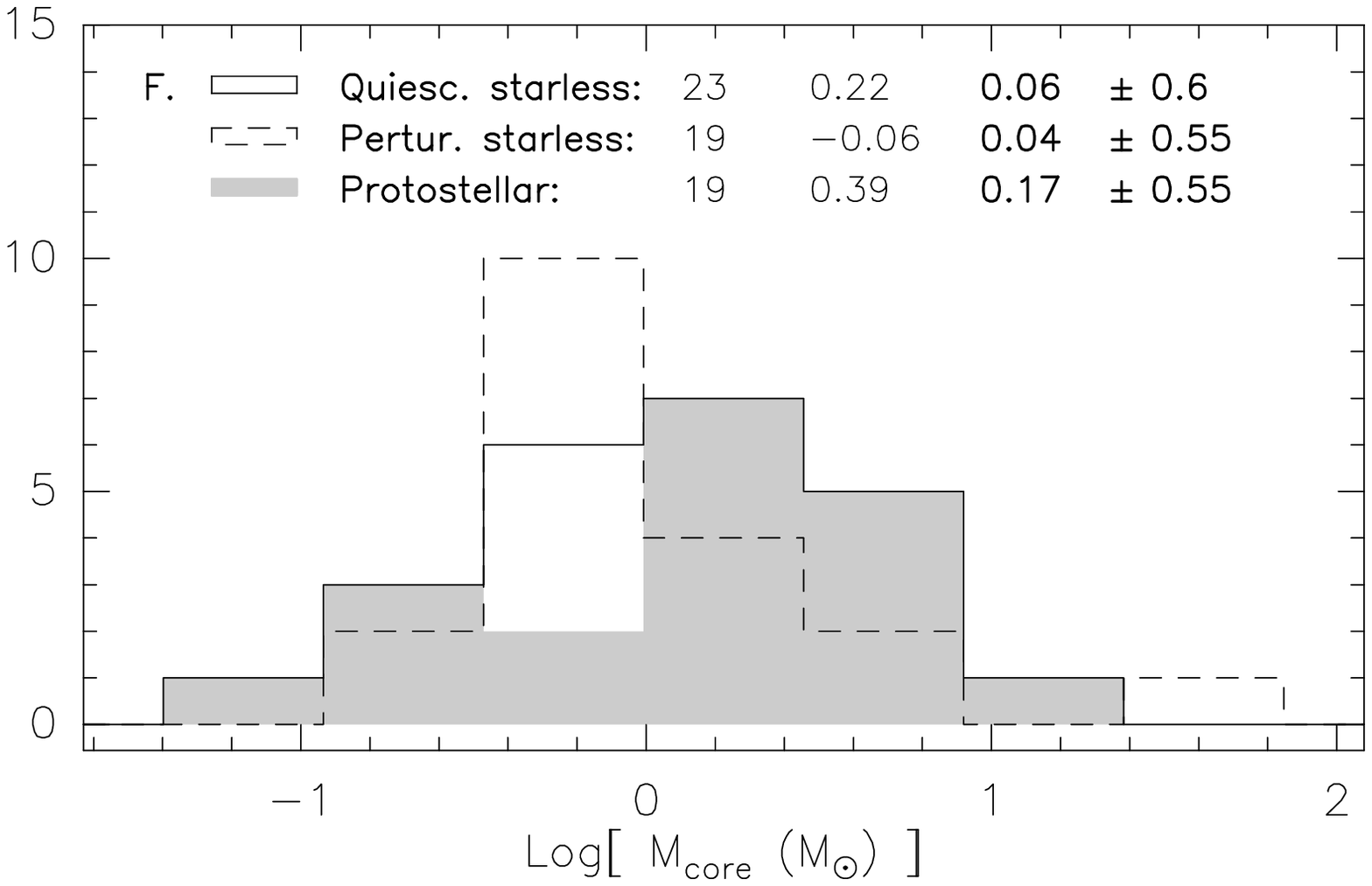, scale=0.34} \\
  \epsfig{file=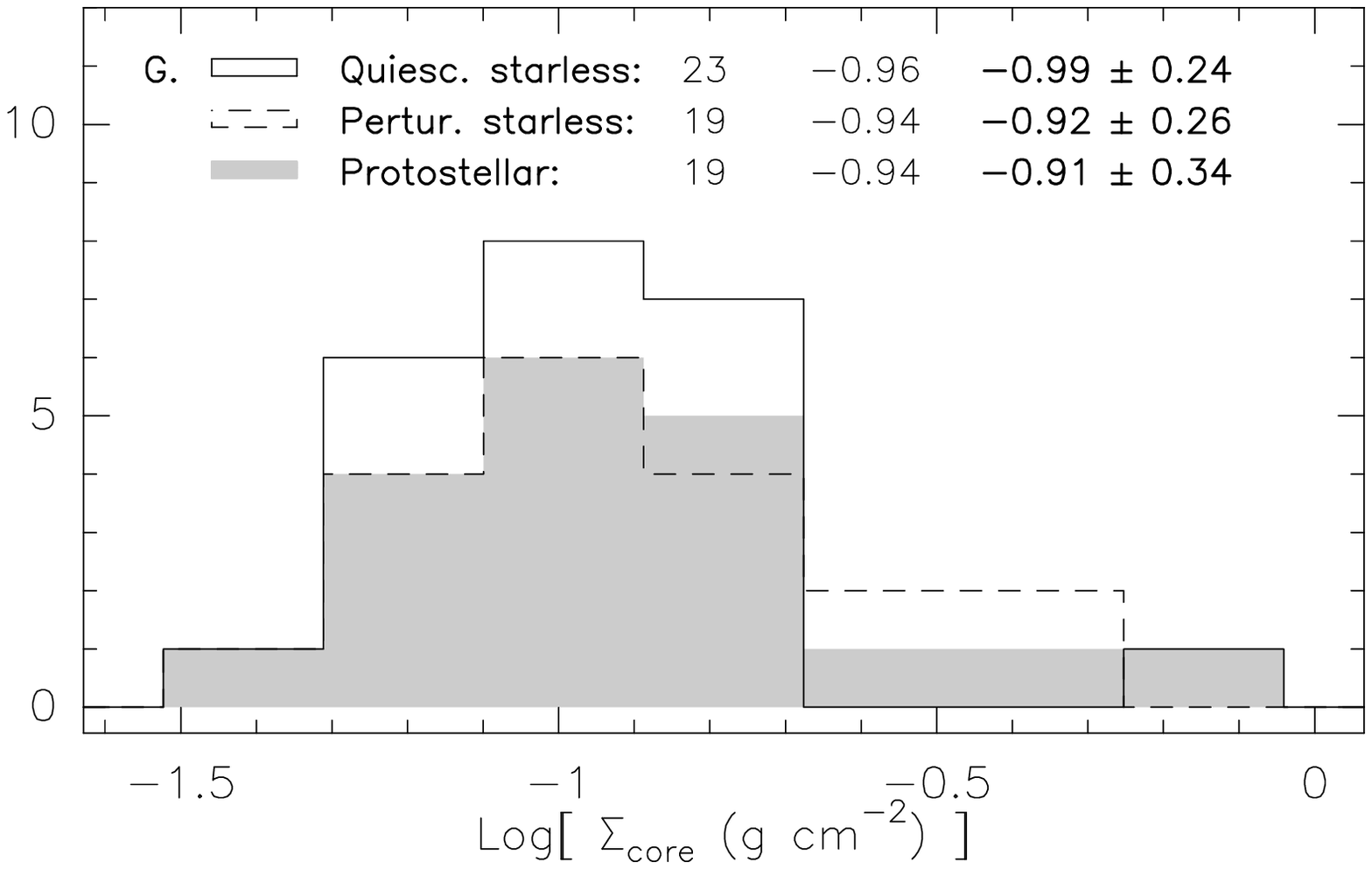, scale=0.34} &
  \epsfig{file=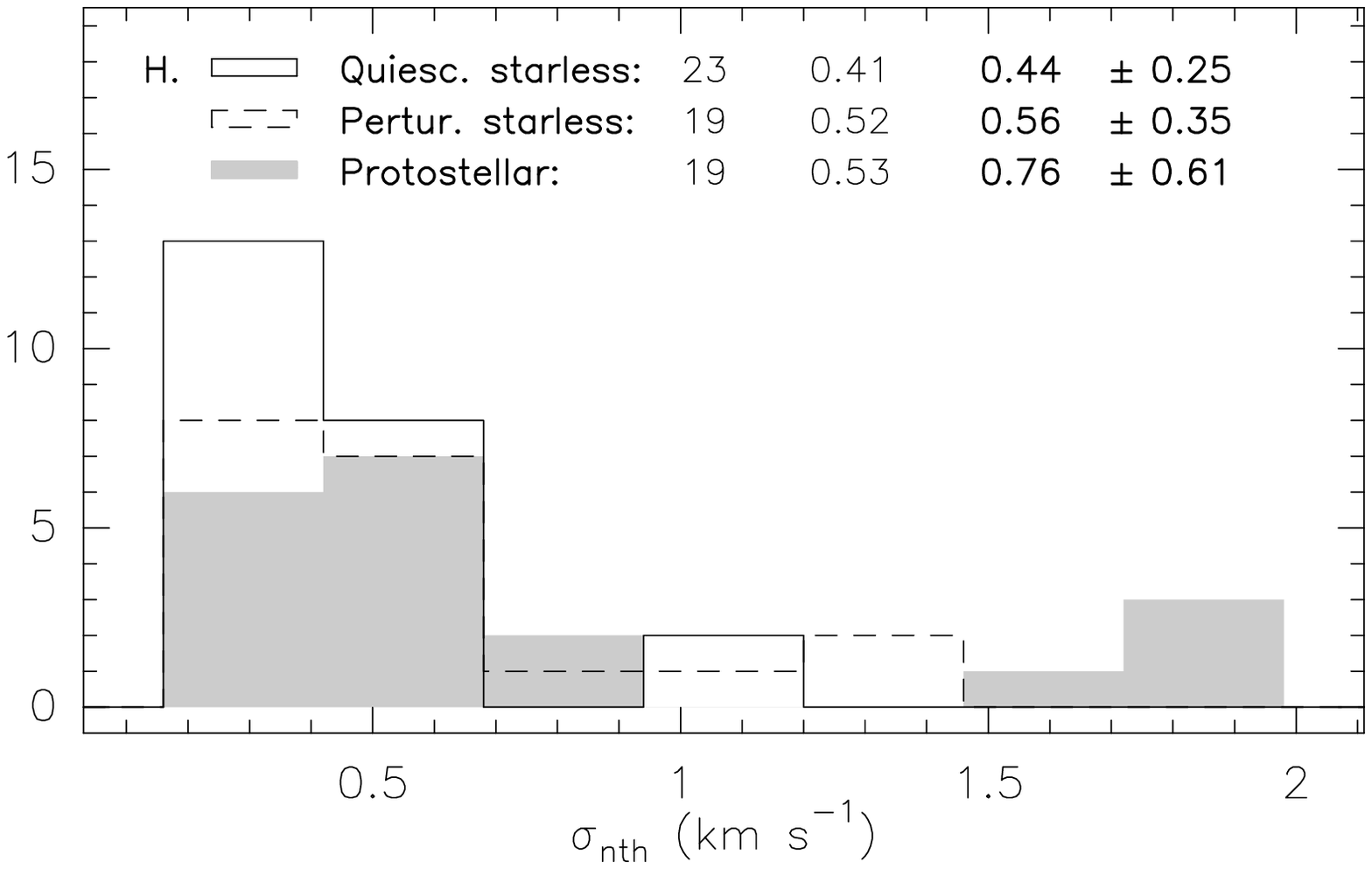, scale=0.34} &
  \epsfig{file=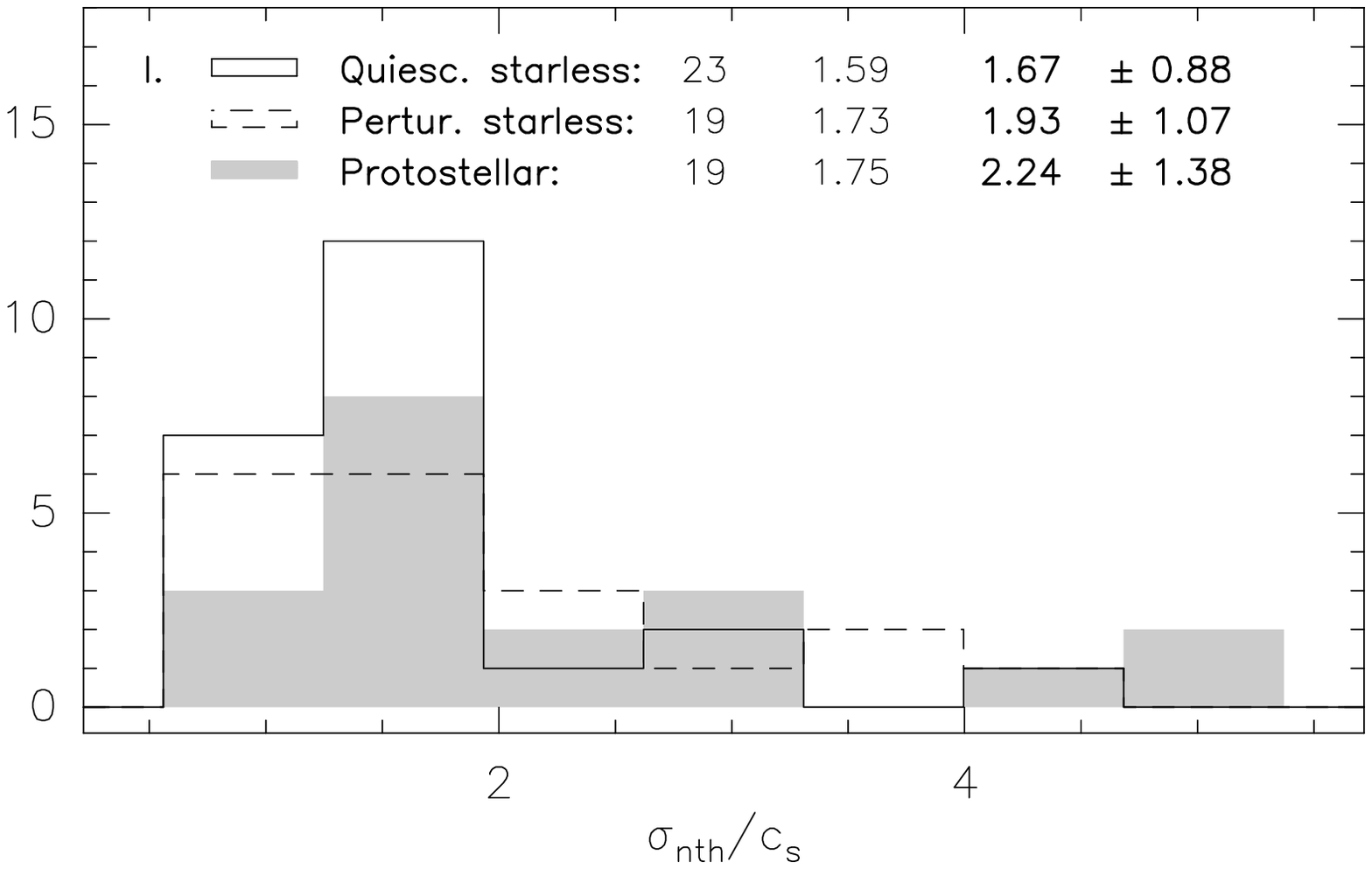, scale=0.34} \\
\end{tabular}
\caption{Distributions of: {\bf (a)} \nh\,(1,1) linewidth, {\bf (b)} opacity, {\bf (c)} deconvolved diametre, {\bf (d)} rotational temperature, {\bf (e)} ammonia column density, {\bf (f)} core mass $M_\mathrm{core}$, {\bf (g)} core surface density $\Sigma_\mathrm{core}$, {\bf (h)} non-thermal dispersion $\sigma_\mathrm{nth}$, and {\bf (i)} non-thermal to thermal velocity dispersion ratio $\sigma_\mathrm{nth}$/$c_s$, for quiescent starless (solid line histrogram), perturbed starless (dashed line histogram), and protostellar (grey filled histogram) cores. The numbers at the top of each panel are the total number of sources used in the histogram, the median, and the mean\,$\pm$\,standard deviation values.}
\label{f:histotypes}
\end{figure*}

For each core, we extracted the spectra averaged over the area assigned by the polygons (shown in Fig.~\ref{f:cores}) in both \nh\,(1,1) and (2,2) lines, and fitted the hyperfine structure of the \nh\,(1,1) transition following the procedure described in Appendix~\ref{a:hfs}, and a Gaussian to the \nh\,(2,2) line (see observed and fitted spectra in Fig.~\ref{f:spectra}). This allowed us to obtain the opacity, velocity, and linewidth for each core\footnote{For four cores it was necessary to fit the ammonia spectra with two different velocity components (see Table~\ref{t:nh3cores} and Fig.~\ref{f:spectra}).}. We derived the rotational temperature and primary beam corrected ammonia column density following the procedure outlined in the appendix of \citet{busquet2009}. All these parameters are listed in Table~\ref{t:nh3cores}. The average values of linewidth, opacity of the \nh\,(1,1) main line, rotational temperature, and ammonia column density of all the 73~cores are $\sim 1.4$~\kms, 1.4, 19~K, and 10$^{15}$~cm$^{-2}$, respectively. Our values of linewidth, rotational temperature and ammonia column density are very similar to the values obtained by \citet{urquhart2011} for the RMS (Red MSX Source) survey, of 2~\kms, 22~K and $2\times10^{15}$~cm$^{-2}$, even using very different observing techniques and sampling different spatial scales (single-dish vs interferometry). If we analyse the distributions of linewidth, opacity, rotational temperature, and column density for the starless and protostellar populations separately (see Fig.~\ref{f:histotypes} panels A to E), the average values that we obtain are (i) $\sim 1.0$~\kms, 1.5, 16.0~K, and 10$^{15}$~cm$^{-2}$ for quiescent starless cores, (ii) $\sim 1.4$~\kms, 1.2, 19.3~K, and 10$^{15}$~cm$^{-2}$ for perturbed starless cores, and (iii) $\sim 1.8$~\kms, 1.2, 21.3~K, and 10$^{15}$~cm$^{-2}$ for protostellar cores. Thus, quiescent starless cores have, as expected, linewidths and rotational temperatures smaller than protostellar cores (see Fig.~\ref{f:histotypes} panels A and D; Table~\ref{t:nh3mean}). The linewidths and temperatures of perturbed starless cores are in between the values measured for quiescent starless and protostellar cores, consistent with the fact that these cores were originally quiescent and then perturbed by the passage of outflows or the high radiation of a nearby bright infrared source. These cores might be similar to the `warm' starless cores reported in \citet{motte2010} using \emph{Herschel} observations towards Cygnus~X, and for some of them a low deuterium fractionation is measured \citep{fontani2011}. Further studies should be carried out to assess the true nature of these perturbed starless cores.

The average values derived for the rotational temperature of quiescent starless and protostellar cores (of 16.0~K and 21.3~K, respectively, corresponding to kinetic temperatures of 18.8~K and 28.8~K, following \citealt{tafalla2004}) can be compared to the values reported in the literature, although derived from observations of lower angular resolution. \citet{olmi2013} studied the dense clumps in the Galactic field $l=59\degr$ (at a distance of 3.6~kpc, similar to the distances of our regions) within the Hi-GAL \emph{Herschel} project \citep{molinari2010}, and derived kinetic temperatures of 17~K for starless cores and 20~K for protostellar cores, following the same trend as we found in our cores. Urquhart \et\ (2011) estimated the kinetic temperatures for clumps associated with young stellar objects and \hii\ regions in the RMS survey to be in the range 20--25~K, also in agreement with our protostellar temperatures. However, it is worth noting that the average rotational temperature of 16~K found in the quiescent starless cores of our clustered massive star-forming regions is significantly larger than the values measured in starless cores of low-mass isolated regions (of about 10~K or even less; \eg\ \citealt{tafalla2002, tafalla2004, rathborne2008, rosolowsky2008, schnee2010}), and of infrared dark clouds with no active star formation ($\sim 10$~K; \eg\ \citealt{ragan2011}). Temperatures of $\sim 15$~K for starless cores are also found by \citet{li2003} in the quiescent cores near the Trapezium cluster. Li \et\ find that the temperatures of the massive quiescent cores can be well explained by the dust being heated by the external UV field from the Trapezium cluster, at a distance of $\sim 1$~pc from the massive cores. This possibility is investigated for the regions in our sample in Sect.~\ref{s:IRNH3}.

\begin{table*}
\caption{Derived parameters for the ammonia cores}
\label{t:nh3extra}
\centering
\begin{tabular}{l c c c c c c c c c c c c}
\hline\hline

&
&$T_\mathrm{kin}$\supb
&$\sigma_\mathrm{th}$\supc
&$\sigma_\mathrm{nth}$\supc
&
&$\sigma_\mathrm{tot}$
&$M_\mathrm{core}$\supd
&$\Sigma_\mathrm{core}$\supe
&$M_\mathrm{vir}$\supf
\\
\texttt{ID}
&Type\supa
&(K)
&(\kms)
&(\kms)
&$\sigma_\mathrm{nth}/c_s$
&(\kms)
&(\mo)
&(g~cm$^{-2}$)
&(\mo)
\\
\hline
01  &P   &16.4  &0.09  &0.32  &1.32  &0.40  &\phn2.56  &0.09  &\phnn5.42 \\
02  &S   &13.3  &0.08  &0.51  &2.36  &0.56  &\phn0.90  &0.11  &\phnn7.19 \\
03  &P   &18.1  &0.09  &0.67  &2.63  &0.71  &\phn0.18  &0.03  &\phn10.26 \\
04  &S*  &17.5  &0.09  &0.60  &2.41  &0.65  &\phn0.18  &0.07  &\phnn5.59 \\
05  &P?  &12.1  &0.08  &0.26  &1.25  &0.33  &\phn1.28  &0.15  &\phnn2.03 \\
06  &P   &66.3  &0.18  &1.59  &3.27  &1.66  &\phn2.76  &0.46  &\phn59.31 \\
07  &P?  &53.9  &0.16  &0.71  &1.61  &0.83  &\phn2.41  &0.18  &\phn17.99 \\
08  &P?  &26.9  &0.11  &0.60  &1.94  &0.67  &\phn2.78  &0.14  &\phn15.76 \\
09  &P   &17.4  &0.09  &0.56  &2.27  &0.62  &\phn5.57  &0.18  &\phn17.10 \\
10  &P   &21.6  &0.10  &0.41  &1.47  &0.49  &\phn0.80  &0.10  &\phnn4.69 \\
11  &S   &16.2  &0.09  &0.21  &0.87  &0.32  &\phn1.81  &0.11  &\phnn1.93 \\
12  &S   &18.5  &0.09  &0.40  &1.56  &0.47  &\phn1.58  &0.12  &\phnn5.80 \\
13  &S   &14.5  &0.08  &0.21  &0.92  &0.31  &\phn0.86  &0.11  &\phnn1.32 \\
14  &S   &11.3  &0.07  &0.17  &0.83  &0.26  &\phn1.11  &0.13  &\phnn0.90 \\
15  &S   &18.9  &0.10  &0.16  &0.63  &0.31  &\phn0.67  &0.06  &\phnn1.16 \\
16  &P   &31.1  &0.12  &0.49  &1.48  &0.59  &\phn2.36  &0.08  &\phn13.23 \\
17  &P?  &21.8  &0.10  &0.33  &1.18  &0.43  &\phn1.54  &0.11  &\phnn4.20 \\
18  &P   &38.0  &0.14  &0.69  &1.89  &0.78  &\phn0.79  &0.07  &\phn16.01 \\
19  &S*  &32.1  &0.12  &0.54  &1.60  &0.64  &\phn1.50  &0.07  &\phn13.19 \\
20  &P   &19.7  &0.10  &0.20  &0.77  &0.33  &\phn1.90  &0.11  &\phnn2.05 \\
21  &S*  &14.6  &0.08  &0.17  &0.76  &0.28  &\phn0.35  &0.10  &\phnn0.66 \\
22  &S   &15.9  &0.09  &0.37  &1.55  &0.44  &\phn3.28  &0.14  &\phnn6.49 \\
23  &P   &21.1  &0.10  &0.40  &1.45  &0.48  &\phn2.87  &0.12  &\phnn7.92 \\
24  &P   &16.8  &0.09  &0.25  &1.04  &0.35  &\phn3.62  &0.15  &\phnn3.42 \\
25  &S*  &22.5  &0.10  &0.49  &1.75  &0.57  &\phn0.91  &0.13  &\phnn6.54 \\
26  &S*  &26.6  &0.11  &0.17  &0.56  &0.35  &\phn0.28  &0.04  &\phnn1.10 \\
27  &P   &34.4  &0.13  &1.87  &5.37  &1.91  &\phn1.70  &0.23  &\phn90.38 \\
28  &S*  &36.6  &0.13  &1.30  &3.62  &1.35  &\phn0.97  &0.10  &\phn49.97 \\
29  &S   &18.9  &0.10  &1.16  &4.48  &1.19  &\phn0.38  &0.13  &\phn22.09 \\
30  &S*  &21.3  &0.10  &1.17  &4.24  &1.20  &\phn0.49  &0.10  &\phn28.71 \\
31  &S   &16.9  &0.09  &0.67  &2.72  &0.71  &\phn0.64  &0.15  &\phnn8.92 \\
32  &P?  &21.8  &0.10  &0.89  &3.21  &0.93  &70.21     &0.35  &108.66 \\
33  &S   &22.0  &0.10  &0.51  &1.83  &0.58  &\phn7.19  &0.10  &\phn22.24 \\
34  &S   &26.2  &0.11  &0.56  &1.84  &0.64  &\phn5.31  &0.07  &\phn26.71 \\
35  &P   &40.5  &0.14  &1.98  &5.23  &2.01  &\phn3.72  &0.91  &\phn75.60 \\
36  &S   &17.9  &0.09  &0.32  &1.28  &0.41  &14.24     &0.66  &\phnn4.96 \\
37  &P?  &19.8  &0.10  &0.36  &1.35  &0.45  &11.64     &0.44  &\phnn6.77 \\
38  &P   &63.2  &0.18  &1.93  &4.07  &1.98  &\phn2.54  &0.18  &135.01 \\
39  &S*  &35.3  &0.13  &1.28  &3.62  &1.33  &\phn5.61  &0.12  &109.15 \\
40  &P?  &34.5  &0.13  &1.09  &3.12  &1.15  &\phn2.76  &0.12  &\phn54.11 \\
41  &S   &32.5  &0.13  &0.55  &1.61  &0.64  &\phn2.55  &0.07  &\phn18.66 \\
42  &S   &27.4  &0.12  &0.54  &1.72  &0.62  &\phn2.35  &0.06  &\phn17.82 \\
43  &S   &27.9  &0.12  &1.03  &3.26  &1.07  &\phn3.33  &0.14  &\phn49.72 \\
44  &P?  &29.0  &0.12  &0.58  &1.82  &0.66  &11.88     &0.12  &\phn33.29 \\
45  &S*  &18.9  &0.10  &0.44  &1.71  &0.51  &\phn6.78  &0.23  &\phn10.54 \\
46  &S*  &14.8  &0.08  &0.36  &1.57  &0.43  &32.56     &0.40  &\phn11.82 \\
47  &P   &24.9  &0.11  &0.90  &3.02  &0.94  &\phn9.81  &0.21  &\phn53.15 \\
48  &P?  &17.8  &0.09  &0.42  &1.68  &0.49  &12.15     &0.21  &\phn13.55 \\
49  &P   &20.6  &0.10  &0.43  &1.59  &0.51  &\phn4.33  &0.17  &\phnn9.43 \\
50  &P?  &16.6  &0.09  &0.56  &2.29  &0.61  &10.63     &0.27  &\phn18.96 \\
51  &S*  &19.4  &0.10  &0.82  &3.14  &0.86  &\phn1.94  &0.27  &\phn17.51 \\
52  &S*  &22.1  &0.10  &0.66  &2.37  &0.72  &\phn0.36  &0.08  &\phnn9.21 \\
53  &P?  &26.1  &0.11  &0.54  &1.77  &0.62  &\phn0.08  &0.04  &\phnn4.03 \\
54  &P   &22.7  &0.10  &0.43  &1.53  &0.52  &\phn0.29  &0.06  &\phnn4.18 \\
55  &S   &14.3  &0.08  &0.29  &1.28  &0.37  &\phn2.59  &0.14  &\phnn3.75 \\
56  &S*  &29.7  &0.12  &0.40  &1.24  &0.52  &\phn2.11  &0.06  &\phnn9.62 \\
57  &S   &15.4  &0.09  &0.26  &1.11  &0.35  &\phn3.41  &0.08  &\phnn4.67 \\
58  &S   &20.4  &0.10  &0.48  &1.79  &0.55  &\phn1.73  &0.05  &\phn14.06 \\
59  &P?  &24.3  &0.11  &0.35  &1.21  &0.46  &\phn0.70  &0.04  &\phnn5.42 \\
\hline
\end{tabular}
\end{table*}
\begin{table*}
\contcaption{}
\centering
\begin{tabular}{l c c c c c c c c c c c c}
\hline\hline

&
&$T_\mathrm{kin}$\supb
&$\sigma_\mathrm{th}$\supc
&$\sigma_\mathrm{nth}$\supc
&
&$\sigma_\mathrm{tot}$
&$M_\mathrm{core}$\supd
&$\Sigma_\mathrm{core}$\supe
&$M_\mathrm{vir}$\supf
\\
\texttt{ID}
&Type\supa
&(K)
&(\kms)
&(\kms)
&$\sigma_\mathrm{nth}/c_s$
&(\kms)
&(\mo)
&(g~cm$^{-2}$)
&(\mo)
\\
\hline
60  &P   &25.2  &0.11  &0.60  &2.00  &0.67  &\phn1.85  &0.05  &\phn21.32 \\
61  &S   &18.3  &0.09  &0.36  &1.39  &0.44  &\phn0.36  &0.06  &\phnn3.15 \\
62  &P?  &20.6  &0.10  &0.74  &2.75  &0.79  &\phn0.52  &0.09  &\phn12.75 \\
63  &S   &14.5  &0.08  &0.41  &1.80  &0.47  &\phn0.29  &0.14  &\phnn2.32 \\
64  &P?  &19.3  &0.10  &0.39  &1.49  &0.47  &\phn0.14  &0.08  &\phnn1.97 \\
65  &P   &23.7  &0.11  &0.46  &1.60  &0.55  &\phn0.11  &0.07  &\phnn2.76 \\
66  &S*  &15.7  &0.09  &0.39  &1.67  &0.46  &\phn0.41  &0.11  &\phnn2.92 \\
67  &S   &15.7  &0.09  &0.17  &0.73  &0.29  &\phn0.15  &0.10  &\phnn0.43 \\
68  &S   &23.6  &0.11  &0.52  &1.79  &0.59  &\phn0.04  &0.04  &\phnn2.78 \\
69  &S   &20.1  &0.10  &0.26  &0.96  &0.37  &\phn0.20  &0.09  &\phnn1.05 \\
71  &S*  &28.2  &0.12  &0.55  &1.75  &0.64  &\phn0.85  &0.15  &\phnn7.25 \\
72  &S*  &22.0  &0.10  &0.18  &0.65  &0.33  &\phn0.81  &0.13  &\phnn1.04 \\
73  &S*  &25.1  &0.11  &0.61  &2.06  &0.68  &\phn0.47  &0.17  &\phnn6.11 \\
74  &P?  &31.9  &0.12  &0.36  &1.07  &0.49  &\phn0.42  &0.09  &\phnn3.05 \\
75  &S*  &36.6  &0.13  &0.30  &0.85  &0.47  &\phn0.81  &0.07  &\phnn3.60 \\
76  &S*  &20.7  &0.10  &0.29  &1.07  &0.40  &\phn2.19  &0.35  &\phnn2.25 \\
77  &P   &25.7  &0.11  &0.17  &0.57  &0.35  &\phn0.18  &0.09  &\phnn0.56 \\
\hline
\end{tabular}
\begin{flushleft}
\supa\ Type of the core: S: quiescent starless, S*: perturbed starless, P: protostellar, and P?: possible protostellar (see Sect.~\ref{s:cores} and Table~\ref{t:nh3types}).\\
\supb\ Kinetic temperature, in K, derived from the rotational temperature and following \citet{tafalla2004}. We emphasise that Tafalla \et\ empirical relation is accurate only for temperatures $\la 20$~K, and thus the kinetic temperature given here should be regarded, for high temperatures, as an approximation.\\
\supc\ Thermal ($\sigma_\mathrm{th}$) and non-thermal ($\sigma_\mathrm{nth}$) velocity dispersion, in \kms.\\
\supd\ Mass of the core derived as $M_\mathrm{core}=(N/X)\mu~m_\mathrm{H}A$, with $\mu$ ($= 2.8$) the mean molecular mass per H molecule, $A$ ($= \pi(\mathrm{size}/2)^2$) the area, and assuming an abundance of \nh\ with respect to H$_2$ ($X$) for all the cores of $4.2\times10^{-8}$ \citep[\eg][]{pillai2006, foster2009, friesen2009, rygl2010, chira2013}.\\
\supe\ Core mass surface density derived as $\Sigma_\mathrm{core}=M_\mathrm{core}/A$.\\
\supf\ Virial mass derived from $M_\mathrm{vir} = 210\,[R/\mathrm{pc}]\,[\Delta v/\mathrm{km~s^{-1}}]^2$~\mo\ assuming uniform density across the core \citep{maclaren1988, estalella1993}. $R$ is the radius of the core (obtained from the diametre size listed in Table~\ref{t:nh3cores}), and $\Delta v$ is the intrinsic linewidth of the \nh\,(1,1) line (listed in Table~\ref{t:nh3cores}).\\
\end{flushleft}
\end{table*}

The linewidths measured in our starless cores have an important non-thermal component, which is almost negligible in low-mass clustered regions \citep[\eg][]{friesen2009, foster2009}. To evaluate this component, we assumed the relation $\sigma_\mathrm{nth} = \sqrt{\sigma_\mathrm{obs}^2- \sigma_\mathrm{th}^2}$
where $\sigma_\mathrm{obs}$ is the measured velocity dispersion of the core ($= \Delta v/(8\ln2)^{1/2}$ for a Gaussian line profile with $\Delta v$ the measured FWHM of the line) and $\sigma_\mathrm{th}$ is the thermal velocity dispersion. For \nh, assuming a Maxwellian velocity distribution, $\sigma_\mathrm{th}$ can be calculated from $\sqrt{k_\mathrm{B}~T_\mathrm{kin}/(\mu~m_\mathrm{H})}$, with $k_\mathrm{B}$ the Boltzmann constant, $\mu$ the molecular weight ($= 17.03$ for ammonia), $m_\mathrm{H}$ the mass of the Hydrogen atom, and $T_\mathrm{kin}$ the kinetic temperature listed in Col.~3 of Table~\ref{t:nh3extra}. In Cols.~5 and 6 of the same Table we list the non-thermal velocity dispersion ($\sigma_\mathrm{nth}$) and the non-thermal velocity dispersion over the isothermal sound speed ($\sigma_\mathrm{nth}/c_s$), where $c_s$ is calculated using a mean molecular weight $\mu= 2.33$. From the distributions shown in Fig.~\ref{f:histotypes} (panels H and I) we see that protostellar cores have a non-thermal component (0.76~\kms, or equivalently $\Delta v_\mathrm{nth}= 1.8$~\kms) higher than the non-thermal component of the quiescent starless (0.44~\kms, or $\Delta v_\mathrm{nth}= 1.0$~kms) and perturbed starless (0.56~\kms, or $\Delta v_\mathrm{nth}=1.3$~\kms) cores. For the $\sigma_\mathrm{nth}/c_s$ we obtain mean ratios of 1.7, 1.9 and 2.2 for quiescent starless, perturbed starless and protostellar cores, respectively (see Table~\ref{t:nh3mean}), indicating that the cores are characterised by moderately supersonic non-thermal motions. In general our cores (both starless and protostellar), associated with intermediate and high-mass star-forming regions, have values of turbulent Mach number $\sim 2$, being greater than those found in Perseus or Ophiucus (with values between 0.5 and 1.5; \eg\ \citealt{andre2007, friesen2009, foster2009}).

From the ammonia column density (listed in Table~\ref{t:nh3cores}), we derived the mass of the core and the core mass surface density, following the procedures explained in the footnotes of Table~\ref{t:nh3extra}. These two quantities are listed in Cols.~8 and 9 of Table~\ref{t:nh3extra}, and their distributions are shown in panels F and G of Fig.~\ref{f:histotypes}. Starless and protostellar cores show similar mean masses ($\sim 1.0$--$1.8$~\mo) and mean surface densities ($\sim 0.10$--$0.12$~g~cm$^{-2}$) (see Table~\ref{t:nh3mean}).

\begin{table*}
\caption{Mean values derived from the \nh\ observations (from Tables~\ref{t:nh3cores} and \ref{t:nh3extra})}
\label{t:nh3mean}
\centering
\begin{tabular}{l c c c c c c c c c c c c c c}
\hline\hline

&$T_\mathrm{rot}$
&$T_\mathrm{kin}$
&size
&$\tau$
&$N_\mathrm{NH_3}$
&$M_\mathrm{core}$
&$\Sigma_\mathrm{core}$
&$\Delta V_\mathrm{(1,1)}$
&$\sigma_\mathrm{nth}$
&$\sigma_\mathrm{nth}$/$c_s$
&$M_\mathrm{vir}$
\\
Core type
&(K)
&(K)
&(pc)
&
&(cm$^{-2}$)
&(\mo)
&(g~cm$^{-2}$)
&(\kms)
&(\kms)
&
&(\mo)
\\
\hline
Quiescent starless	&16.0	&18.8	&0.063	&1.5		&10$^{15.0}$	&1.15	&0.10	&1.03	&0.44	&1.67	&\phn5.1		\\
Perturbed starless	&19.3	&24.2	&0.056	&1.2		&10$^{15.0}$	&1.09	&0.12	&1.36	&0.56	&1.93	&\phn7.0		\\
Protostellar			&21.3	&28.8	&0.063	&1.2		&10$^{15.0}$	&1.47	&0.12	&1.81	&0.76	&2.24	&11.7		\\
All sample			&18.8	&23.8	&0.065	&1.4		&10$^{15.0}$	&1.36	&0.12	&1.36	&0.57	&1.91	&\phn7.8		\\
\hline
\end{tabular}
\end{table*}

\subsection{Ammonia correlations}\label{s:correlations}

\begin{figure*}
\begin{center}
\begin{tabular}[b]{c c c}
\vspace{0.5cm}
  \epsfig{file=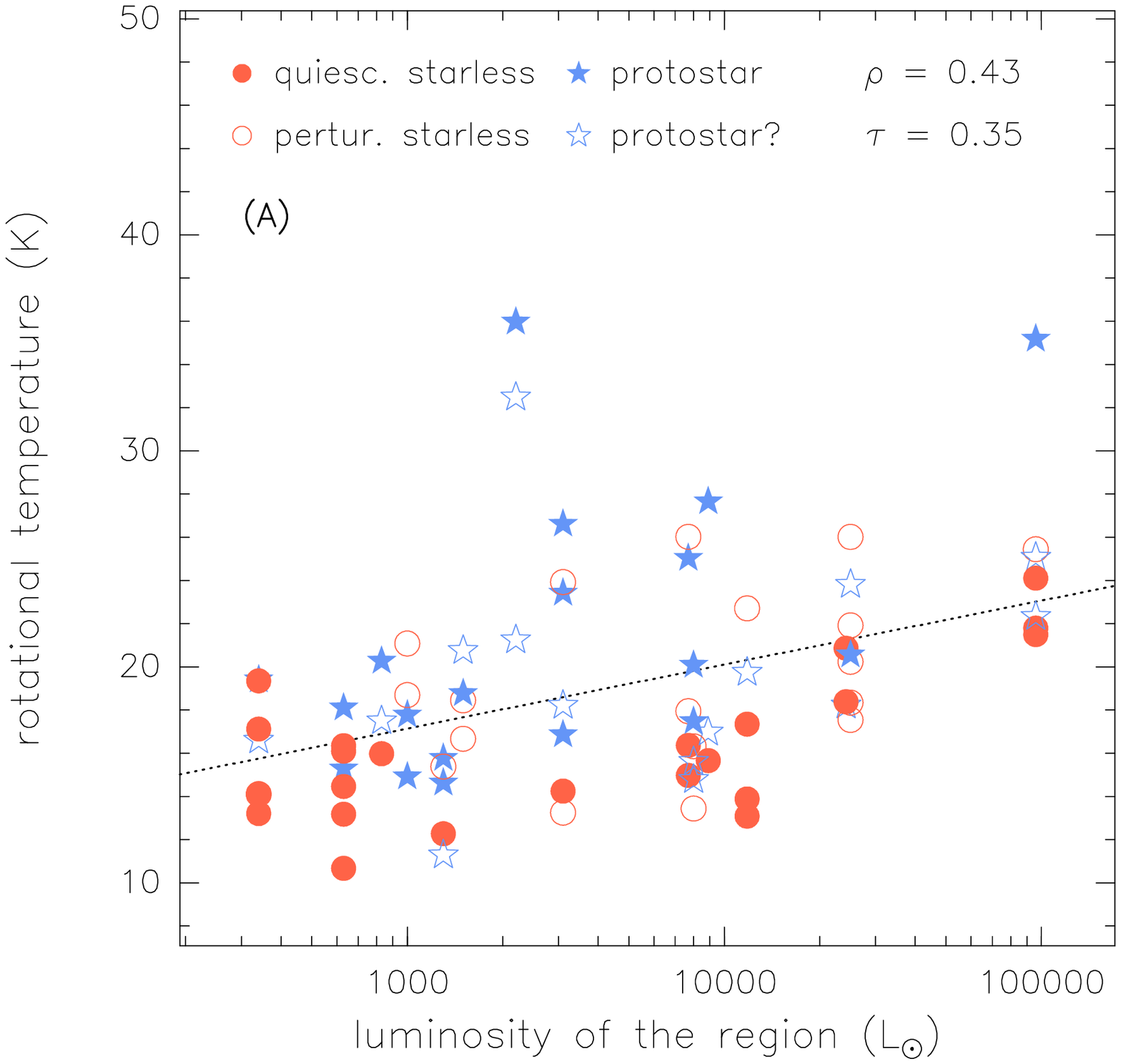, scale=0.3} &
  \epsfig{file=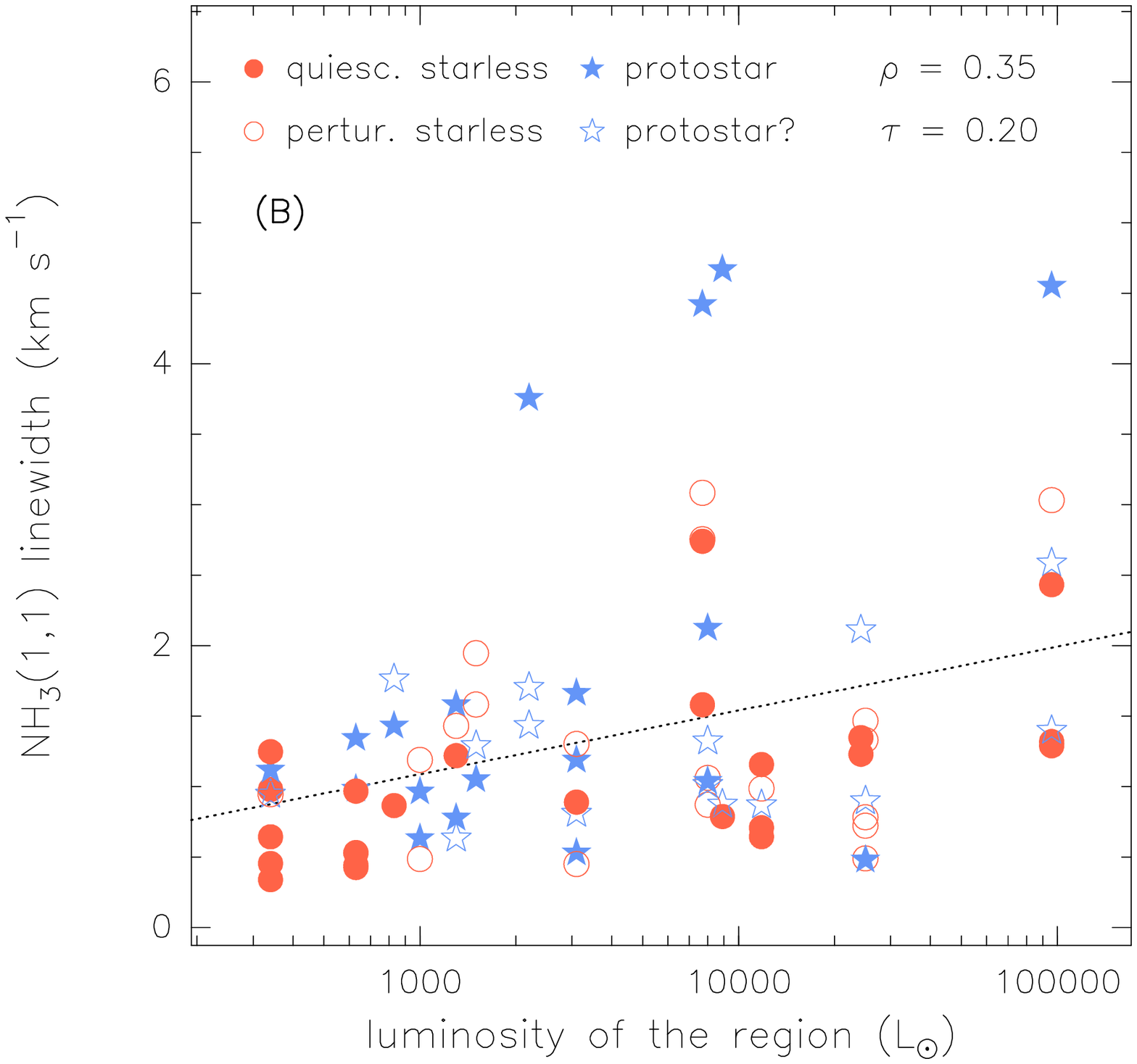, scale=0.3} &
  \epsfig{file=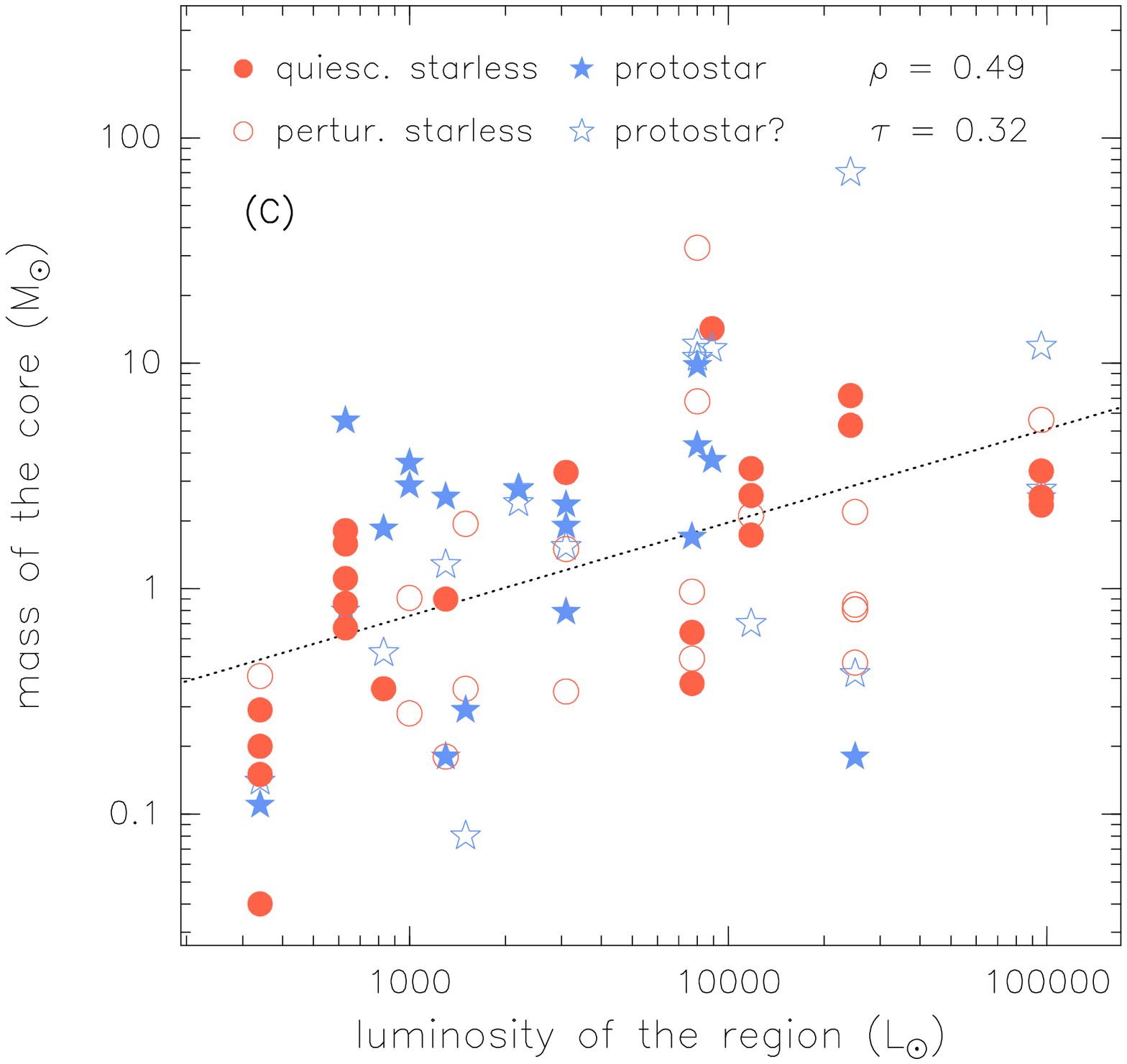, scale=0.3} \\
  \epsfig{file=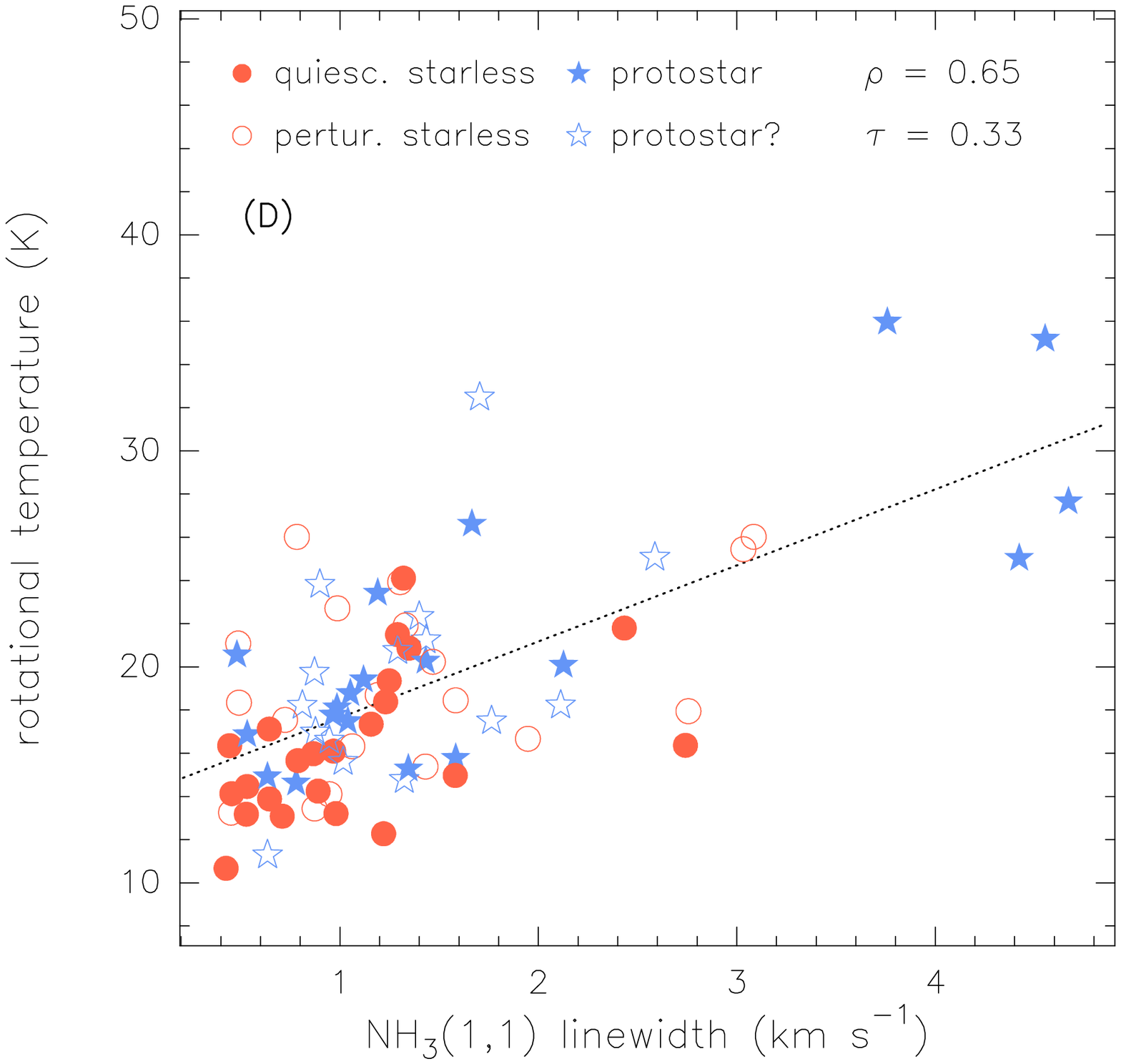, scale=0.3} &
  \epsfig{file=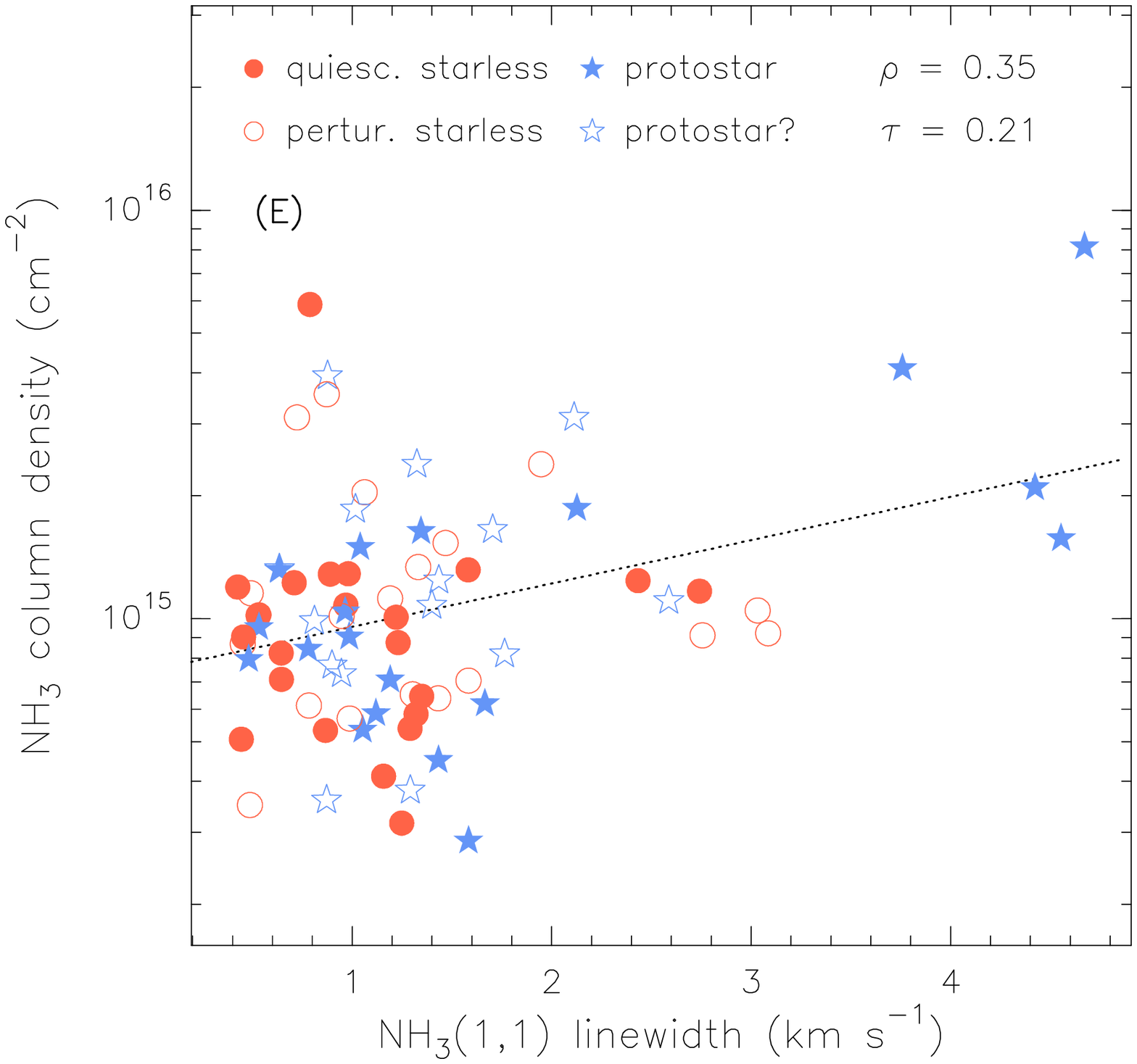, scale=0.3} &
  \epsfig{file=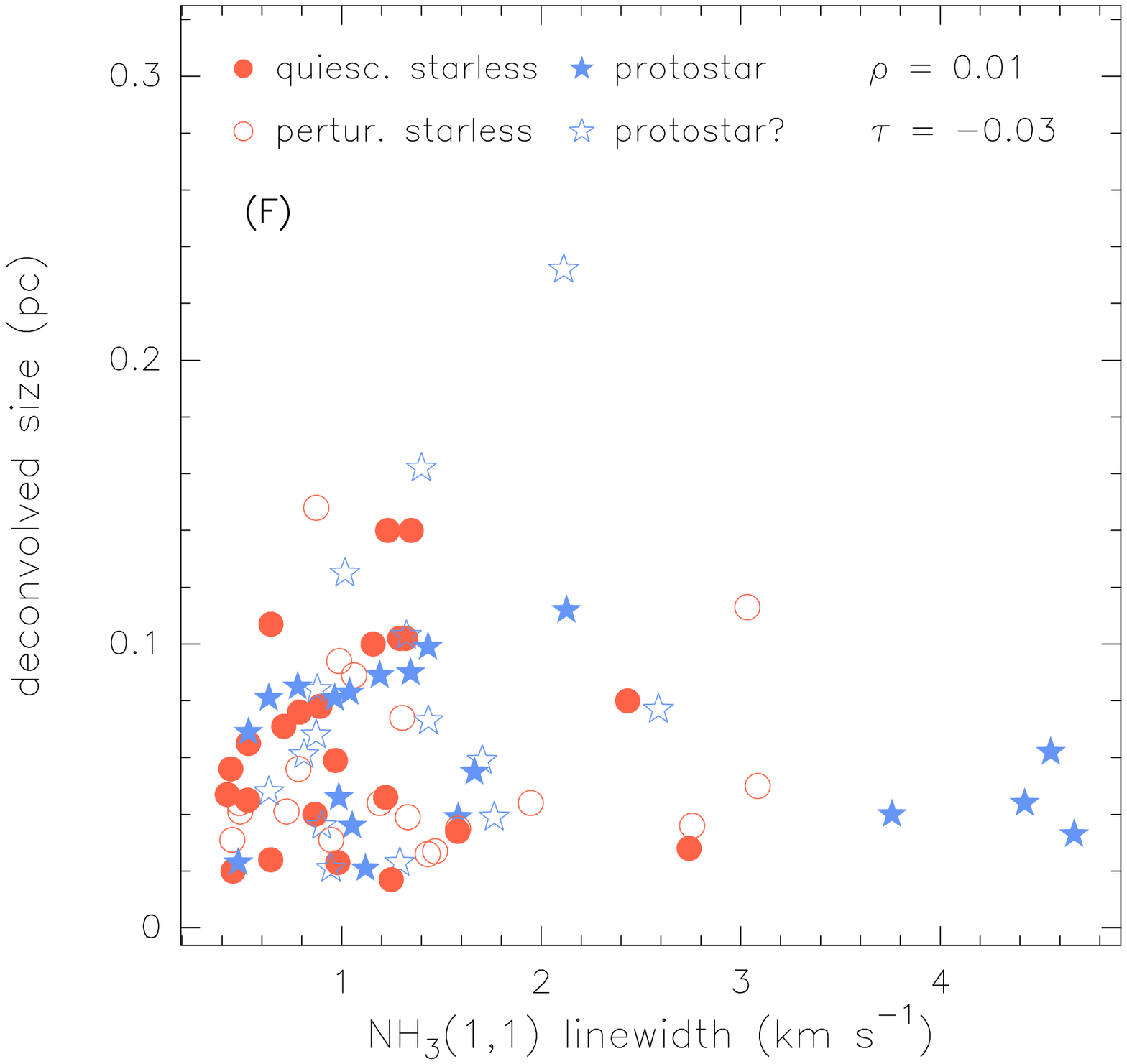, scale=0.3} \\
\end{tabular}
\caption{Scatter plots showing the correlations with respect to the bolometric luminosity of {\bf (a)} the rotational temperature, {\bf (b)} the \nh\,(1,1) linewidth, and {\bf (c)} the core mass, in the top panels; and the correlations with respect to the linewidth of {\bf (d)} the rotational temperature, {\bf (e)} the ammonia column density, and {\bf (f)} the core diametre. `Quiescent starless' and `perturbed starless' cores are shown in red filled and open circles respectively; while `protostellar' and `tentatively protostellar' cores are shown as blue filled and open stars, respectively (see Sect.~\ref{s:cores} for details). The black dotted lines show the result of least-squares fits to the data (see Sect.~\ref{s:correlations}), with the Pearson correlation coefficient ($\rho$) shown in the top right corner, together with Kendall's $\tau$ correlation coefficient.}
\label{f:correlations}
\end{center}
\end{figure*}

As presented in Sect.~\ref{s:dense}, we note some differences in the parameters derived for the starless and protostellar cores. In particular, an increase of the temperature and linewidth as we move toward more evolved objets (protostellar cores). In a search for a physical relation between the derived parameters, we plotted several quantities against others, shown as six scatter plots in Fig.~\ref{f:correlations}. In these plots we show the distribution of starless and protostellar core samples in red-circles and blue-stars, respectively. We performed some statistical tests to search for possible correlations between the physical quantities. In the top right corner of each plot, we give Kendall's $\tau$ rank correlation coefficient and the Pearson correlation coefficient ($\rho$) to a least-squares fit (dotted lines). We find that in five of the six plots there is a possible correlation, with $\tau$ coefficients $\sim 0.3$, and $\rho$ in the range 0.4--0.6.

In the top panels of Fig.~\ref{f:correlations}, we show the relation between the temperature, the linewidth and the mass of the core with the bolometric luminosity of the region, while in the bottom panels, we compare the temperature, the ammonia column density and the size of the core with the observed linewidth. We note that we are not able to derive a bolometric luminosity for each individual ammonia core but a single luminosity of the whole region\footnote{To estimate the bolometric luminosity $L_\mathrm{bol}$ (listed in Table~\ref{t:nh3sources}), we compiled the flux densities at different frequencies from 2MASS, \emph{Spitzer}/IRAC+MIPS, WISE, MSX, IRAS and JCMT (SCUBA; \citealt{difrancesco2008}) catalogues and followed the methodology explained in \citet{palau2013}.}. Thus, the plots in Fig.~\ref{f:correlations} must be interpreted as a correlation between the properties of the cores within a region with respect to the luminosity of the region (which is in general dominated by the brightest mid-IR source). We find a correlation in five of the six panels (panels A to E of Fig.~\ref{f:correlations}). Similar trends between the same parameters have been reported from single-dish observational studies \citep[\eg][]{churchwell1990, myers1991, jijina1999, rygl2010, urquhart2011, sepulveda2011}. Our correlations found in ammonia at higher angular resolution thus confirm that these relations take place also at the smaller spatial scales ($\sim 0.05$~pc) of individual cores, and are not an observational (average) artifact due to the poor angular resolution of single-dish observations when studying clumps of sizes $\sim 0.1$--10~pc. It is worth noting that in panel B (linewidth vs luminosity) there are four cores with linewidths $\sim 4$~\kms, located well above the trend found for the other cores. These four cores (AFGL\,5142--1; 19035+0641--1; 20126+4104--1; G75.78+0.34--1) are known to be associated with powerful outflows that could perturb the motions of the dense gas close to the powering source \citep[see][]{zhang2002, cesaroni2005, sanchezmongephd2011, sanchezmonge2013}. Finally, in panel F, we investigate a possible relation between the size of the core and its linewidth. Since the publication of the Larson's scaling relations (Larson 1981), several authors \citep[\eg][]{myers1983, casellimyers1995, pirogov2003} found, from single-dish studies, that larger dense cores have broader lines, as expected in virialised cores. In our sample, however, we find no correlation (with $\tau= -0.03$ and $\rho= 0.01$) between these two quantities. A possible explanation for this could be that a number of cores suffer an important injection of turbulence, \eg\ from the passage of powerful outflows, which should affect the gas motions of the cores only at the scales of our sample ($\sim 0.05$~pc; see \eg\ \citealt{palau2007b, liu2010, nakamurali2011}), and is diluted at the larger scales studied in single-dish observations. Alternatively, the large linewidths might indicate that the cores are collapsing. In Sect.~\ref{s:dynamics}, we discuss in more detail the dynamics of the cores in our sample, and relate it to previous single-dish studies.

\begin{figure}
\begin{center}
\begin{tabular}[b]{c}
  \epsfig{file=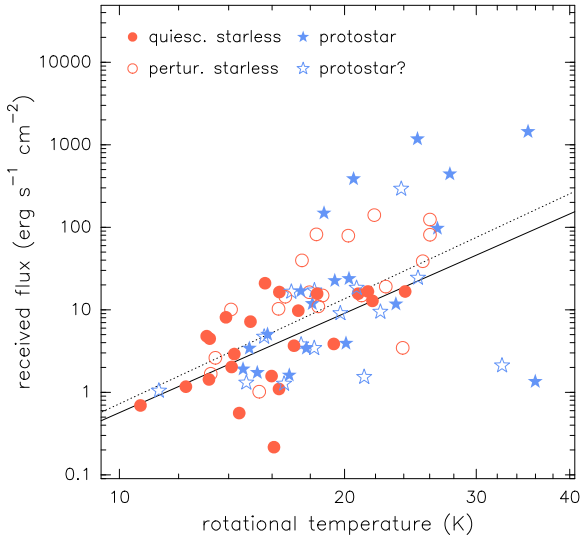, scale=0.42} \\
\end{tabular}
\caption{Received flux for each ammonia core versus the rotational temperature (see Sect.~\ref{s:IRNH3} for details). The dotted line shows a least-squares fit to the data, and the solid line corresponds to the relation $L_\mathrm{bol}/(4\pi(\Delta_\mathrm{MIR-NH_3})^2)=\sigma_\mathrm{SB}\,T_\mathrm{rot}^4$, with $\sigma_\mathrm{SB}$ the Stephan-Boltzmann constant. Symbols are as in Fig.~\ref{f:correlations}.}
\label{f:IRNH3}
\end{center}
\end{figure}

\section{Discussion}\label{s:discussion}

\subsection{Temperature of the gas and IR radiation}\label{s:IRNH3}

In Sect.~\ref{s:correlations}, we reported a correlation between the temperature of the cores and the luminosity of the region (see Fig.~\ref{f:correlations} panel A). These higher temperatures can result from various heating processes: (i) feedback from low-mass stars, \eg\ molecular outflows, as found in some cores of AFGL\,5142 or 20293+5932 \citep[\eg][]{zhang2002, palau2007b}; (ii) shocks from convergent flows \citep[\eg][]{csengeri2011a, beltran2012}; (iii) feedback from high-mass stars, \eg\ strong radiation \citep[\eg][]{li2003}. In the following, we test the feedback from massive stars as a possible agent for the increase of temperature in the dense cores.

Our interferometric observations permit the study of the spatial distribution of the cores, and their physical properties, with respect to, \eg\ the (strong) infrared sources and hence the massive stars. In our regions, the bolometric luminosity is generally dominated, at mid-infrared (MIR) and far-infrared (FIR) wavelengths, by a single strong source associated with the most massive star. We assume that the main source of radiation that can heat the surrounding gas is the massive star detectable in the MIR. To determine if there is a relation between the temperature of the cores and the proximity of the MIR source, we searched the MSX Point Source Catalog \citep{price1999} for the coordinates of the MIR sources and calculated the MSX to \nh\ core separation ($\Delta_\mathrm{MIR-NH_3}$; using the coordinates listed in Table~\ref{t:nh3cores} for our cores). Using the luminosity and position of the most massive (radiating dominant) source and the separation to each ammonia dense core, we derived the incident flux received by each core as $L_\mathrm{bol}/(4\pi(\Delta_\mathrm{MIR-NH_3})^2)$. Note that we are measuring a projected separation, which is a lower limit to the true separation. Thus, for all cores, we actually have an upper limit on the received flux, which affects the absolute normalization of the derived relation and increases the scatter of the data points in this relation. In Fig.~\ref{f:IRNH3}, we plot the received flux versus the rotational temperature. If the radiation of the most massive star dominates the heating of its surroundings, we expect that the cores closer to the most massive source, \ie\ with a large received flux, will have higher temperatures. This is consistent with the trend shown in Fig.~\ref{f:IRNH3}, where the highest temperatures correspond to the cores with a higher received flux. The black solid line corresponds to the relation $L_\mathrm{bol}/(4\pi(\Delta_\mathrm{MIR-NH_3})^2)=\sigma_\mathrm{SB}\,T_\mathrm{rot}^4$, with $\sigma_\mathrm{SB}$ the Stephan-Boltzmann constant, while the dotted line corresponds to a least-squares fit with a correlation coefficient of 0.6. For a given bolometric luminosity, the temperature is a factor of $\sim 3$ times lower for a core located at a distance $\sim 0.5$~pc, than for a similar core located at $< 0.05$~pc.

Summarising, the radiation deposited in the environment by the most luminous star of the cluster can be responsible for the temperatures measured in the cores of our clustered regions. Thus, the large temperatures measured for our starless cores, in comparison with the temperatures measured in low-mass star-forming regions, can be interpreted as external heating by the radiation of the most massive (forming) stars of the cluster.

\begin{figure}
\begin{center}
\begin{tabular}[b]{c}
  \epsfig{file=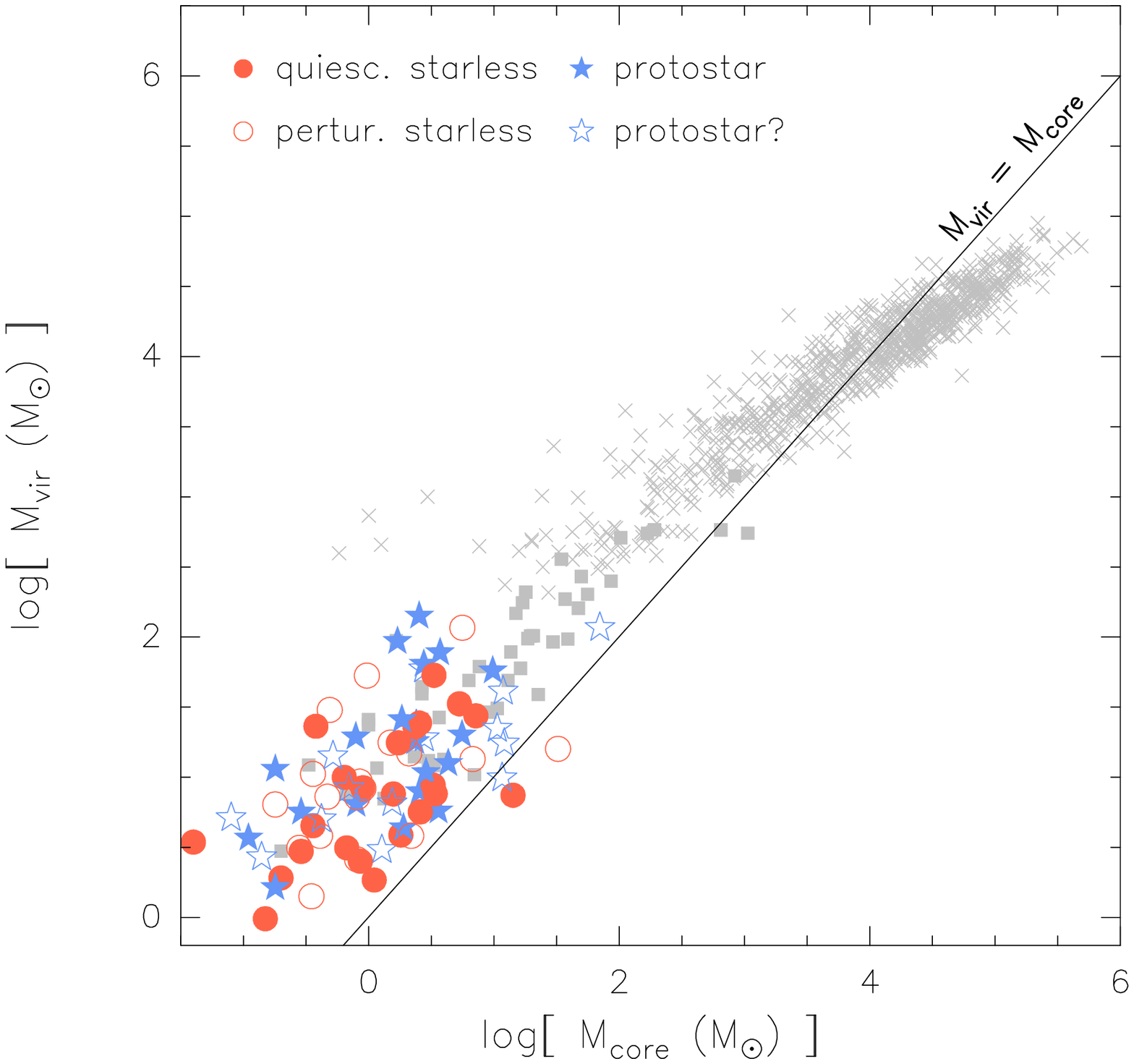, scale=0.42} \\
\end{tabular}
\caption{Scatter plot showing the relation between the virial mass $M_\mathrm{vir}$ (derived as $M_\mathrm{vir}=210\,R\,\Delta v^2$; \citealt{maclaren1988}) and the mass of the core $M_\mathrm{core}$. Blue-star and red-circle symbols correspond to the protostellar and starless cores studied in this work (symbols as in Fig.~\ref{f:correlations}). Grey squares correspond to single-dish ammonia data \citep{sepulveda2011} and grey crosses correspond to $^{13}$CO single-dish data \citep{romanduval2010}. The solid line corresponds to virialised cores with $M_\mathrm{vir} = M_\mathrm{core}$.}
\label{f:Mvir}
\end{center}
\end{figure}

\begin{figure}
\begin{center}
\begin{tabular}[b]{c}
  \epsfig{file=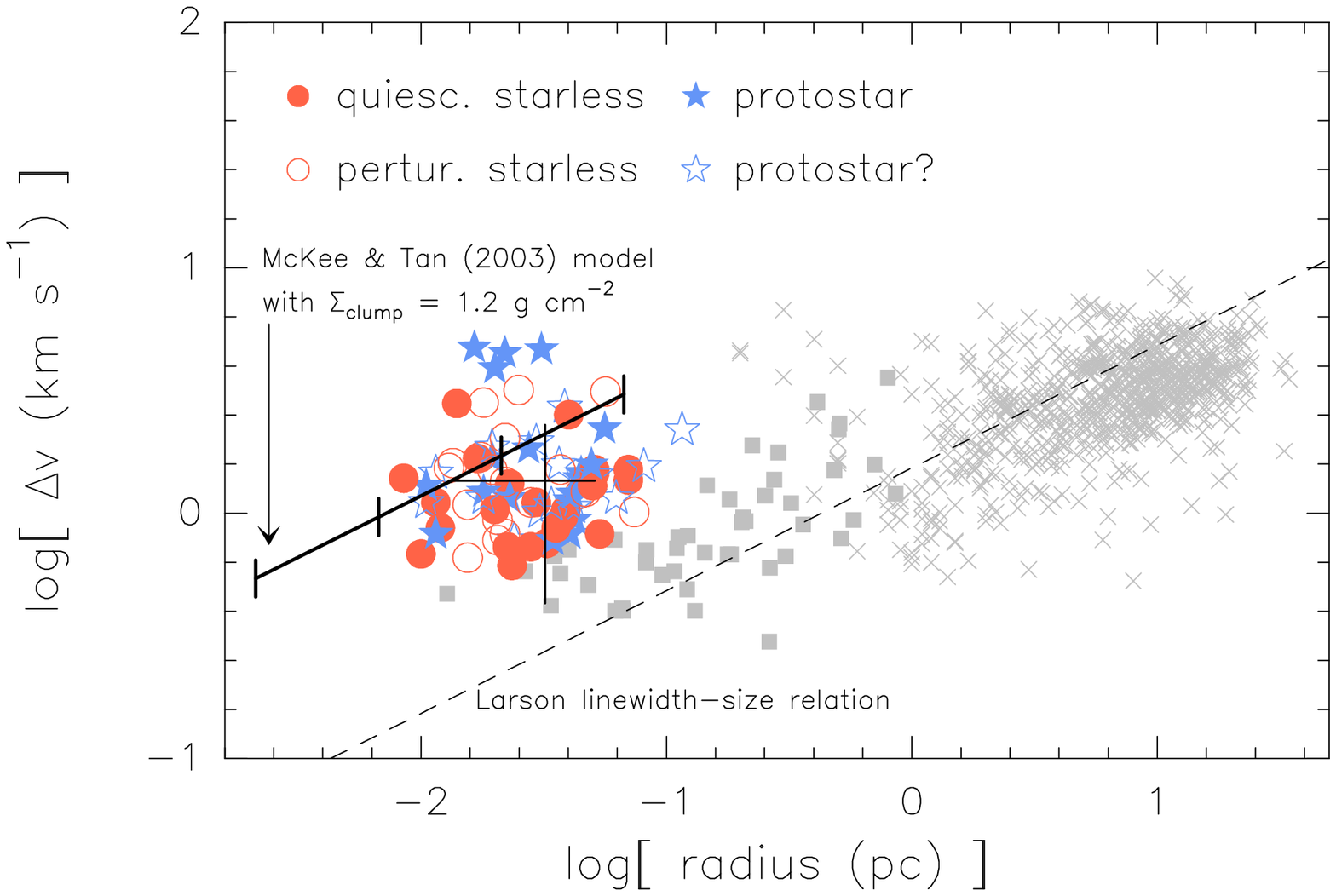, scale=0.42} \\
\end{tabular}
\caption{Variation of the linewidth, $\Delta v$, with core radius, following \citet{csengeri2011b}. Red and blue symbols correspond to the VLA ammonia data shown in this paper (see Fig.~\ref{f:correlations} for details), grey squares correspond to single-dish ammonia data \citep{sepulveda2011}, and grey crosses correspond to $^{13}$CO single-dish data \citep{romanduval2010}. The dashed line indicates a Larson like linewidth-size relation ($\Delta v \propto \mathrm{size}^{0.38}$). The thick solid line correspond to the turbulent regulated quasi-static model \citep{mckeetan2003}, with a surface mass density of 1.3~g~cm$^{-2}$ and a core mass of 0.1 to 100~\mo\ (each interval corresponds to 0.1, 1, 10, and 100~\mo). The black cross marks the mean and dispersion values of the linewidth ($1.36\pm0.93$~\kms) and radius ($0.032\pm0.019$~pc) for the cores studied in this work.}
\label{f:linesize}
\end{center}
\end{figure}

\subsection{Dynamical state of the cores}\label{s:dynamics}

We estimated the virial mass of each core as $M_\mathrm{vir}=210\,[R/\mathrm{pc}]\,[\Delta v/\mathrm{km~s^{-1}}]^2$~\mo\ (\citealt{maclaren1988}; see also \citealt{bertoldimckee1992}), where the numerical value accounts for an uniform density distribution across the core, and $R$ and $\Delta v$ are the radius of the core and the \nh\,(1,1) linewidth, respectively (both listed in Table~\ref{t:nh3cores}). Note that this estimate considers the simplest case of virial equilibrium where only gravity and velocity dispersion of the gas are taken into account, thus neglecting the effects of, \eg\ external pressure and magnetic fields. The calculated virial masses are listed in Col.~10 of Table~\ref{t:nh3extra}. In Figure~\ref{f:Mvir}, we present a plot of virial mass, $M_\mathrm{vir}$, against the mass of the core derived from the ammonia emission, $M_\mathrm{core}$. In Fig.~\ref{f:Mvir}, we have also included the ammonia clumps studied by \citet{sepulveda2011} with the Haystack~37\,m single-dish telescope (grey squares), and the $^{13}$CO clouds of \citet{romanduval2010} observed with the FCRAO~14\,m single-dish telescope (grey crosses). These two samples permit to trace clumps at scales $\sim 0.1$--10~pc. From the plot we find that the trend outlined by this population of clumps is nicely followed by our \nh\ core population sampling lower masses, with the slope of the trend close to 1. This, in general, favors a scenario of a self-similar, self-gravitating, virialised hierarchy of structures from clumps to cores. The fact that the virial masses of almost all the cores at 0.05~pc (studied in this work) are several times larger than the core masses ($M_\mathrm{vir}/M_\mathrm{core}\sim 10$)\footnote{Note that the offset seen in Fig.~\ref{f:Mvir} can not be easily explained in terms of the uncertainties of the observational quantities involved in the plot. The two main sources of uncertainty are the ammonia abundance and the linewidth. For the ammonia abundance we assumed a mean value of $(4.2\pm2.7)\times10^{-8}$. This value is obtained from different surveys \citep{pillai2006, foster2009, friesen2009, rygl2010, chira2013}, over a total of 122 clumps of different star forming regions. Thus, we expect an uncertainty of $\sim 60$\% for the mass of the core. The linewidth is obtained from the spectral line fit described in Sect.~\ref{s:dense}, which provides errors of a few tenths of \kms, corresponding to a 10--30\% uncertainty in linewidth. These observational errors alone are expected to affect the data randomly instead of introducing a systematic offset, and thus are not likely the cause of the observed offset of about one order of magnitude.} suggests that the initial assumptions in our virial analysis might be too simplistic and that other mechanisms of support should be taken into account (\eg\ external pressure, magnetic fields, large scale velocity gradients; see \citealt{maclowklessen2004, mckeeostriker2007} and references therein). The turbulence-regulated quasi-static scenario proposed by \citet{mckeetan2002, mckeetan2003} considers the effects of the external pressure and an internal pressure that is derived from a mixture of turbulence and large-scale magnetic fields, to characterise the evolution of dense cores until the formation of massive stars. Thus, a virialised core that is in pressure equilibrium with its natal self-gravitating clump has a mean velocity dispersion, $\sigma_\mathrm{vir}$, and a core radius, $R_\mathrm{vir}$, defined by
\begin{equation}
\bigg[\frac{\sigma_\mathrm{vir}}{\mathrm{km~s}^{-1}}\bigg] = 1.09\,\bigg[\frac{M_\mathrm{core}}{60~M_{\sun}}\bigg]^{1/4}\,\bigg[\frac{\Sigma_\mathrm{clump}}{\mathrm{g~cm}^{-2}}\bigg]^{1/4},
\end{equation}
\begin{equation}
\bigg[\frac{R_\mathrm{vir}}{\mathrm{pc}}\bigg] = 0.057\,\bigg[\frac{M_\mathrm{core}}{60~M_{\sun}}\bigg]^{1/2}\,\bigg[\frac{\Sigma_\mathrm{clump}}{\mathrm{g~cm}^{-2}}\bigg]^{-1/2},
\end{equation}
where the numerical values apply to the fiducial cores of \citet{mckeetan2003} (see also \citealt{tan2013}) with a power-law density distribution $\rho \propto r^{-k_\rho}$ with $k_\rho=1.5$. $M_\mathrm{core}$ corresponds to the mass of the core, and $\Sigma_\mathrm{clump}$ is the mass surface density of the clump where the core is embedded, assumed to be equal to the external pressure \citep{mckeetan2003}. In Fig.~\ref{f:linesize}, we follow the approach of \citet{csengeri2011b} to compare our observations (linewidth and core size) to the turbulent core model of \citet{mckeetan2003} (represented by the thick black line). The turbulent dispersion and the size of the cores of the model have been computed for a 0.1 to 100~\mo\ core mass, and adopting $\Sigma_\mathrm{clump}=1.2$~g~cm$^{-2}$ as an average mass surface density of the clump, which has been estimated for our regions from single-dish (sub)millimetre continuum observations (see Table~\ref{t:clumps} for details). From Fig.~\ref{f:linesize}, we find that the cores studied at scales of 0.05~pc have a mean and dispersion linewidth and radius of $1.36\pm0.93$~\kms\ and $0.032\pm0.019$~pc, respectively. These values, indicated by a black cross in Fig.~\ref{f:linesize}, are slightly below the predictions of the turbulent core model, a result recently found also by \citet{csengeri2011b} in a smaller sample of cores in the Cygnus-X region. This suggests that, when taking into account the external pressure effect at scales $\sim 0.05$~pc, either other forms of support (\eg\ strong magnetic fields at core scales) are required to stabilise the cores, or the cores are close to collapse or even collapsing. Observations to gather information on the magnetic field as well as on the internal kinematics of the cores should definitively confirm this picture of the \nh\ dense cores at 0.05~pc in our sample.

\begin{table}
\caption{Parameters derived for the clumps associated with the 15 massive star-forming regions}
\label{t:clumps}
\centering
\begin{tabular}{c c c c c c c}
\hline\hline

&$S_\nu$\supb
&Mass\supc
&FWHM\supd
&diametre\supd
&$\Sigma_\mathrm{clump}$\supe
\\
\#\supa
&(Jy)
&(\mo)
&(\arcsec)
&(\arcsec)~/~(pc)
&(g~cm$^{-2}$)
\\
\hline
\phn1.	&\phn1.2	&\phnn85		&32.4	&30.5~/~0.27		&0.32	\\
\phn2.	&15.3	&\phn278		&28.7	&25.1~/~0.22		&1.55	\\
\phn3.	&\phn6.8	&\phn123		&44.2	&41.9~/~0.37		&0.25	\\
\phn4.	&21.1	&\phn384		&40.2	&37.7~/~0.33		&0.94	\\
\phn5.	&\phn4.5	&\phnn64		&30.1	&26.6~/~0.21		&0.40	\\
\phn6.	&\phn7.3	&\phn199		&31.9	&28.7~/~0.31		&0.57	\\
\phn7.	&32.9	&1160		&30.0	&26.5~/~0.32		&2.97	\\
\phn8.	&13.3	&\phn201		&34.5	&31.5~/~0.25		&0.85	\\
\phn9.	&28.3	&2290		&37.8	&35.1~/~0.65		&1.46	\\
10.		&15.2	&\phn341		&45.0	&42.8~/~0.42		&0.53	\\
11.		&15.8	&\phn174		&56.5	&54.7~/~0.37		&0.34	\\
12.		&\phn6.7	&\phn253		&38.5	&35.9~/~0.45		&0.33	\\
13.		&\phn4.5	&\phn145		&34.6	&31.6~/~0.37		&0.29	\\
14.		&\phn5.2	&\phnn17		&16.5	&\phn8.8~/~0.03	&4.30	\\
15.		&67.1	&\phn185		&41.6	&39.2~/~0.13		&2.78	\\
\hline
\end{tabular}
\begin{flushleft}
\supa\ Number of the region as indicated in Table~\ref{t:nh3sources}.\\
\supb\ Flux density at 850~$\mu$m obtained from the SCUBA Legacy Project \citep{difrancesco2008}, except for: 00117+6412 (region \#1: 1.2~mm from IRAM~30\,m, \citealt{sanchezmonge2008}); 20081+3122 (region \#7: 850~$\mu$m from JCMT, \citealt{thompson2006}); 22198+6336 (region \#14: 850~$\mu$m from JCMT, \citealt{jenness1995}).\\
\supc\ Mass of the clump measured from the JCMT at 850~$\mu$m or the IRAM~30\,m at 1.2~mm, assuming a dust temperature of 20~K, a dust (+gas) mass opacity coefficient of 0.0175~cm$^{2}$~g$^{-1}$ at 850~$\mu$m and 0.0090~cm$^{2}$~g$^{-1}$ at 1.2~mm \citep{ossenkopfhenning1994}, and a gas-to-dust mass ratio of 100.\\
\supd\ Observed diametre at full width half maximum obtained (FWHM) from a Gaussian fit, and deconvolved diametre derived taking into account a main beam for the SCUBA observations at 850~$\mu$m of 14\arcsec. Our sizes obtained from a Gaussian fit are slightly different to those reported by \citet{palau2013}, which were estimated by measuring the observed diametre directly on the image, and using a beam size of 23\arcsec\ for deconvolution, which affects mainly the sources with the smallest sizes. \\
\supe\ Clump mass surface density derived as $\Sigma_\mathrm{clump}=\mathrm{Mass}/\mathrm{Area}$ with Area $=\pi\,(\mathrm{diametre/2})^2$.\\
\end{flushleft}
\end{table}

\section{Conclusions}\label{s:conclusions}

We present the results of VLA observations of the \nh\,(1,1) and (2,2) inversion transition lines towards a sample of 15 massive star-forming regions. We obtained an average spatial resolution of $\sim 0.05$~pc, and analysed the ammonia emission in a consistent way. Our main conclusions can be summarised as follows:

\begin{itemize}
\item[-] We identified a total of 73 ammonia dense cores using the two-dimensional CLUMPFIND algorithm. On average we found of the order of 5--6 cores in each field (of $\sim 2$~arcmin, or 1.5~pc at a distance of 2.5~kpc), with sizes $\sim 0.06$~pc. The cores were classified as quiescent starless, perturbed starless, and protostellar based on its association with infrared emission and taking into account other star formation signposts (\eg\ outflows, masers, centimetre continuum emission) reported in the literature.

\item[-] The average values of linewidth, opacity, rotational temperature, and ammonia column density for the 73 dense cores are $\sim 1.4$~\kms, 1.4, 19~K, and $10^{15}$~cm$^{-2}$, respectively. We found that protostellar cores have larger temperatures and linewidths ($\sim 21.3$~K and 1.8~\kms) than quiescent starless cores ($\sim 16.0$~K and 1.0~\kms). Similarly, cores classified as perturbed starless have temperatures and linewidths ($\sim 19.3$~K and 1.4~\kms) slightly larger than quiescent starless cores, but smaller than cores harboring a protostar.

\item[-] The average temperatures derived for the quiescent starless cores in our clustered regions ($\sim 16$~K) are larger than those measured in starless cores in more isolated regions ($\sim 10$~K). Our starless and protostellar cores show important non-thermal supersonic components ($\sigma_\mathrm{nth}\sim 0.44$~\kms\ and 0.76~\kms, respectively) and $\sigma_\mathrm{nth}/c_s\sim 2.0$, well above to the close-to-thermal lines measured in isolated star-forming regions.

\item[-] We found that the temperature and \nh\,(1,1) linewidth of the cores are correlated with the luminosity of the region, suggesting that the feedback of the luminous massive (forming) stars can affect the surrounding molecular environment.

\item[-] We evaluated a possible relation between the temperature of the core and its separation with the most important radiating source in the region, in terms of the incident flux received by the core. The highest temperatures are measured for those cores located close to the brightest MIR source (separations $< 0.05$~pc), with the temperature decaying by a factor of $\sim 3$ when the cores are located at distances $\sim 0.5$~pc. Thus, the large temperatures found in starless cores in massive star-forming regions, can be interpreted as external heating by the radiation of most the massive star of the cluster.

\item[-] We compared the virial masses of our cores (at scales $\sim 0.05$~pc) with those of clumps with sizes $\sim 0.1$--10~pc. A simple virial equilibrium analysis seems to suggest a trend between $M_\mathrm{vir}$ and core/clump masses favoring a scenario of a self-similar, self-graviting, virialised hierarchy of structures from clumps to cores. A closer inspection of the dynamical state at core scales, taking into account external pressure \citep{mckeetan2003}, indicates that either another form of support (\eg\ from magnetic field) is required to stabilise the cores, or that the cores are close to collapse or even collapsing.

\end{itemize} 

Overall, our data suggest that the cores forming in clustered environments harboring intermediate/high-mass stars are more affected/perturbed than cores in clustered low-mass star-forming regions, and thus the initial conditions for star formation in these environments seem to be different to those typically assumed in low-mass star-forming regions.

\section*{Acknowledgments}

We are grateful to James di Francesco for support with SCUBA data. We thank the anonymous referee for his/her constructive criticisms. The figures of this paper have been done in GREG of the GILDAS software package developed at the Institut de Radioastronomie Millim\'etrique (IRAM) and the Observatoire de Grenoble. A.P. is supported by a JAE-Doc CSIC fellowship co-funded with the European Social Fund under the program `Junta para la Ampliaci\'on de Estudios', by the Spanish MICINN grant AYA2011-30228-C03-02 (co-funded with FEDER funds), and by the AGAUR grant 2009SGR1172 (Catalonia). G.B. is funded by an Italian Space Agency (ASI) fellowship under contract number I/005/11/0.


\appendix

\section[]{HyperFine Structure (HfS) fitting}\label{a:hfs}

\subsection{Parameters of the fit}

The usual fitting of the hyperfine structure of a transition, for instance, the inversion transition NH$_3$ $(J,K)=(1,1)$, estimates four free parameters of the transition, namely 
\begin{enumerate}
\renewcommand{\labelenumi}{\roman{enumi})}
\item $\tau_m$, the optical depth of the main component, 
\item $A\times\tau_m$, product of the amplitude ($A=f[J_\nu(T_\mathrm{ex})-J_\nu(T_\mathrm{bg})]$, where $f$ is the filling factor, $J_\nu(T)$ is the Planck function in units of temperature, $T_\mathrm{ex}$ is the excitation temperature, and $T_\mathrm{bg}$ is the background temperature) times the optical depth of the main component, 
\item $v_\mathrm{LSR}$, the central velocity of the main component, and 
\item $\Delta v$, the linewidth of the hyperfine components,
\end{enumerate}
with the usual assumptions that $T_\mathrm{ex}$, $v_\mathrm{LSR}$, and $\Delta v$ are the same for all the hyperfine components.

However, some of the parameters are ill-conditioned in some cases. In the optically thin limit, $\tau_m\ll1$, only an upper limit for the value of $\tau_m$ can be obtained. However, the parameter $A\times\tau_m$ gives the peak intensity of the line (provided that the hyperfine components are unresolved, or, in the case NH$_3$, the hyperfine magnetic components that form part of the main component), and is well constrained. This parameter, in the Rayleigh-Jeans approximation, is necessary to calculate the column density of emitting molecules. In the optically thick limit, $\tau_m\gg1$, only lower limits for $\tau_m$ and $A\times\tau_m$ can be obtained (and consequently, a lower limit for the column density). 

A better fit parameter than $\tau_m$, with well defined values in the optically thin and optically thick limits, is $1-e^{-\tau_m}$. This parameter always has values between 0 (optically thin limit, $\tau_m\ll1$) and 1 (optically thick limit, $\tau_m\gg1$). Similarly, $A\left(1-e^{-\tau_m}\right)$ is a better fit parameter than $A\times\tau_m$, since it is well constrained in all cases.

The Hyperfine Structure\footnote{HfS is freely available and can be downloaded from http://www.am.ub.edu/$\sim$robert/hfs/hfs.html} (hereafter HfS) fitting procedure used to fit the NH$_3$ $(J,K)=(1,1)$ lines in this paper, fits the four parameters 
\begin{enumerate}
\renewcommand{\labelenumi}{\roman{enumi})}
\item $1-e^{-\tau_m}$,
\item $A\left(1-e^{-\tau_m}\right)$,
\item $v_\mathrm{LSR}$, 
\item $\Delta v$.
\end{enumerate}
The values and uncertainties for the parameters needed to derive physical properties of the emitting gas, $\tau_m$ and $A\times\tau_m$ (or their limits in the two limiting cases) can be easily derived from the values and uncertainties for the parameters of the fit.

\subsection{Fitting procedure}

The fitting strategy was similar to that used by \citet{estalella2012}. HfS samples the parameter space of dimension four, defined by the parameters $1-e^{-\tau_m}$, $A\left(1-e^{-\tau_m}\right)$, $v_\mathrm{LSR}$, and $\Delta v$, to find the minimum value of $\chi^2$,
\begin{equation}
\chi^2\equiv\sum_{i=1}^n
\left[
\frac{y_i^\mathrm{obs}-y_i^\mathrm{mod}(p_1,\ldots,p_m)}{\sigma_i}
\right]^2,
\end{equation}
where $y_i^\mathrm{obs}$ are the observed line intensities for the $n$ spectral channels, $y_i^\mathrm{mod}(p_1,\ldots,p_m)$ are the model line intensities, depending on $m$ free parameters $p_k$, and $\sigma_i$ are the errors of the observations.

If the model is a good fit to the observations, $\chi^2$ has a minimum value given by 
\begin{equation}
\chi^2_\mathrm{min}\simeq n-m. 
\end{equation}
This can be used to estimate the observational error, $\sigma$ (the same for all points), if unknown,
\begin{equation}
\sigma^2\simeq\frac{1}{n-m}\sum_{i=1}^n
\left[y_i^\mathrm{obs}-y_i^\mathrm{mod}\right]^2.
\end{equation}

Several sampling methods of the $m$-dimensional parameter space are possible, \ie\ regular grid, random, Halton sequence \citep{halton1964}. We adopted a Halton quasi-random sequence because it samples the parameter space more evenly than a purely random sequence, and the convergence of the fitting procedure to the minimum of $\chi^2$ is faster. 

\begin{table}
\caption{Values of $\Delta(m,\alpha)$ for calculating the parameter uncertainties, where $m$ is the number of fitted parameters simultaneously, and $\alpha$ is the significance level, given in percent and in the equivalent number of sigmas for a Gaussian error distribution.}
\label{t:hfs}
\begin{center}
\begin{tabular}{rrrrr}
\hline\hline
&
\multicolumn{4}{c}{$\alpha$}\\
\cline{2-5}
& 
\multicolumn{1}{c}{68.23\%}& 
\multicolumn{1}{c}{90.00\%}& 
\multicolumn{1}{c}{99.00\%}& 
\multicolumn{1}{c}{99.90\%}\\ 
\multicolumn{1}{c}{$m$}& 
\multicolumn{1}{c}{($1\sigma$)}& 
\multicolumn{1}{c}{($1.64\sigma$)}& 
\multicolumn{1}{c}{($2.57\sigma$)}& 
\multicolumn{1}{c}{($3.30\sigma$)}\\ 
\hline
 1&  1.00&  2.71&  6.64& 10.83\\
 2&  2.30&  4.60&  9.21& 13.82\\
 3&  3.53&  6.25& 11.34& 16.27\\
 4&  4.72&  7.78& 13.28& 18.46\\
 5&  5.89&  9.24& 15.09& 20.52\\
 6&  7.04& 10.64& 16.81& 22.46\\
 7&  8.17& 12.02& 18.48& 24.32\\
 8&  9.30& 13.36& 20.09& 26.12\\
 9& 10.42& 14.68& 21.67& 27.87\\
10& 11.53& 15.99& 23.21& 29.59\\
\hline
\end{tabular}
\end{center}
\end{table}

\subsection{Parameter error estimation}

Once the best fit parameter values for which $\chi^2=\chi^2_\mathrm{min}$ are found, the uncertainty in the fitted parameters, $\sigma(p_k)$, $(k=1\ldots m)$, can be estimated as the projection over each parameter axis of the confidence region of the parameters, the region of the $m$-dimensional parameter space for which $\chi^2$ does not  exceed the minimum value by a certain amount, $\chi^2_\mathrm{min}+\Delta$.

Following \citet{avni1976} and \citet{walljenkins2003}, the probability 
\begin{equation}
\mathrm{Prob}\,[\chi^2-\chi^2_\mathrm{min}\le\Delta(m,\alpha)]=\alpha,
\end{equation}
is that of a chi-square distribution with $m$ degrees of freedom, and $\alpha$ is the significance level ($0<\alpha<1$). Thus, $\Delta(m,\alpha)$ is the increment in $\chi^2$ such that if the observation is repeated a large number of times, a fraction $\alpha$ of times the values of the fitted parameters will be inside the confidence region, i.e.\  in the interval $p_k\pm\sigma(p_k)$. For instance, in the case of this paper, $m=4$, and for a significance level of 0.68 (equivalent to 1 $\sigma$ for a Gaussian error distribution), the value of the increment in $\chi^2_\mathrm{min}$ is $\Delta(4,0.68)=4.72$ (\citealt{lampton1976}; see Table \ref{t:hfs}).

Assuming that the model is a good fit to the observations, \ie\ $\chi^2_\mathrm{min}\simeq n-m$, the condition 
\begin{equation}
\chi^2=\chi^2_\mathrm{min}+\Delta(m,\alpha),
\end{equation}
can be written as
\begin{equation}
\frac{\chi^2}{\chi^2_\mathrm{min}}\simeq1+\frac{\Delta(m,\alpha)}{n-m}.
\end{equation}
This expression is useful since it can be given in terms of the rms fit residual, $\epsilon$,
\begin{equation}
\epsilon^2\equiv
\frac
{\sum_{i=1}^n\left[(y_i^\mathrm{obs}-y_i^\mathrm{mod})/\sigma_i\right]^2}
{\sum_{i=1}^n 1/\sigma_i^2}=
\frac{\chi^2}{\sum_{i=1}^n 1/\sigma_i^2},
\end{equation}
for which we obtain that the confidence region is given by the parameter values that increase the rms fit residual to
\begin{equation}
\epsilon\simeq\epsilon_\mathrm{min}
\sqrt{1+\frac{\Delta(m,\alpha)}{n-m}}.
\end{equation}
This last expression can be used even when the errors of the observations are unknown.

\section[]{Velocity, linewidth and spectra images}\label{a:extra}

\begin{table*}
\caption{Classification of ammonia cores in protostellar or starless}
\label{t:nh3types}
\centering
\begin{tabular}{l c c c c c c c c l}
\hline\hline
\texttt{ID}~Region\supa
&\#
&Type\supb
&IRAC
&MIPS/WISE
&mm
&cm
&outflow
&maser
&comments
\\
\hline
01~00117+6412   &01 &P   &point  &-      &yes  &no   &no    &no    & \\
02              &02 &S   &dark   &-      &yes  &no   &no    &no    & \\
03              &03 &P   &point  &yes    &yes  &yes  &yes   &yes   & \\ 
04              &04 &S*  &dark   &-      &yes  &no   &no    &yes   & \\
05              &05 &P?  &point  &-      &no   &no   &no    &no    &no mm/outflow/maser associated \\

06~AFGL5142     &01 &P   &point  &yes    &yes  &yes  &yes   &yes   & \\
07              &02 &P?  &point  &-      &no   &no   &OA    &no    &core along outflow axis \\
08              &03 &P?  &point  &-      &no   &no   &no    &no    & \\ 

09~05345+3157NE &01 &P   &point  &-      &yes  &yes  &no?   &no    & \\
10              &02 &P   &point  &yes    &yes  &no   &yes   &yes   & \\
11              &03 &S   &dark   &-      &no   &no   &no    &no    & \\ 
12              &04 &S   &dark?  &yes?   &yes  &no   &no    &no    &point IR source at the border, but dark \\
13              &05 &S   &dark   &-      &yes  &no   &no    &no    & \\
14              &06 &S   &dark   &-      &yes  &no   &no    &no    & \\
15              &07 &S   &dark   &-      &no   &no   &no    &no    & \\

16~05358+3543NE &01 &P   &ext?   &-      &yes  &no   &yes   &no    & \\
17              &02 &P?  &point  &-      &no   &no   &OA    &no    &core along outflow axis \\
18              &03 &P   &point  &yes    &yes  &yes  &yes   &yes   & \\ 
19              &04 &S*  &dark   &-      &no   &no   &OA    &no    &near CO outflow lobe \\
20              &05 &P   &point  &-      &no   &no   &yes   &no    & \\
21              &06 &S*  &dark   &-      &no   &no   &OA    &no    &near SiO outflow lobe \\
22              &07 &S   &dark   &-      &no   &no   &no    &no    & \\

23~05373+2349   &01 &P   &point  &-      &yes  &yes  &yes   &no    & \\
25              &02 &S*  &dark?  &yes?   &faint&no   &no    &no    &between two strong IR sources \\

27~19035+0641   &01 &P   &point  &yes    &     &yes  &      &      & \\
28              &02 &S*  &dark   &-      &     &no   &      &      &close to bright IRAC source \\
29              &03 &S   &dark   &-      &     &no   &      &      & \\ 
30              &04 &S?  &ext    &-      &     &no   &      &      & \\
31              &05 &S   &dark   &-      &     &no   &      &      & \\

32~20081+3122   &01 &P?  &point  &-      &yes? &no   &OA    &no    &mm source 4\arcsec\ to the south of the polygon \\
33              &02 &S   &dark   &-      &no   &no   &no    &no    & \\

35~20126+4104   &01 &P   &point  &yes    &yes  &yes  &yes   &yes   & \\
36              &02 &S   &dark   &-      &yes  &no   &no    &no    & \\
37              &03 &P?  &point  &-      &no   &no   &no    &no    & \\ 

38~G75.78+0.34  &01 &P   &point  &yes    &yes  &yes  &yes   &yes   & \\
39              &02 &S*  &ext    &-      &no   &no   &no    &no    &near strong IRAC source \\
40              &03 &P?  &point  &-      &no   &no   &no    &no    & \\ 
41              &04 &S   &dark   &-      &no   &no   &no    &no    & \\
43              &05 &S   &dark   &-      &no   &no   &no    &no    & \\
44              &06 &P?  &point  &-      &no   &no   &no    &no    & \\

45~20293+3952   &01 &S*  &dark?  &-      &no   &no   &no    &no    &near strong IRAC source \\
46              &02 &S*  &dark?  &-      &yes  &no   &OA    &no    &perturbed by outflow \\
47              &03 &P   &point  &yes    &yes  &no   &yes   &yes   & \\ 
48              &04 &P?  &point  &-      &yes  &no   &no    &no    & \\
49              &05 &P   &point  &yes    &yes  &no   &yes   &no    & \\
50              &06 &P?  &point  &-      &     &no   &no    &      & \\

51~20343+4129   &01 &S*  &ext    &-      &yes  &no   &no    &no    &near strong IRAC source, perturbed? \\
52              &02 &S*  &dark   &-      &yes  &no   &OA?   &no    &near strong IRAC source, perturbed? \\
53              &03 &P?  &dark?  &-      &yes  &yes? &OA?   &no    &near strong IRAC source, perturbed? \\ 
54              &04 &P   &point  &yes    &faint&yes  &no    &no    & \\

55~22134+5834   &01 &S   &dark   &-      &     &no   &      &no    & \\
56              &02 &S*  &dark?  &-      &     &no   &      &no    &near strong IRAC source, perturbed? \\
57              &03 &S   &dark   &-      &     &no   &      &no    & \\ 
58              &04 &S   &dark   &-      &     &no   &      &no    & \\
59              &05 &P?  &point  &-      &     &no   &      &no    & \\
\hline
\end{tabular}
\end{table*}
\begin{table*}
\contcaption{}
\centering
\begin{tabular}{l c c c c c c c c l}
\hline\hline
\texttt{ID}~Region\supa
&\#
&Type\supb
&IRAC
&MIPS/WISE
&mm
&cm
&outflow
&maser
&comments
\\
\hline
60~22172+5549N  &01 &P   &point  &yes    &yes  &no   &yes   &no    & \\
61              &02 &S   &dark   &-      &no   &no   &no    &no    & \\
62              &03 &P?  &point  &yes    &faint&no   &OA    &no    &core along outflow axis \\ 

63~22198+6336   &01 &S   &dark   &-      &no   &no   &no    &no    & \\
64              &02 &P?  &point  &-      &no   &no   &no    &no    & \\
65              &03 &P   &point  &yes    &yes  &yes  &yes   &yes   & \\ 
66              &04 &S*  &ext    &-      &no   &no   &OA    &no    &core along outflow axis \\
67              &05 &S   &dark   &-      &no   &no   &no    &no    & \\
68              &06 &S   &dark   &-      &no   &no   &no    &no    & \\
69              &07 &S   &dark   &-      &yes  &no   &no    &no    & \\
70              &08 &S   &dark   &-      &     &no   &      &no    & \\

71~CepA         &01 &S*  &dark   &-      &     &no   &OA    &no    &core close to outflow axis \\
72              &02 &S*  &dark   &-      &     &no   &no    &no    & \\
73              &03 &S*  &dark   &-      &     &no   &OA    &no    &core close to outflow axis \\ 
74              &04 &P?  &point  &-      &     &no   &OA    &no    &core close to outflow axis \\
75              &05 &S*  &ext    &-      &     &no   &OA    &no    &core close to outflow axis \\
76              &06 &S?  &ext    &-      &     &no   &no    &no    & \\
77              &07 &P   &point  &yes    &     &no   &OA    &no    &core close to outflow axis \\
\hline
\end{tabular}
\begin{flushleft}
\supa\ References: We have made use of the \emph{Spitzer}/IRAC, \emph{Spitzer}/MIPS, and WISE images to characterise the IR properties of the cores at wavelengths ranging from 3.6~$\mu$m to 24~$\mu$m. For each region we have searched the literature for information on millimeter continuum, centimetre continuum, outflow and maser emission. The references used (for each region) are:\\
00117+6412: \citet{palau2010, busquetphd2010, sanchezmongephd2011};\\
AFGL5142: \citet{zhang2002, zhang2007, busquetphd2010, sanchezmongephd2011, busquet2011, palau2011, palau2013};\\
05345+3157NE: \citet{fontani2008, fontani2009, fontani2012a};\\
05358+3543NE: \citet{beuther2002c, beuther2007, leurini2007, sanchezmongephd2011};\\
05373+2349: \citet{rodriguez1980, varricatt2010, molinari1996, molinari2002, molinari2008, gutermuth2009, khanzadyan2011};\\
19035+0641: \citet{sanchezmongephd2011};\\
20081+3122: \citet{su2009, kumar2004};\\
20126+4104: \citet{cesaroni1997, cesaroni1999, cesaroni2005, ketozhang2010, busquetphd2010};\\
G75.78+0.34: \citet{sanchezmongephd2011, sanchezmonge2013};\\
20293+3952: \citet{beuther2002d, beuther2004a, beuther2004b, palau2007a};\\
20343+4129: \citet{palau2007b, fontani2012b};\\
22134+5834: \citet{busquetphd2010, sanchezmongephd2011};\\
22172+5549N: \citet{fontani2004, palau2011, palau2013};\\
22198+6336: \citet{sanchezmonge2010, sanchezmongephd2011, palau2011, palau2013};\\
CepA: \citet{torrelles1986, patel2005}.\\
\supb\ Type of the core: S: quiescent starless, S*: perturbed starless, P: protostellar, P?: tentatively protostellar (see Sect.~\ref{s:cores}).\\
\end{flushleft}
\end{table*}

In Table~\ref{t:nh3types}, we summarize the infrared properties of each core, together with the information of other star formation signposts reported in the literature. In Col.~4 of the table, we classify each core as `quiescent starless' (S), `perturbed starless' (S*), `protostellar' ( P) or `tentatively protostellar' (P?).

In Fig.~\ref{f:mom12}, we present the first-order (intensity weighted mean $v_\mathrm{LSR}$) and second-order (intensity weighted mean linewidth) moment maps for the 15 massive star-forming regions studied in this work. Note that the second-order moment maps have been corrected to show directly the linewidth ($\Delta v$) instead of the dispersion ($\sigma_v$), by using the relation $\Delta v = \sqrt{8\ln2}\,\sigma_v$.

In Fig.~\ref{f:spectra}, we present the observed and fitted spectra of the \nh\,(1,1) and \nh\,(2,2) transitions for all the cores identified and listed in Table~\ref{t:nh3cores}.

\clearpage
\begin{figure*}
\centering
\begin{tabular}[b]{c c c c c}
\vspace{0.5cm}
  \epsfig{file=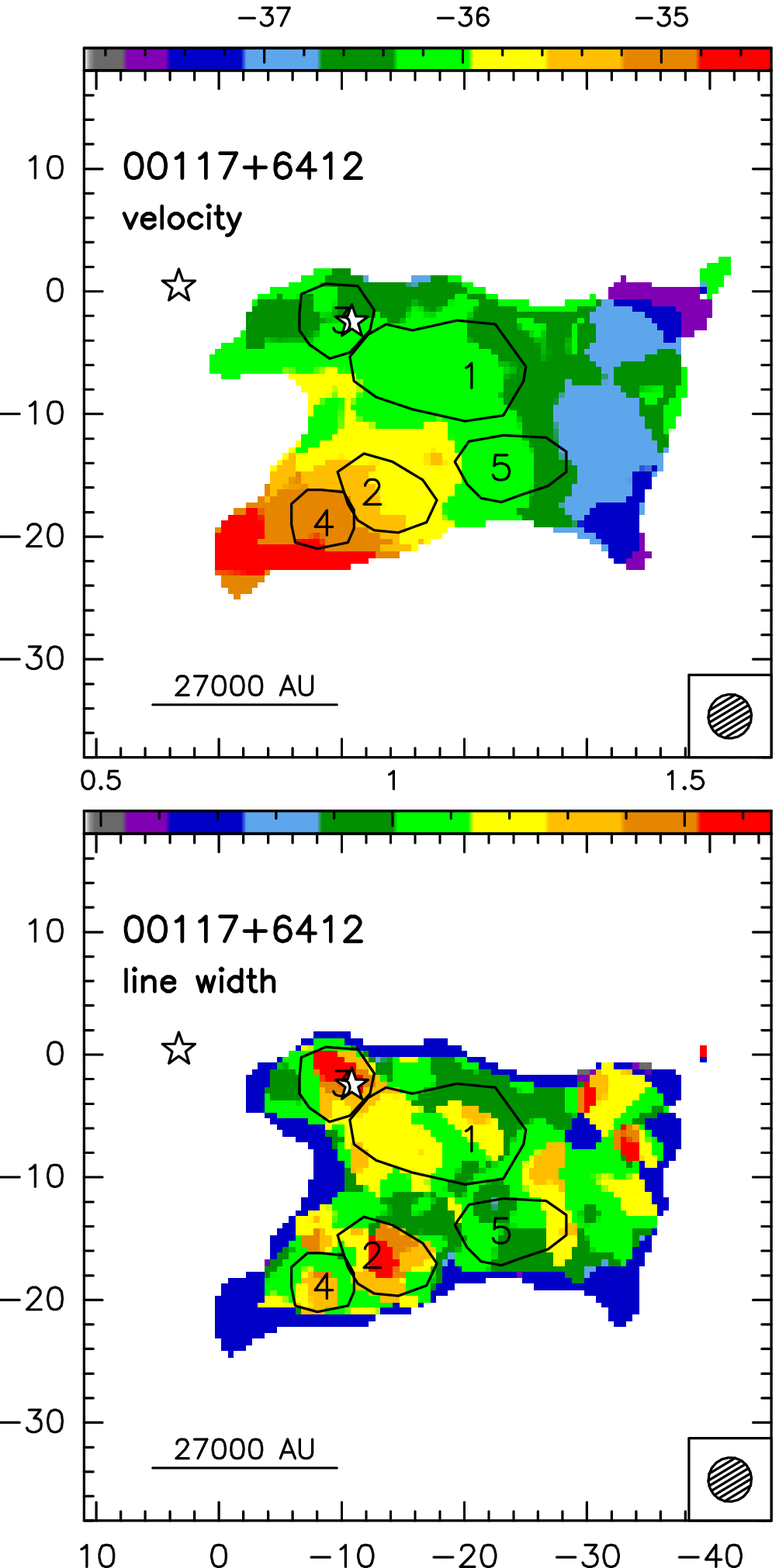, scale=0.5} &&
  \epsfig{file=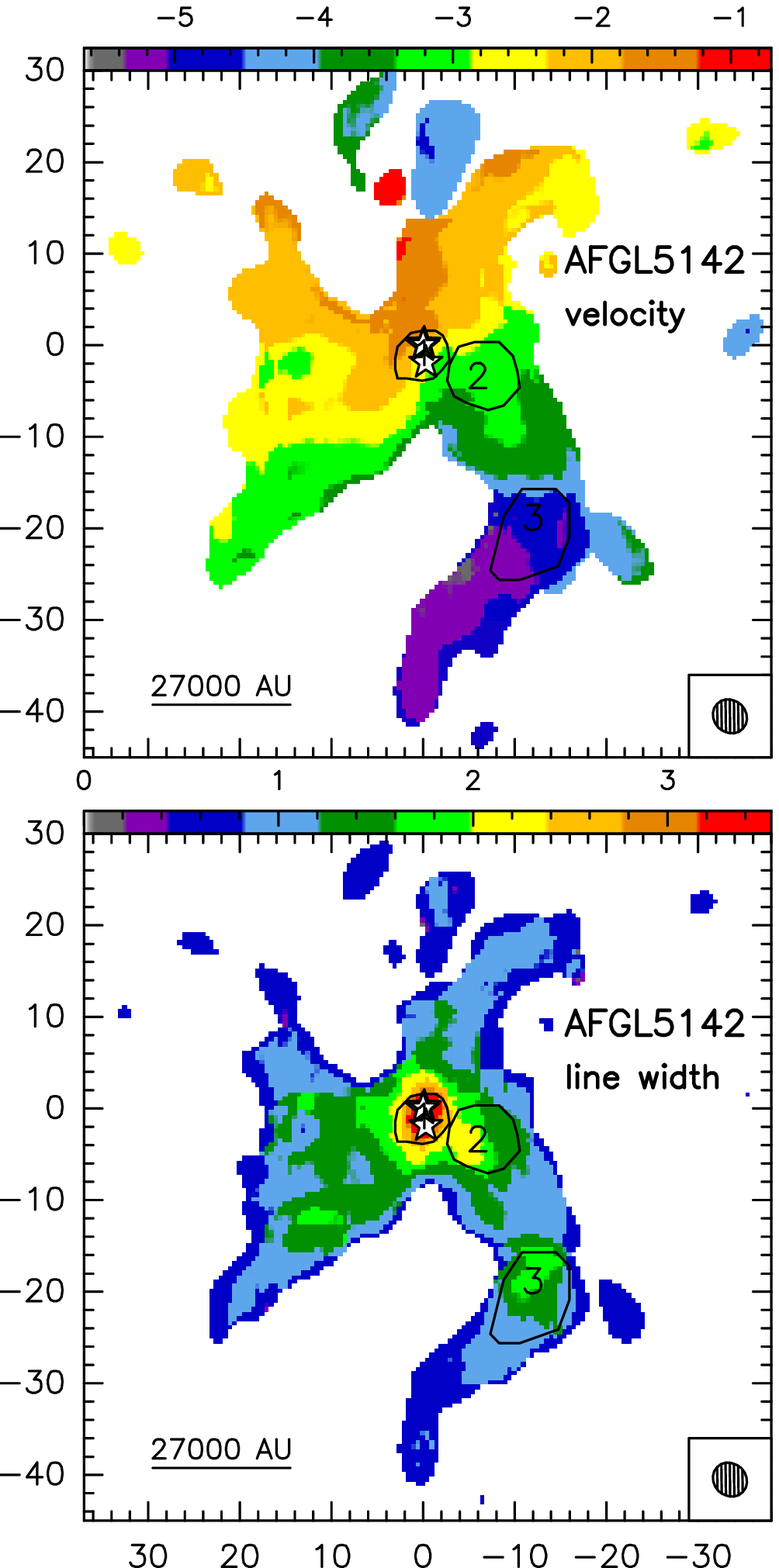, scale=0.5} &&
  \epsfig{file=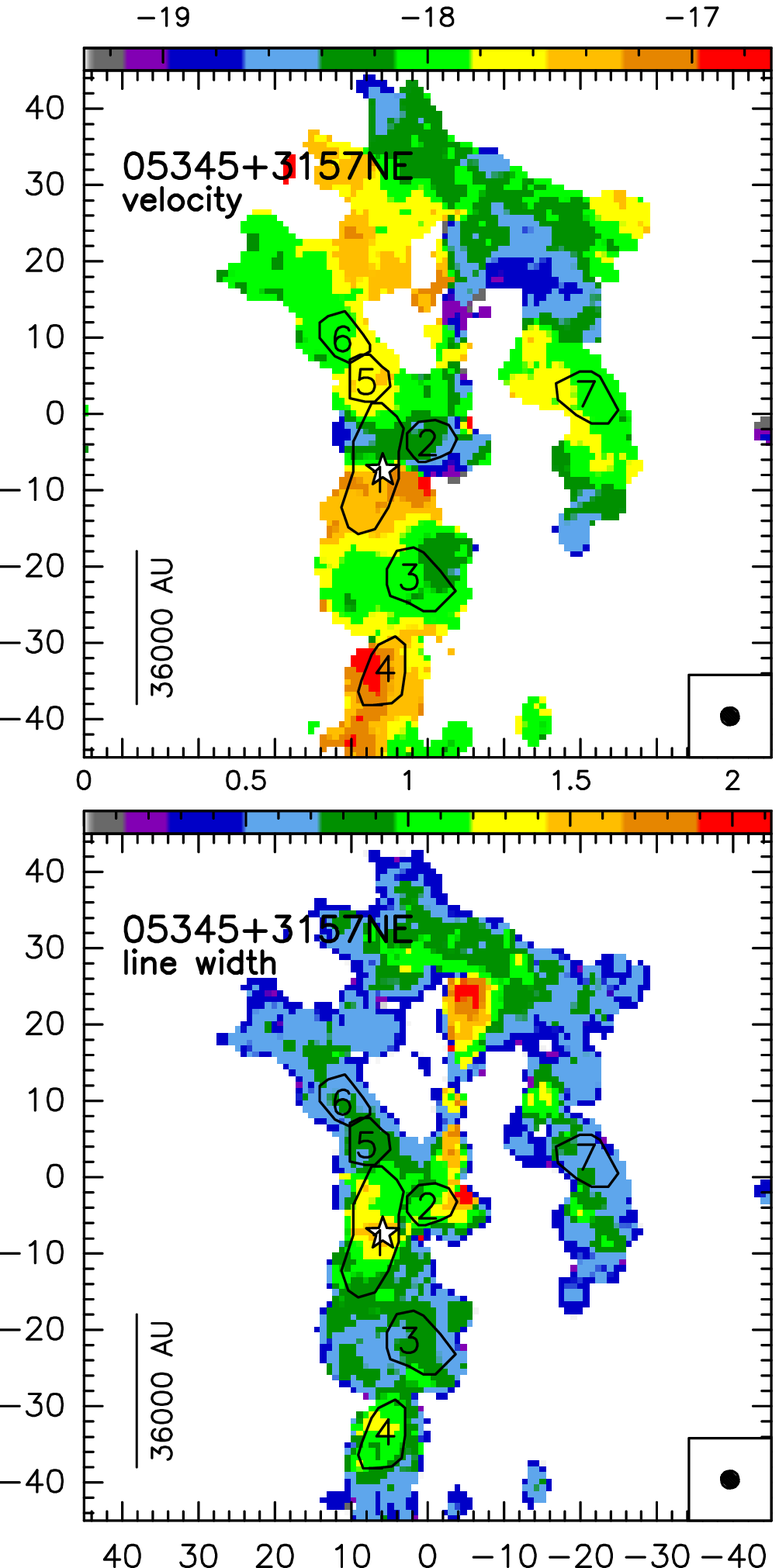, scale=0.5} \\
  \cline{1-5}  && && \\
  \epsfig{file=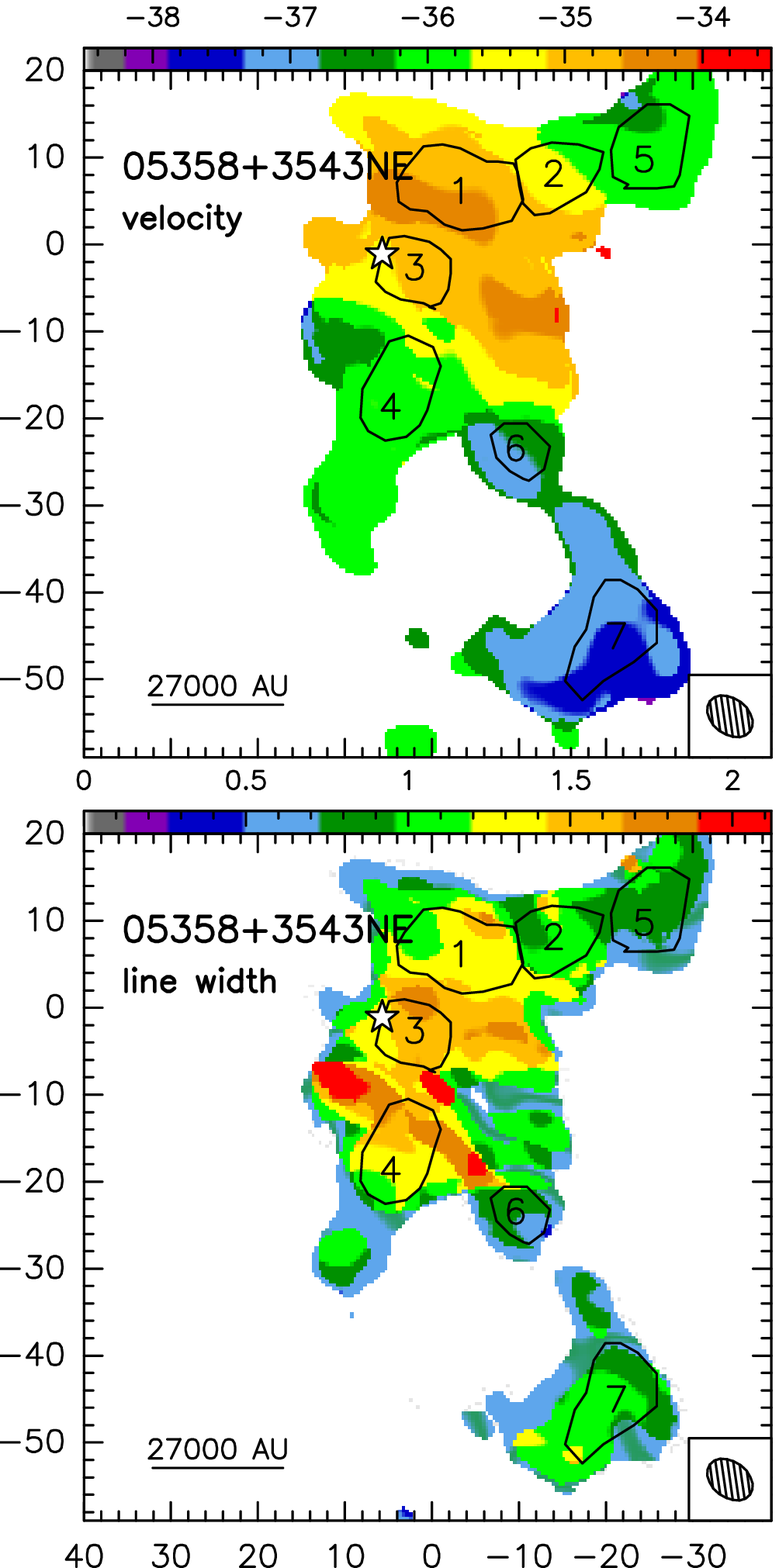, scale=0.5} &&
  \epsfig{file=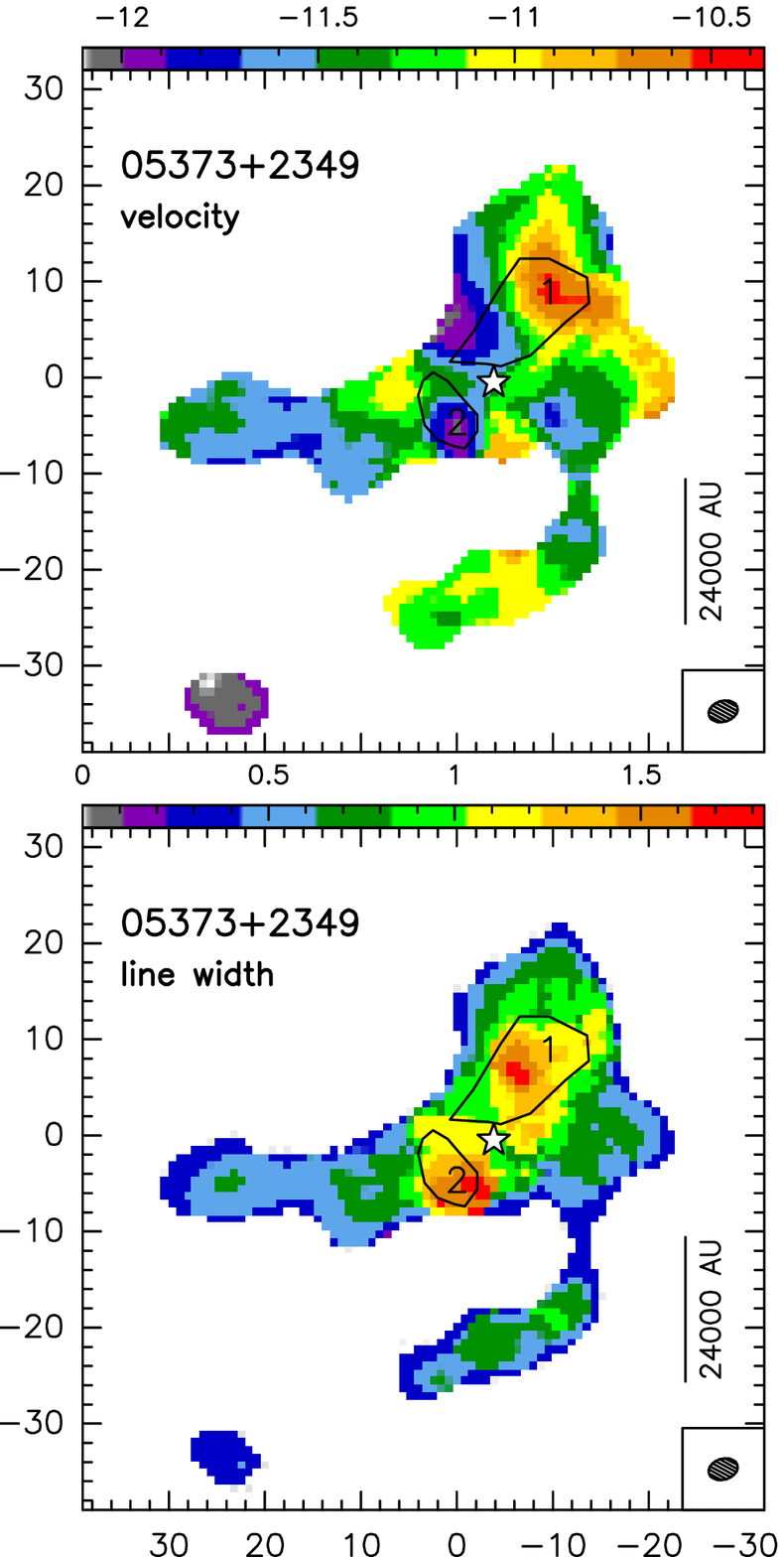, scale=0.5} &&
  \epsfig{file=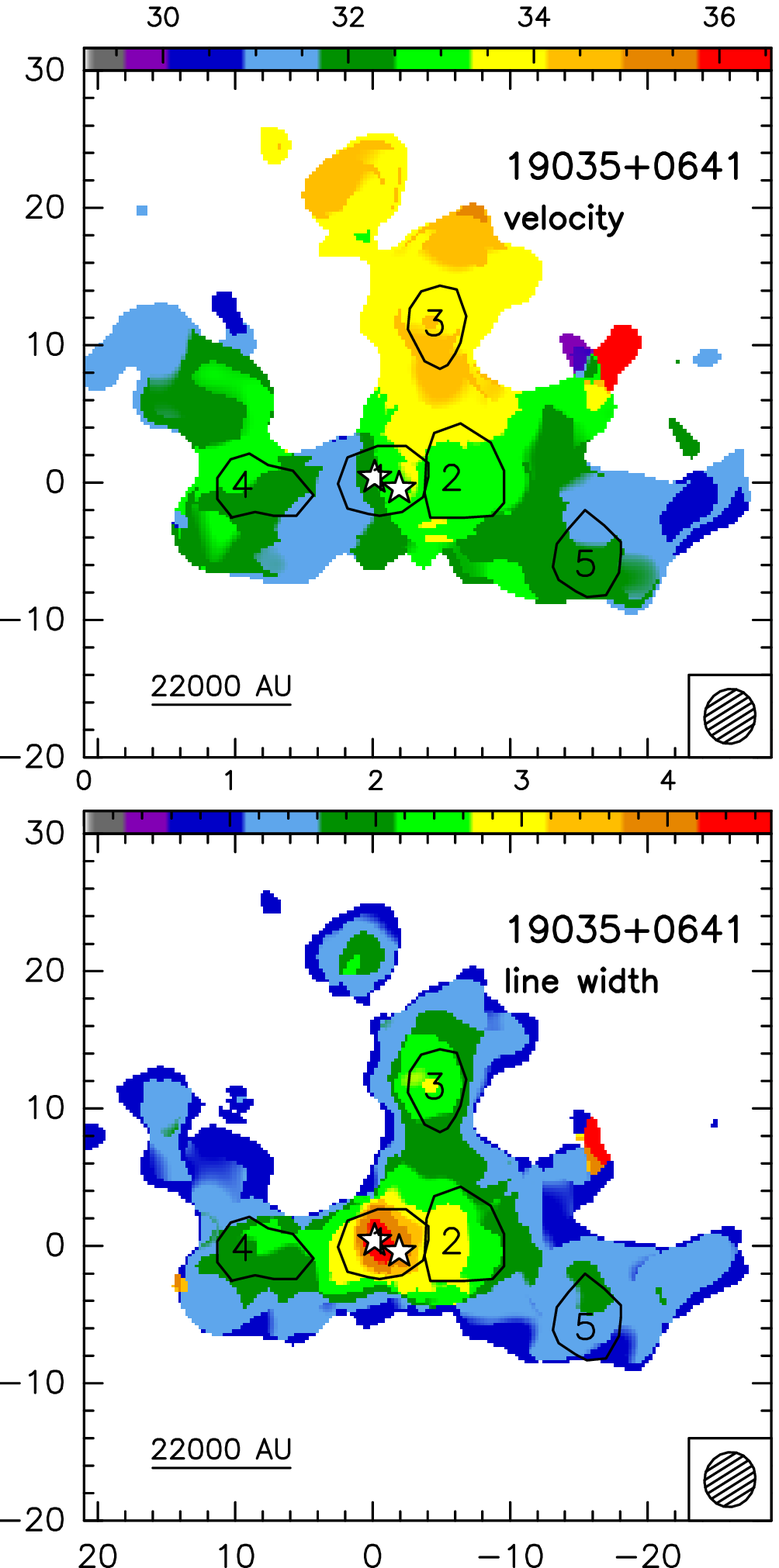, scale=0.5} \\
\end{tabular}
\caption{Velocity and linewidth maps (in units of \kms) obtained from the first and second-order moment maps. Note that the second-order moment map has been multiplied by $\sqrt{8\,\ln2}$ to obtain linewidth instead of dispersion. For each region we show two panels corresponding to the intensity weighted mean $v_\mathrm{LSR}$ (top panel) and to the intensity weighted mean velocity linewidth (bottom panel). Colour scales are indicated in the top of each panel. Symbols are as in Fig.~\ref{f:cores}.}
\label{f:mom12}
\end{figure*}
\begin{figure*}
\centering
\begin{tabular}[b]{c c c c c}
\vspace{0.5cm}
  \epsfig{file=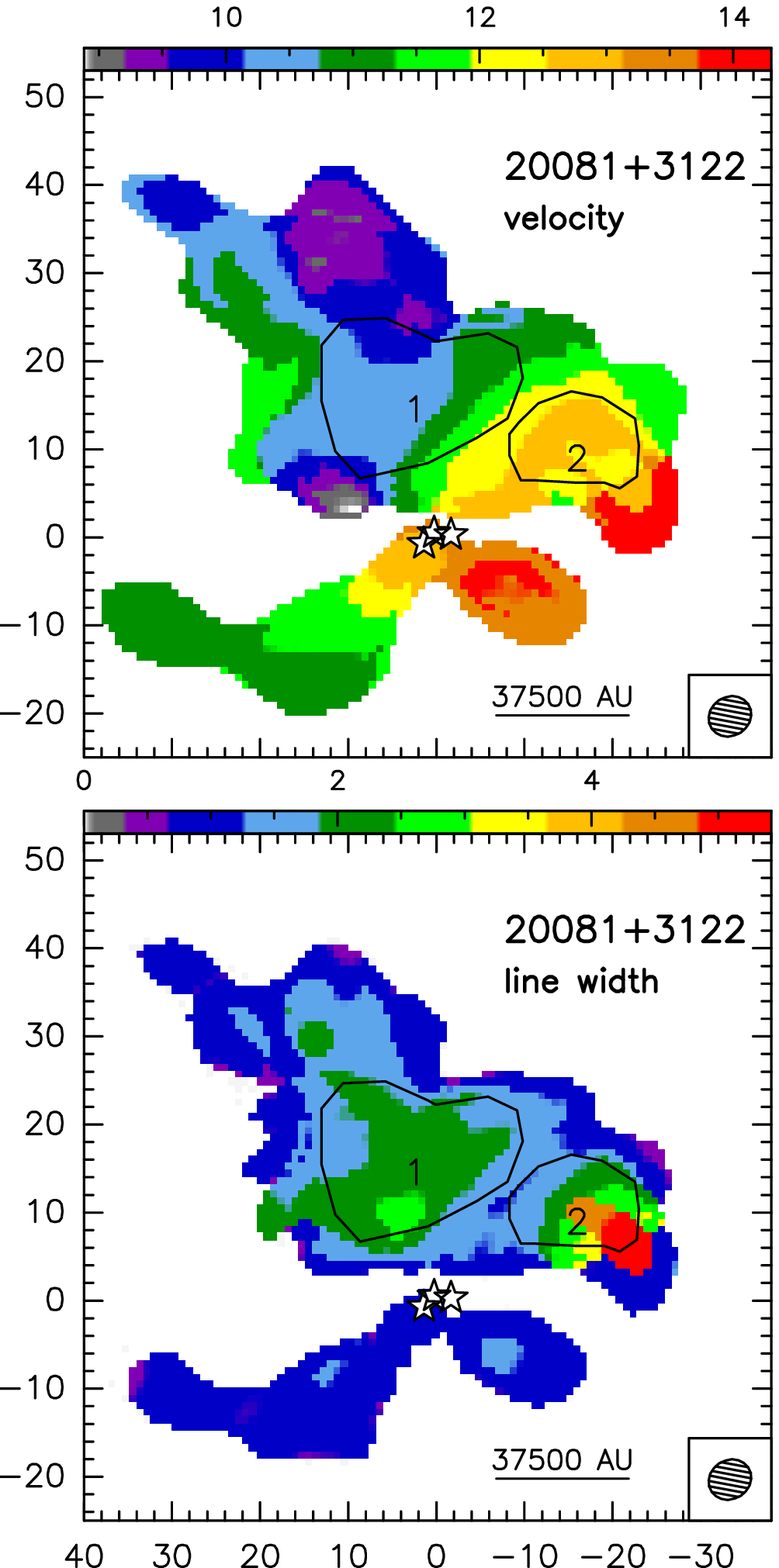, scale=0.5} &&
  \epsfig{file=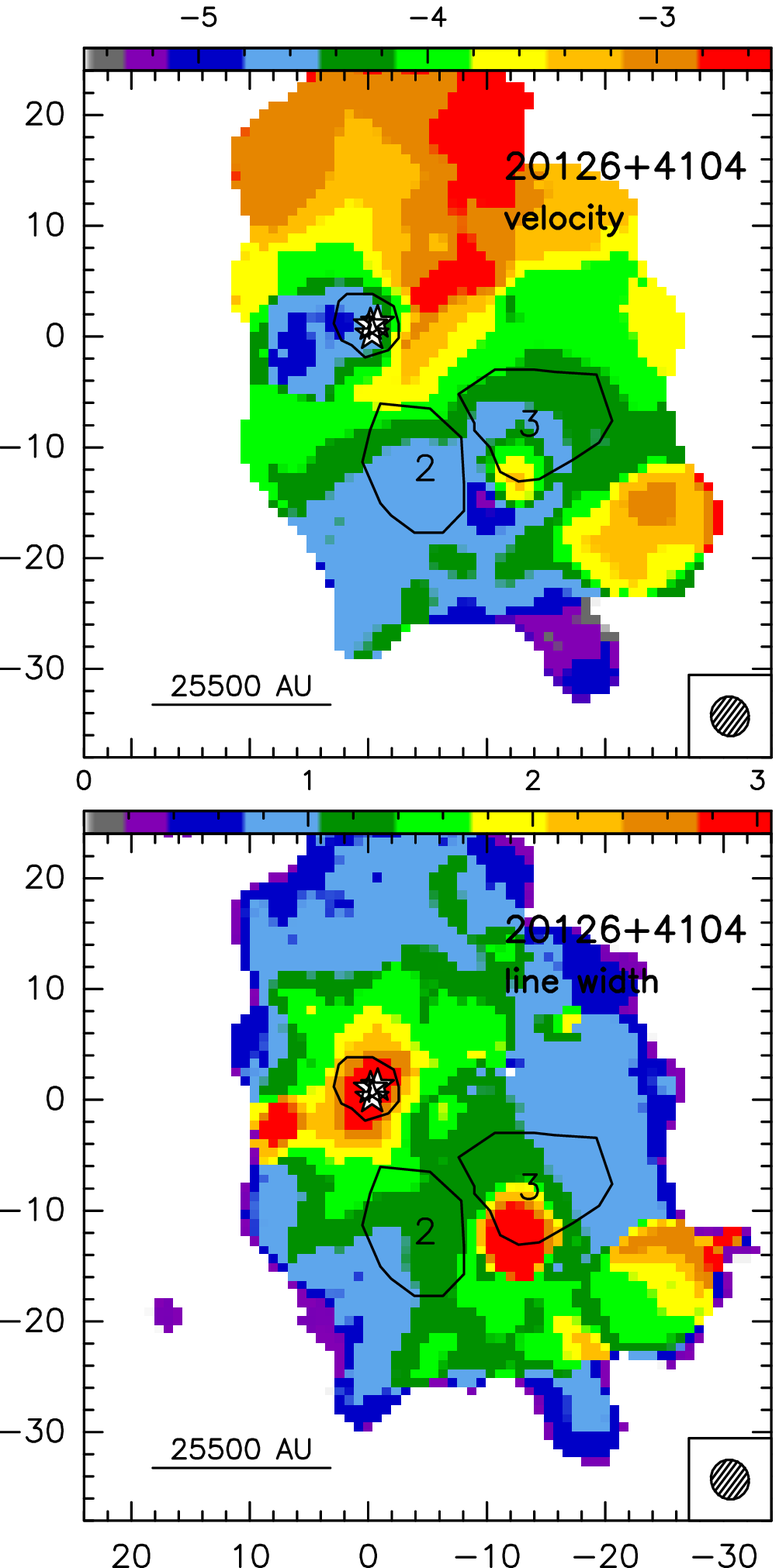, scale=0.5} &&
  \epsfig{file=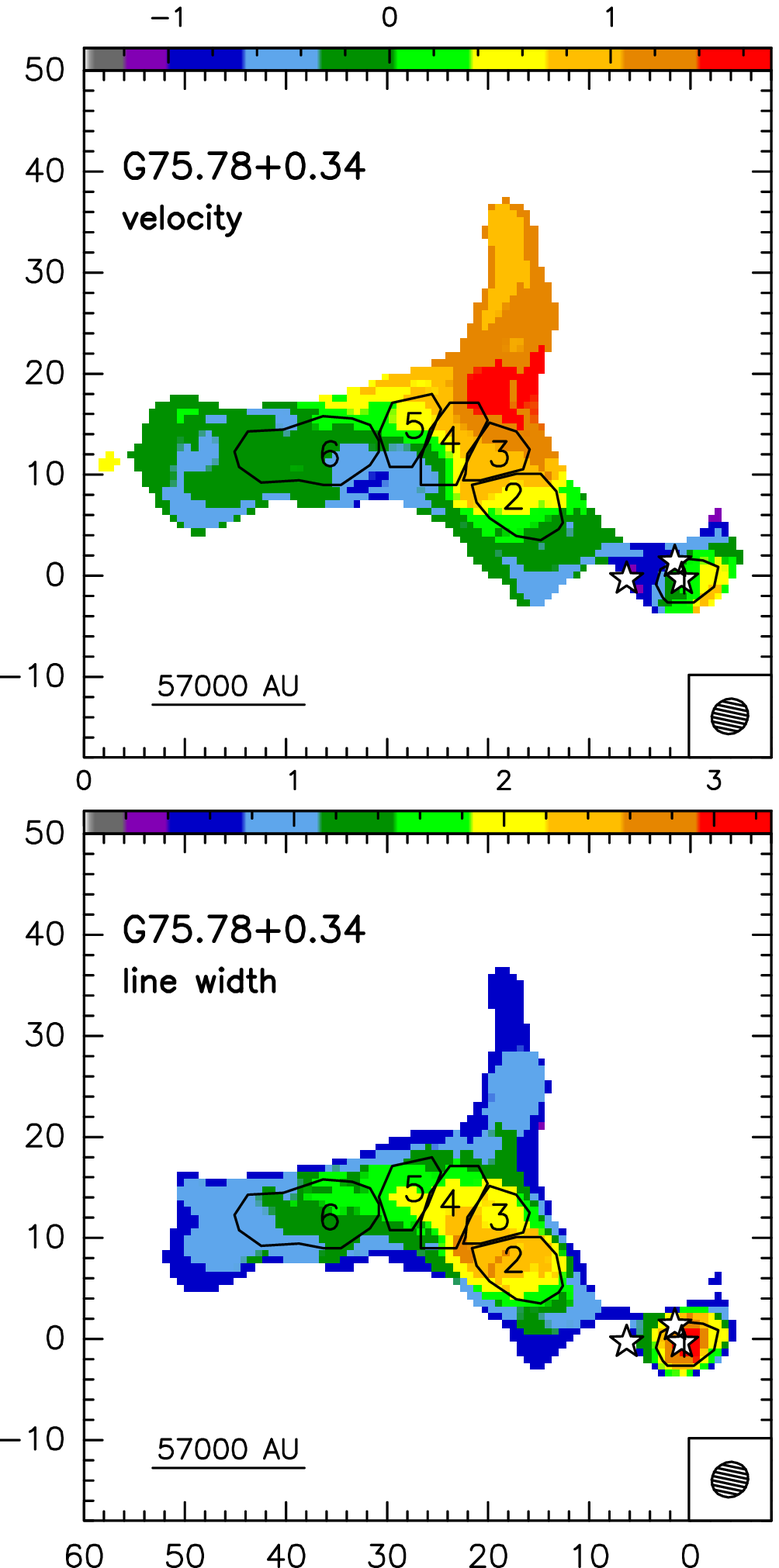, scale=0.5} \\
  \cline{1-5}  && && \\
  \epsfig{file=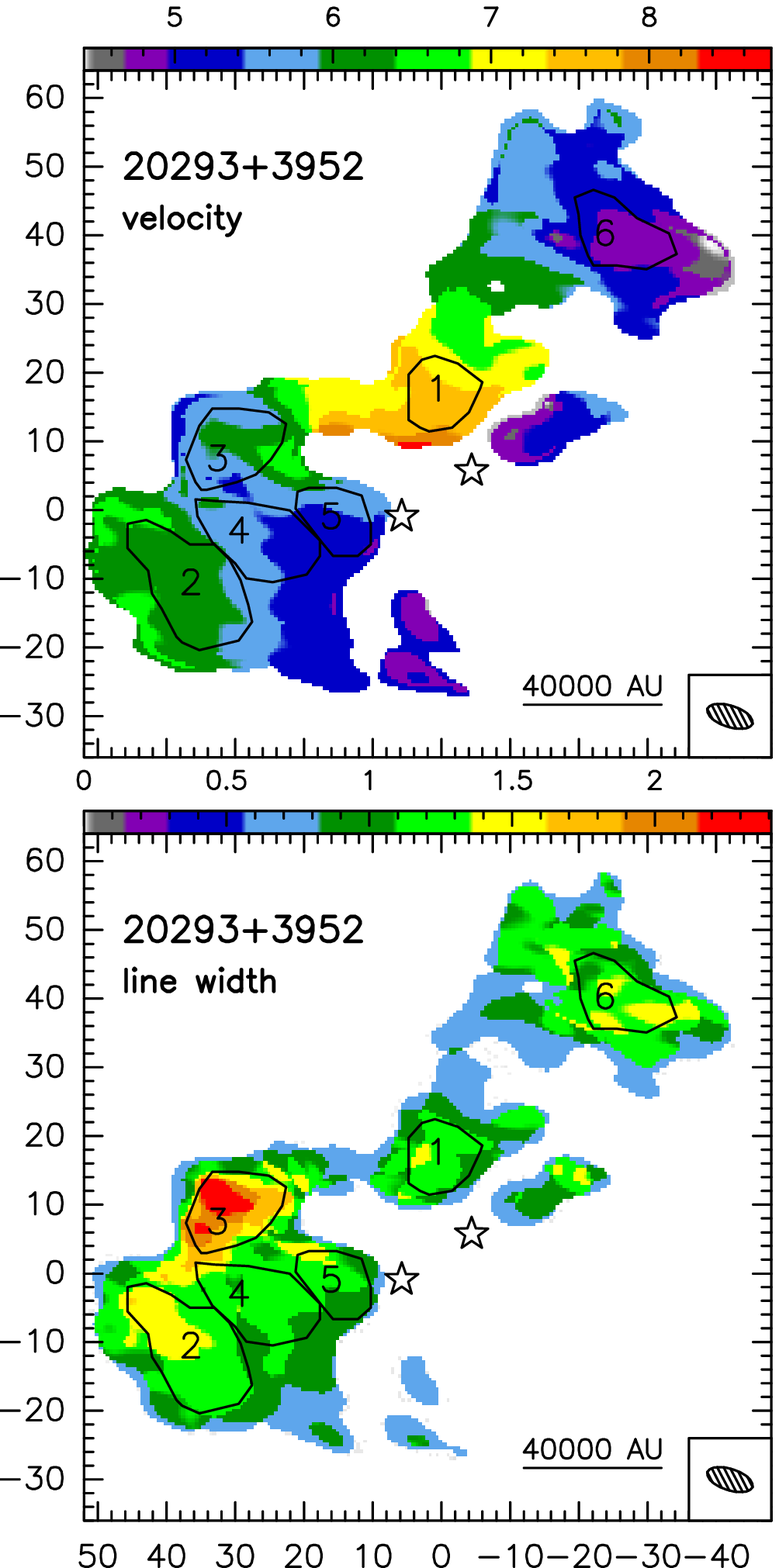, scale=0.5} &&
  \epsfig{file=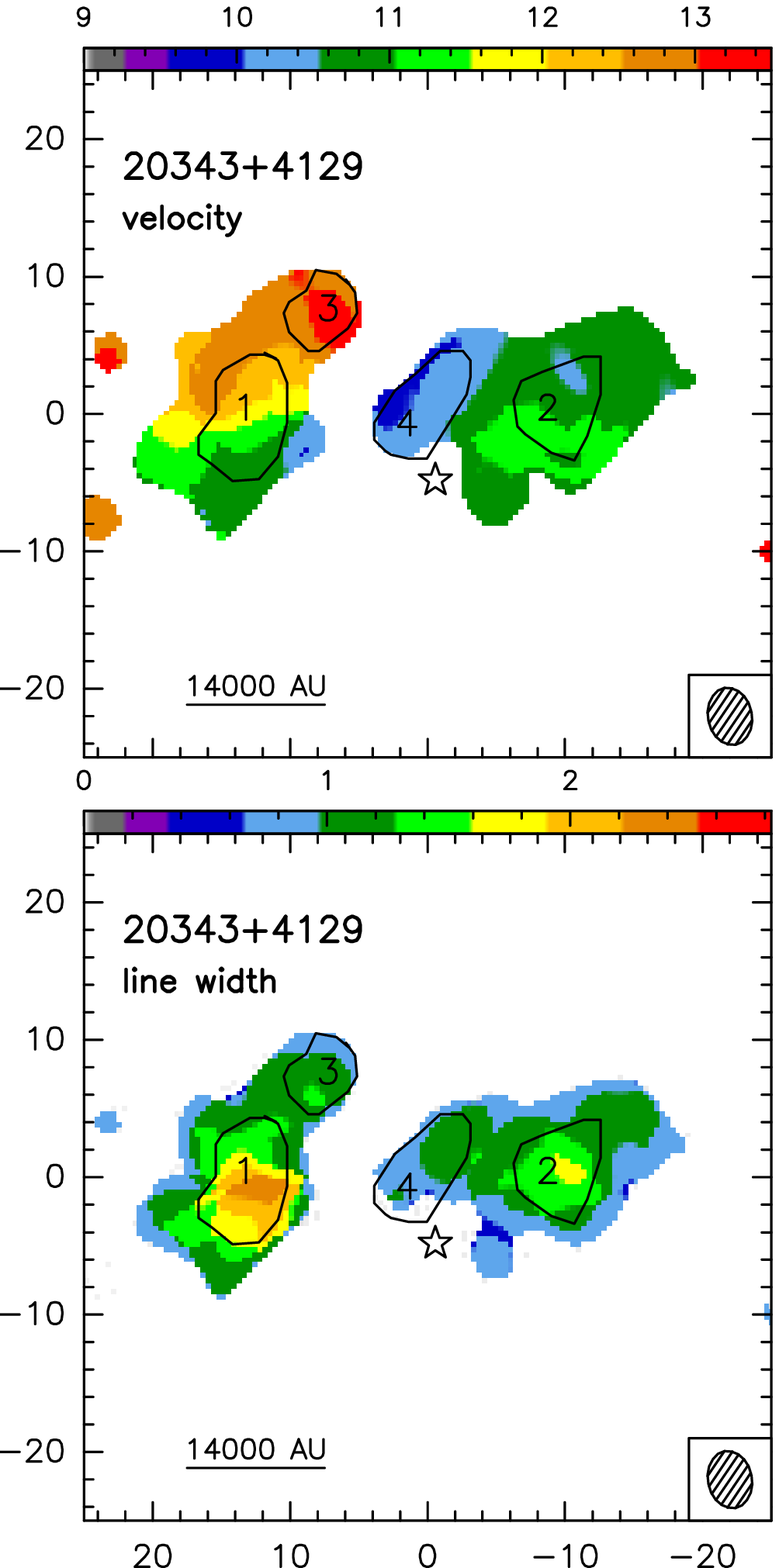, scale=0.5} &&
  \epsfig{file=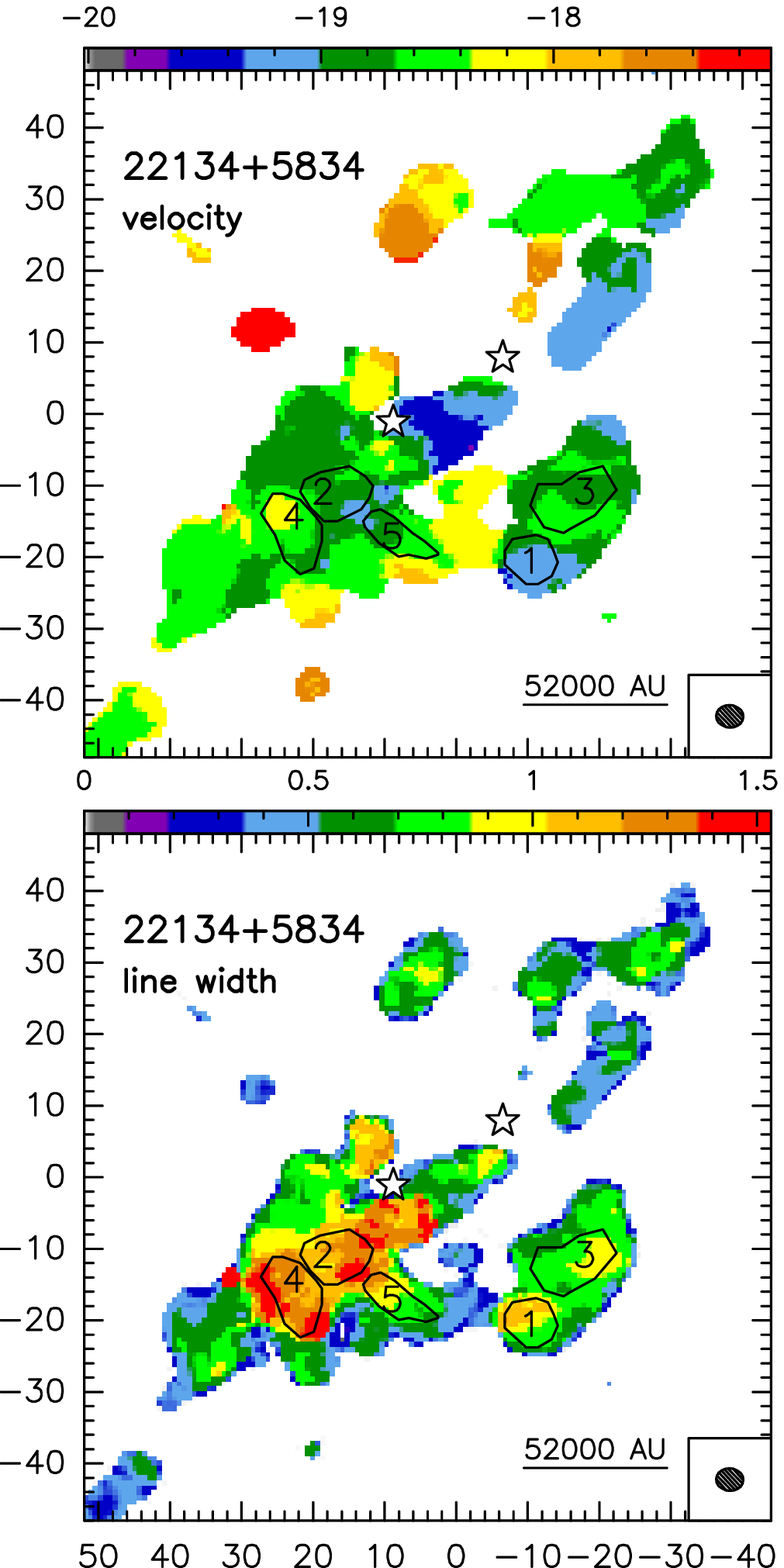, scale=0.5} \\
\end{tabular}
\contcaption{}
\end{figure*}
\begin{figure*}
\centering
\begin{tabular}[b]{c c c c c}
\vspace{0.5cm}
  \epsfig{file=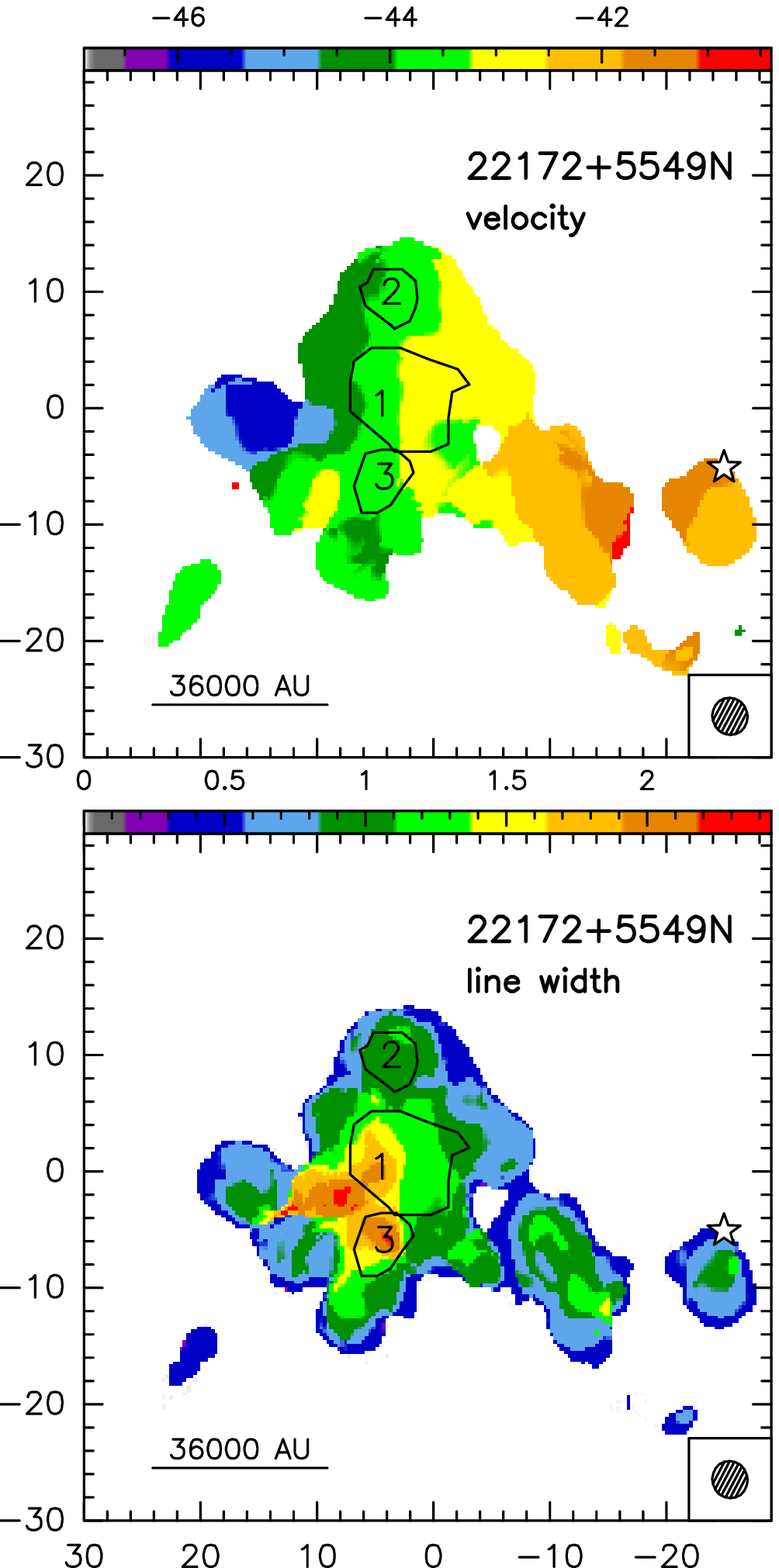, scale=0.5} &&
  \epsfig{file=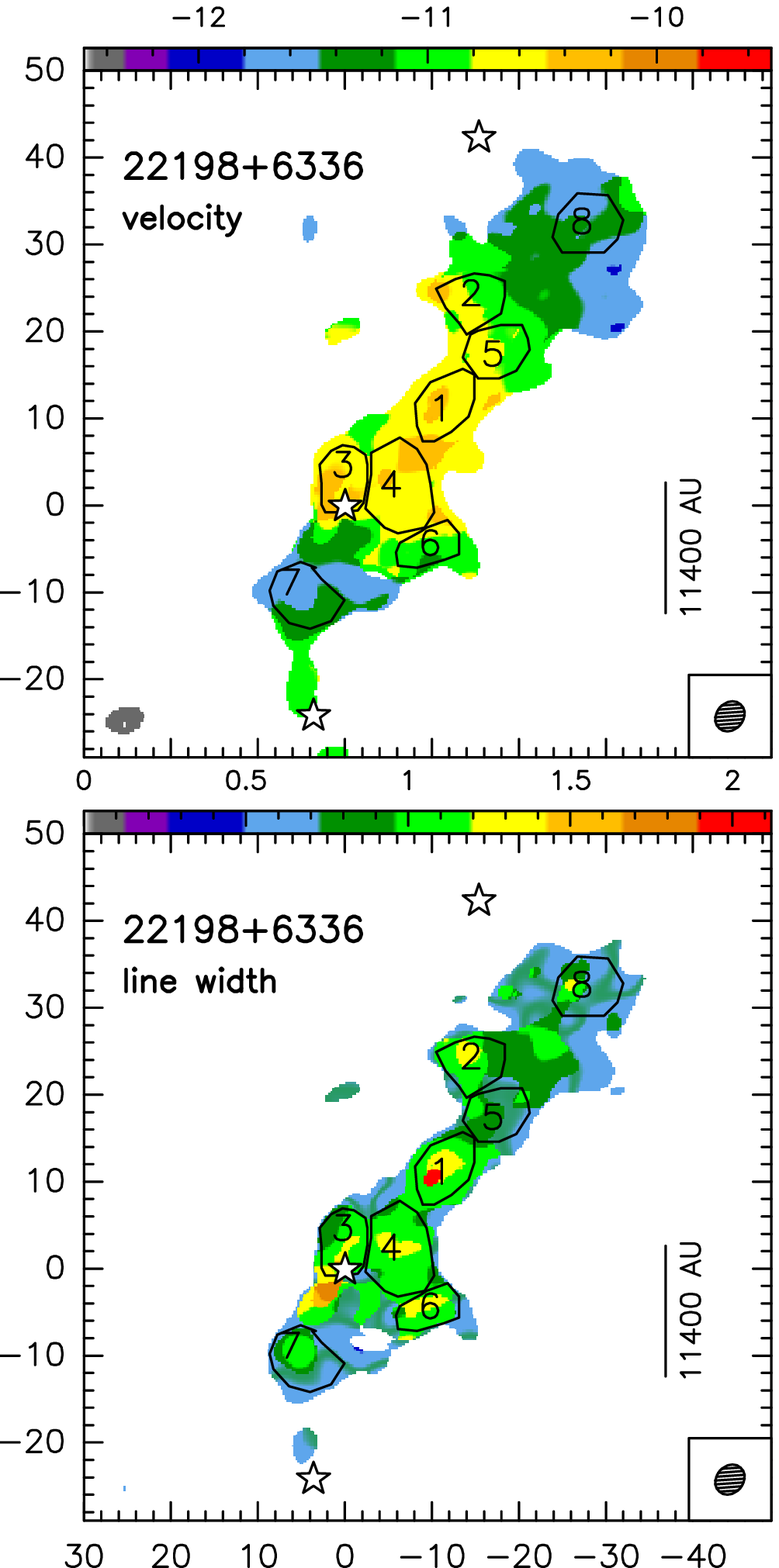, scale=0.5} &&
  \epsfig{file=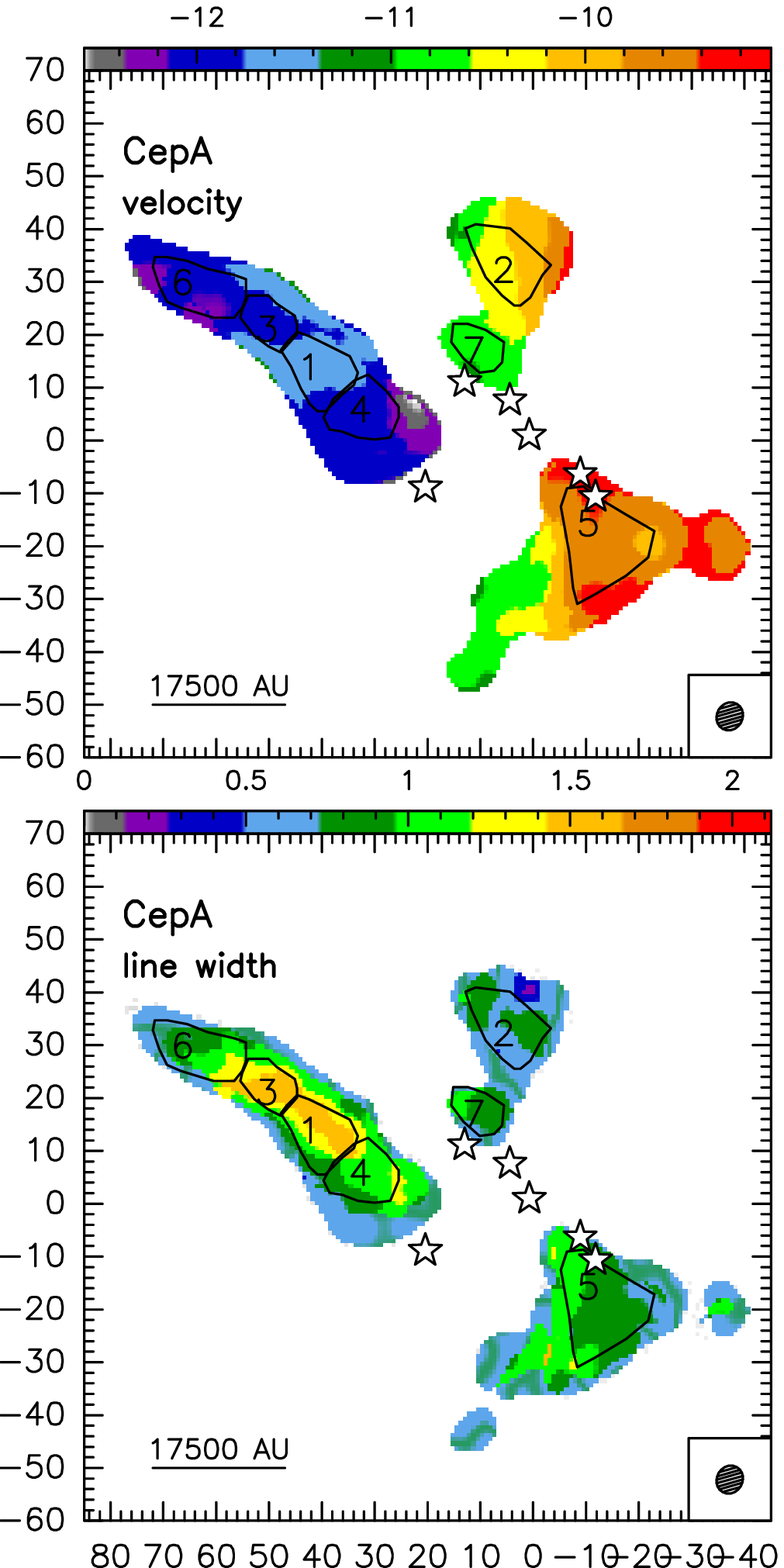, scale=0.5} \\
\end{tabular}
\contcaption{}
\end{figure*}

\begin{figure*}
\begin{center}
\begin{tabular}[b]{c}
\vspace{0.5cm}
  \epsfig{file=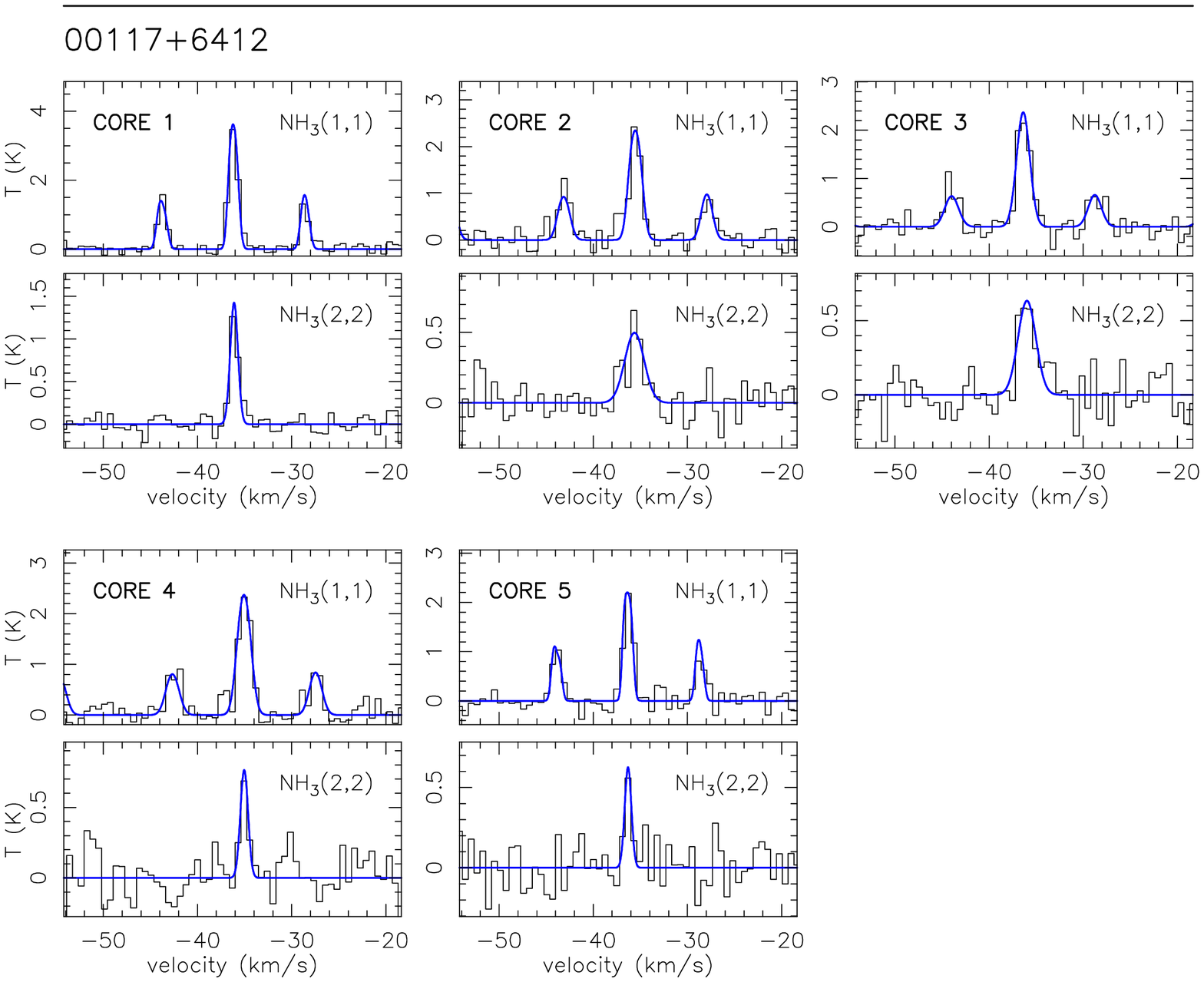, scale=0.85} \\
  \epsfig{file=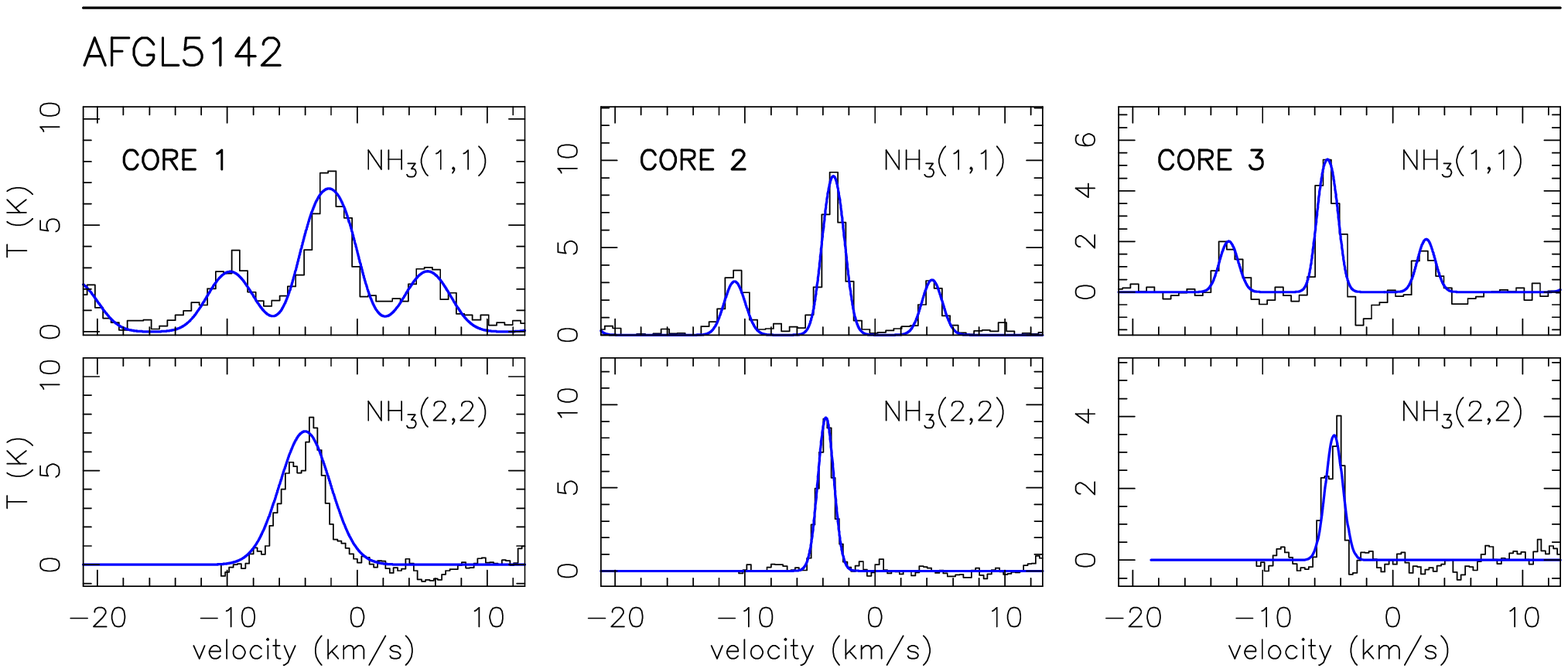, scale=0.85} \\
\end{tabular}
\caption{Observed (black histogram) and fitted (thick blue line) spectra of the \nh\,(1,1) and (2,2) lines for the cores listed in Table~\ref{t:nh3cores}. Vertical axis corresponds to brightness temperature (in K), and horizontal axis corresponds to velocity (in \kms).}
\label{f:spectra}
\end{center}
\end{figure*}
\begin{figure*}
\begin{center}
\begin{tabular}[b]{c}
\vspace{0.5cm}
  \epsfig{file=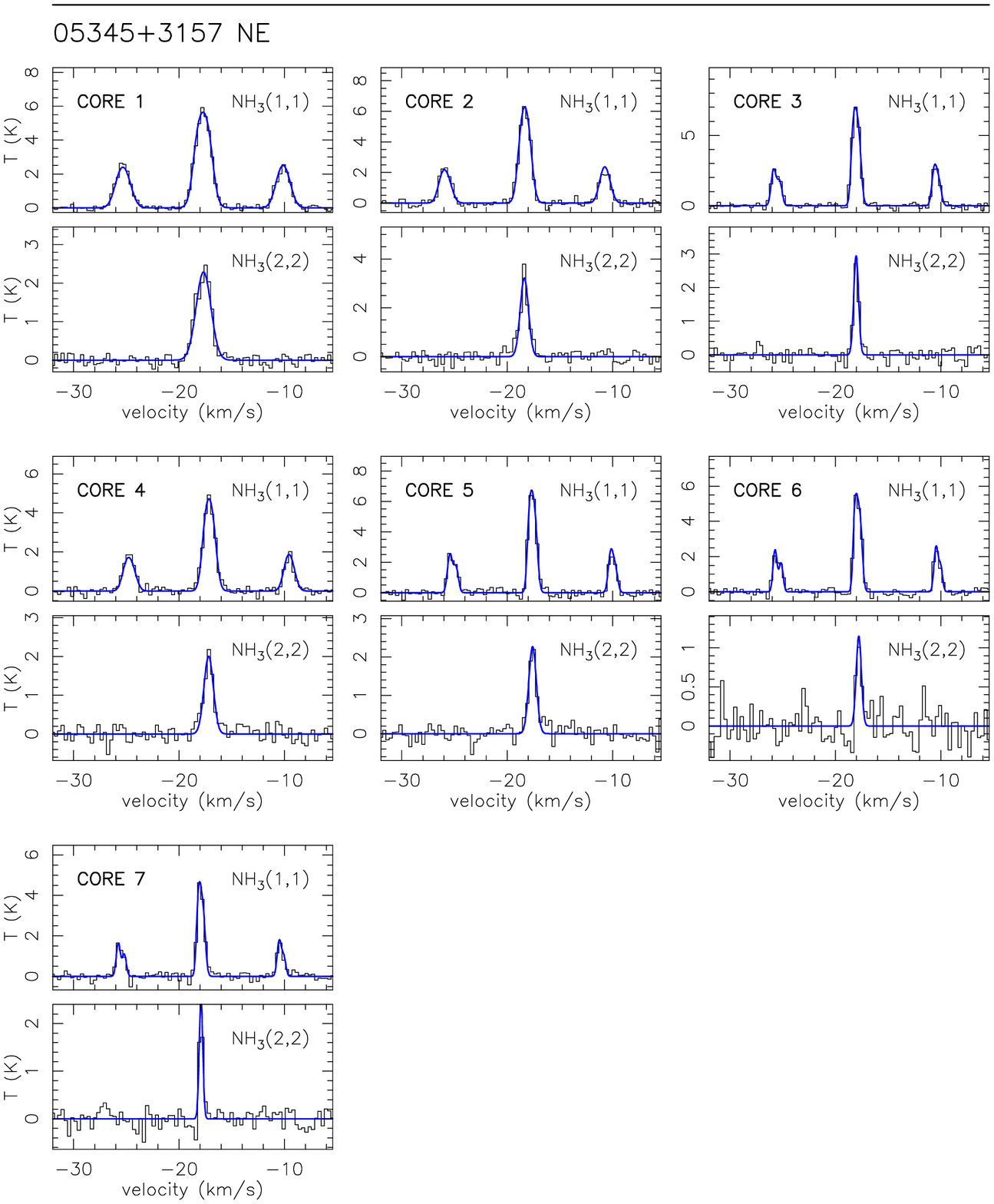, scale=0.85} \\
\end{tabular}
\contcaption{}
\end{center}
\end{figure*}
\begin{figure*}
\begin{center}
\begin{tabular}[b]{c}
\vspace{0.5cm}
  \epsfig{file=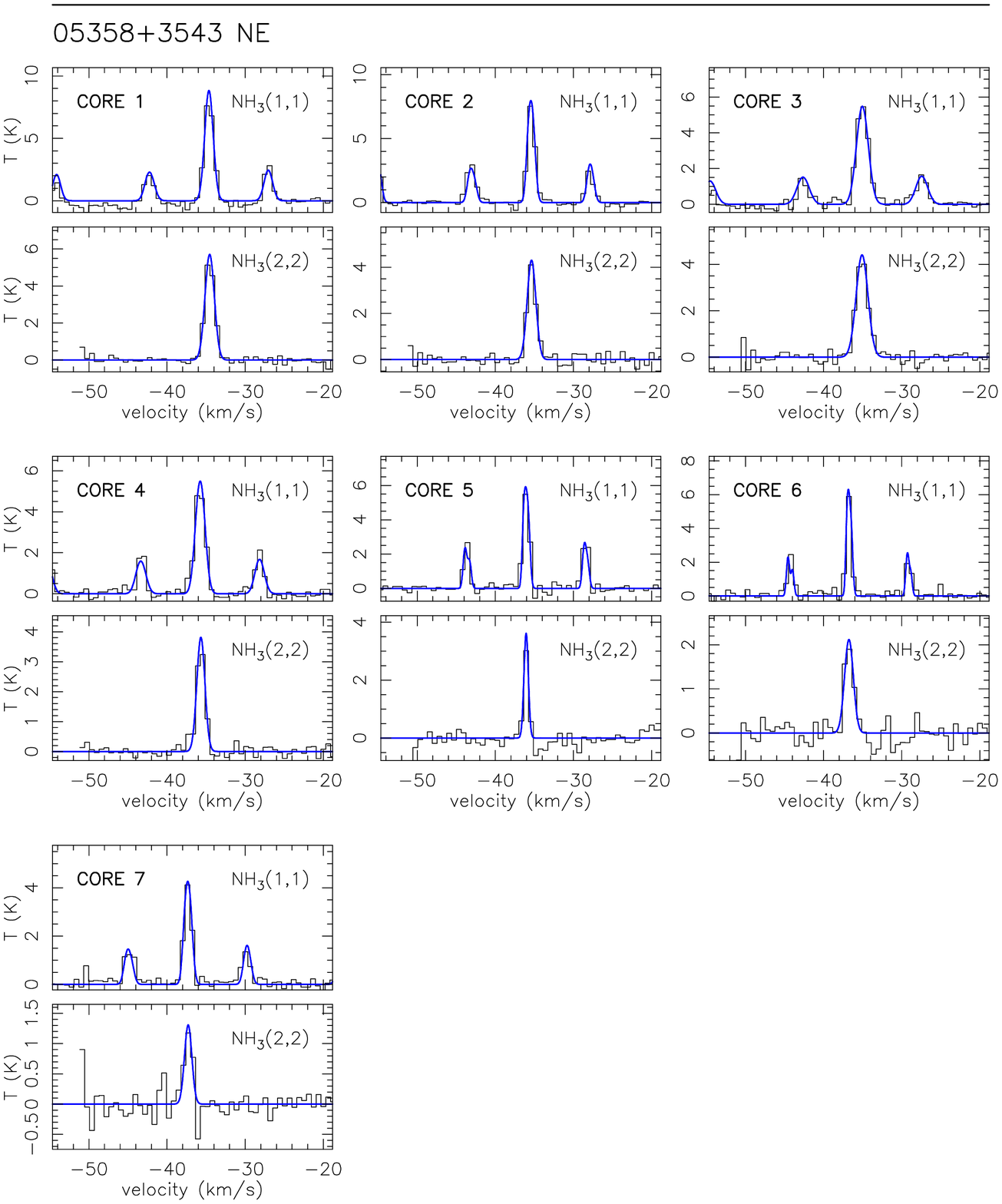, scale=0.85} \\ 
\end{tabular}
\contcaption{}
\end{center}
\end{figure*}
\begin{figure*}
\begin{center}
\begin{tabular}[b]{c}
\vspace{0.5cm}
  \epsfig{file=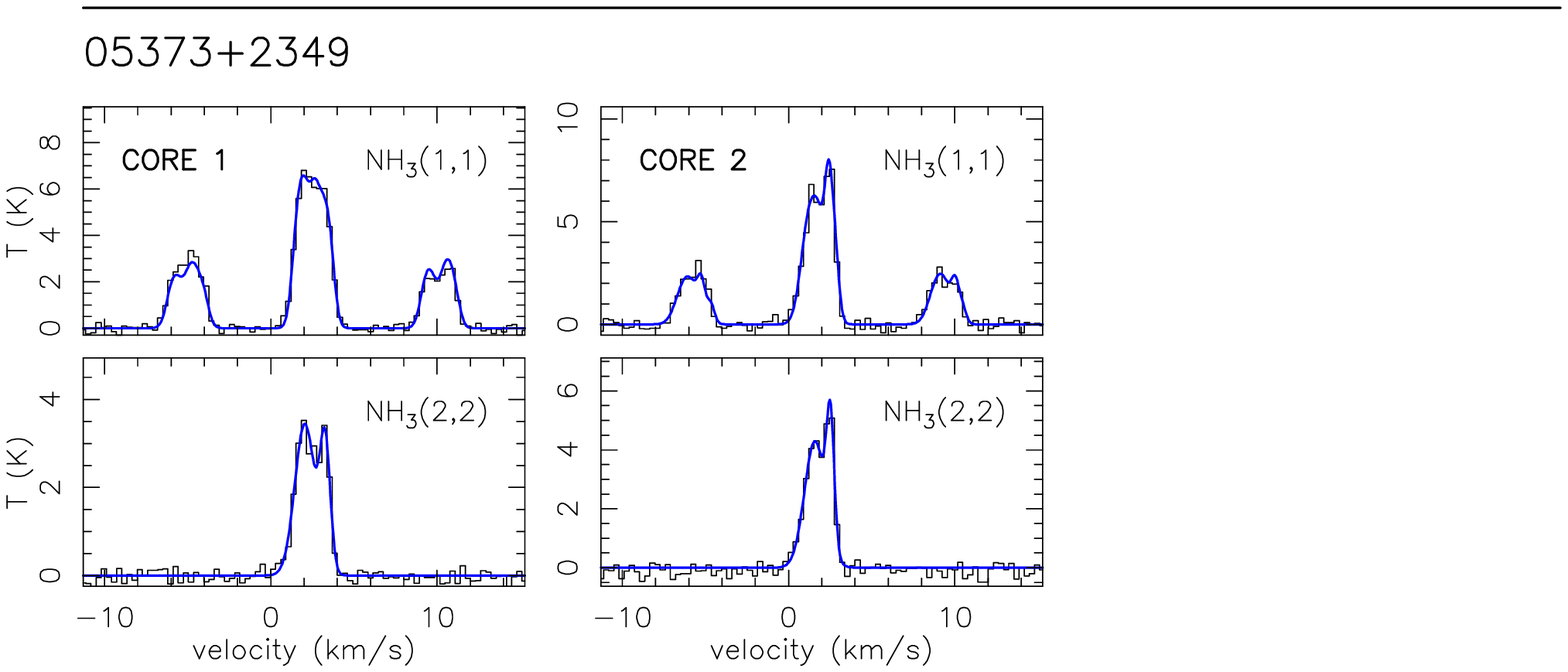, scale=0.85} \\ 
  \epsfig{file=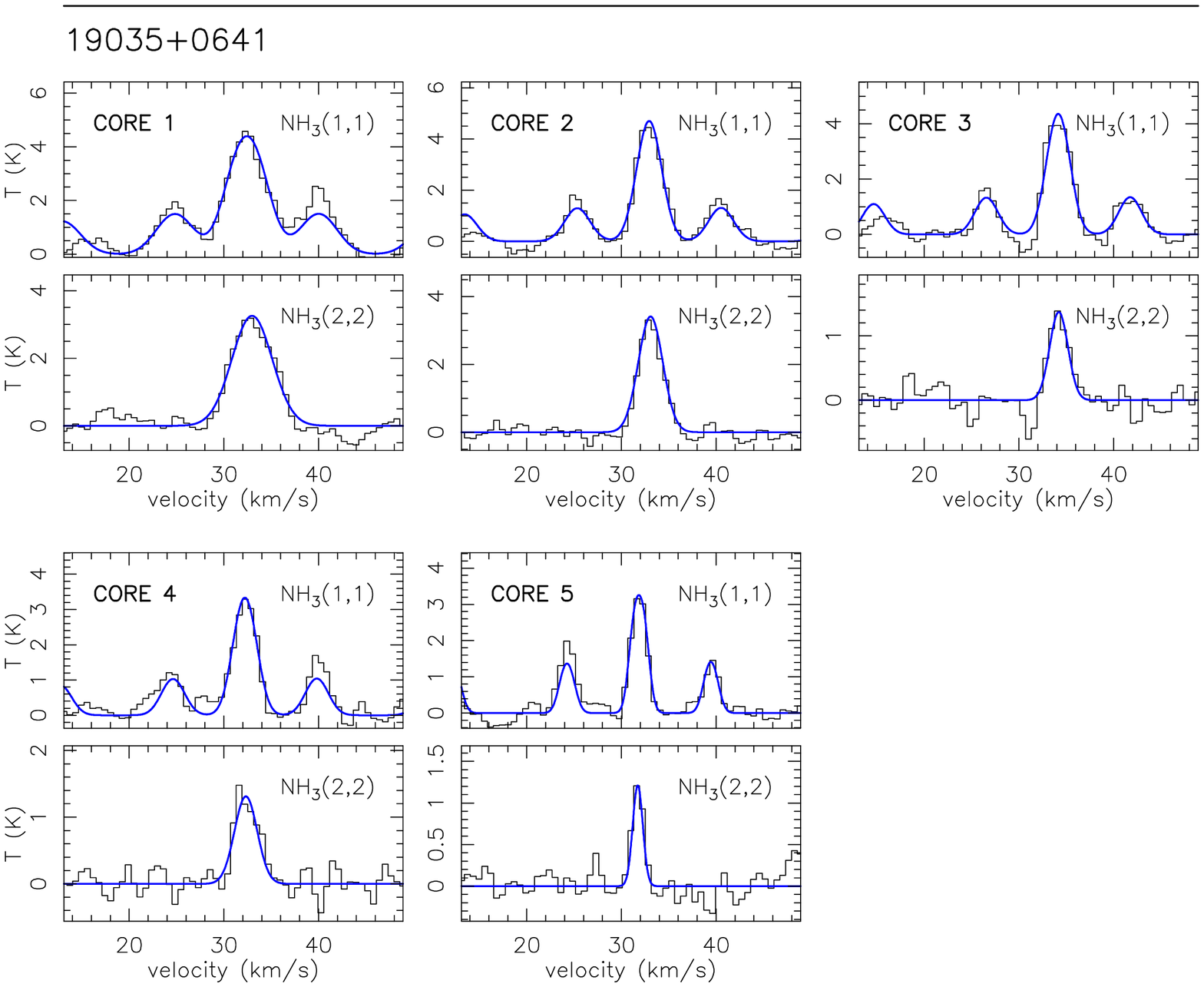, scale=0.85} \\ 
\end{tabular}
\contcaption{}
\end{center}
\end{figure*}
\begin{figure*}
\begin{center}
\begin{tabular}[b]{c}
\vspace{0.5cm}
  \epsfig{file=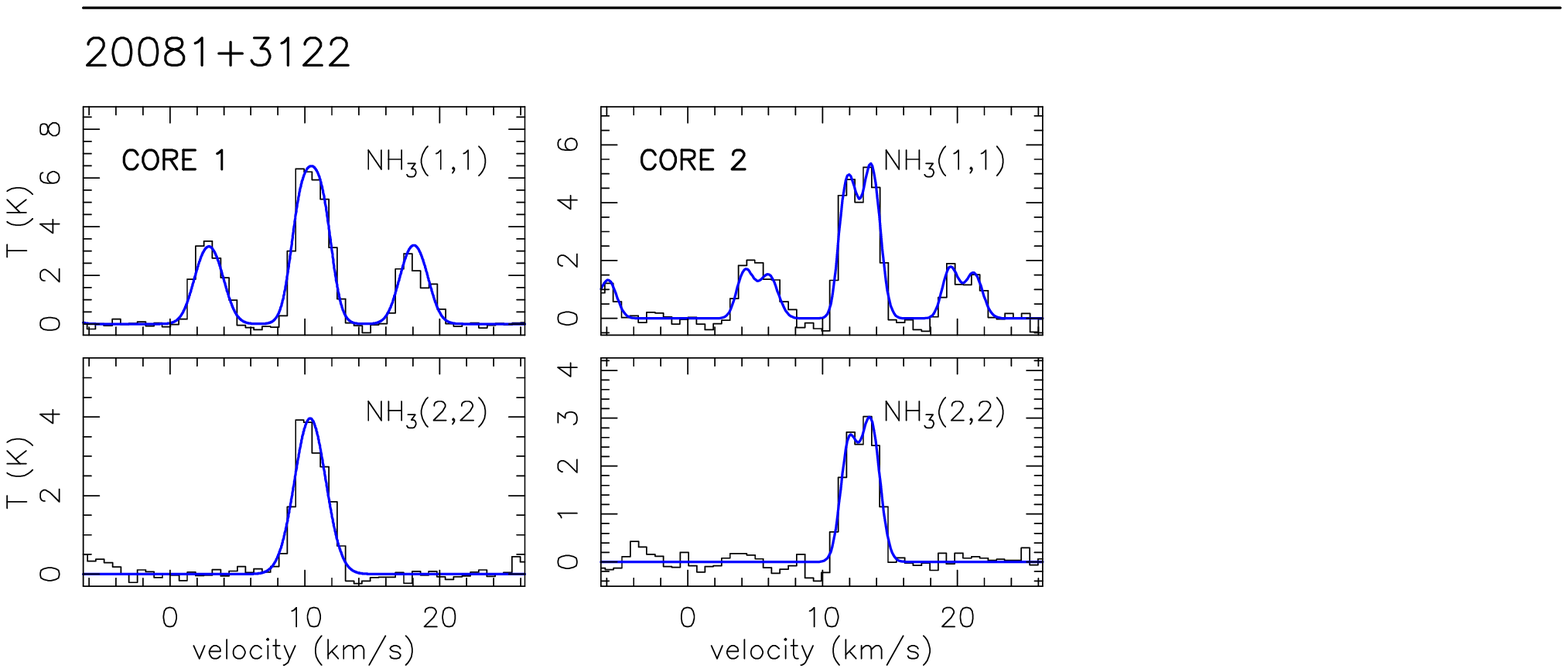, scale=0.85} \\ 
  \epsfig{file=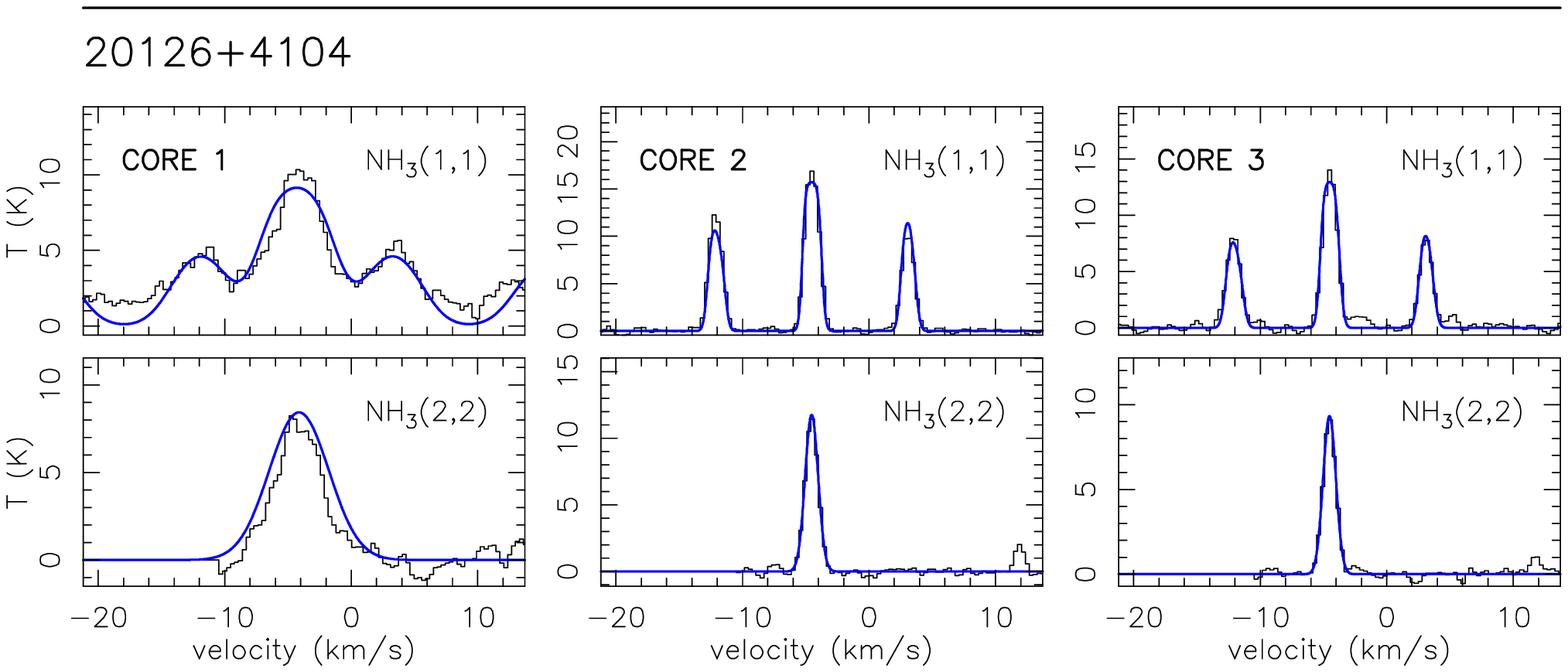, scale=0.85} \\ 
\end{tabular}
\contcaption{}
\end{center}
\end{figure*}
\begin{figure*}
\begin{center}
\begin{tabular}[b]{c}
\vspace{0.5cm}
  \epsfig{file=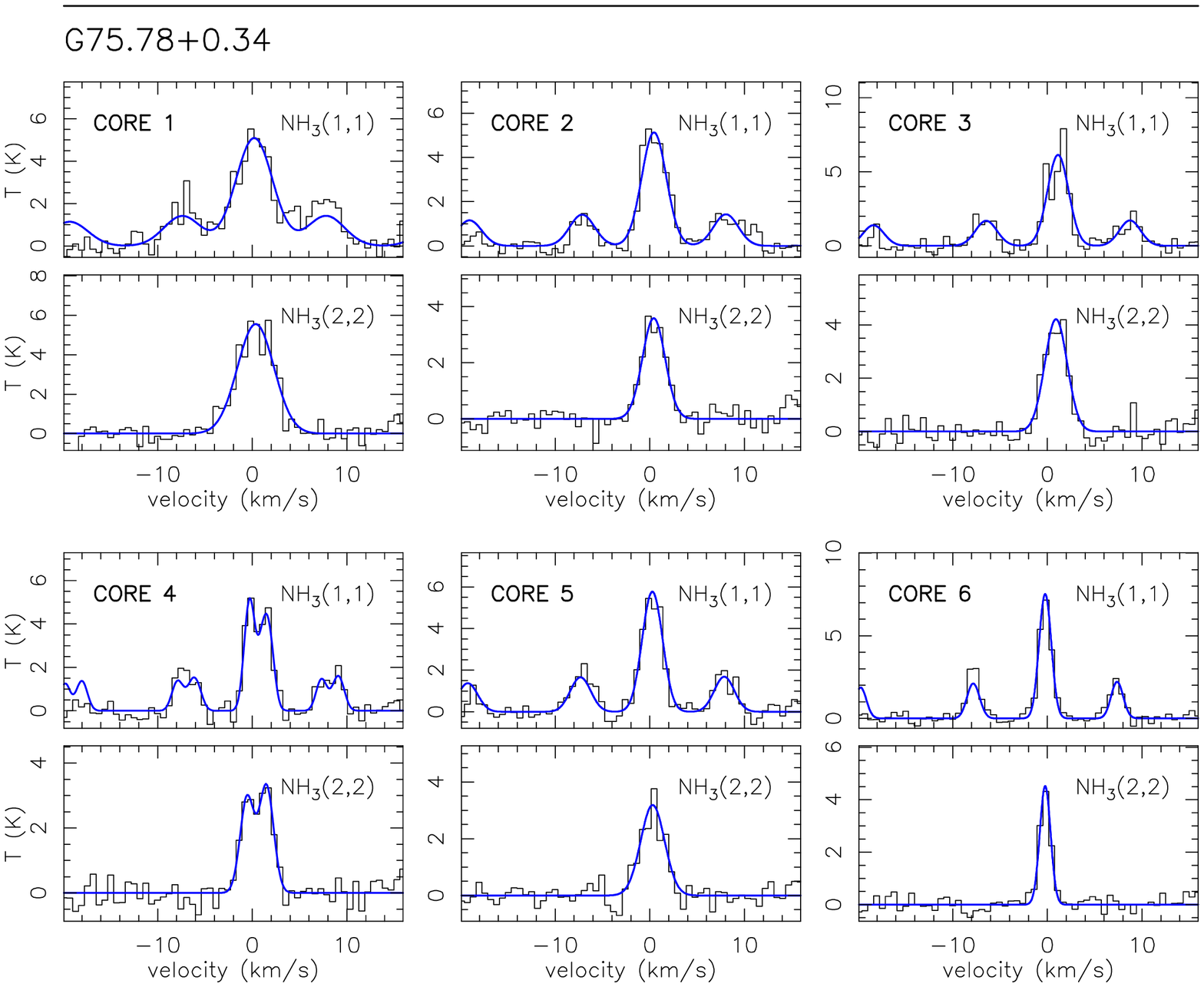, scale=0.85} \\ 
\end{tabular}
\contcaption{}
\end{center}
\end{figure*}
\begin{figure*}
\begin{center}
\begin{tabular}[b]{c}
\vspace{0.5cm}
  \epsfig{file=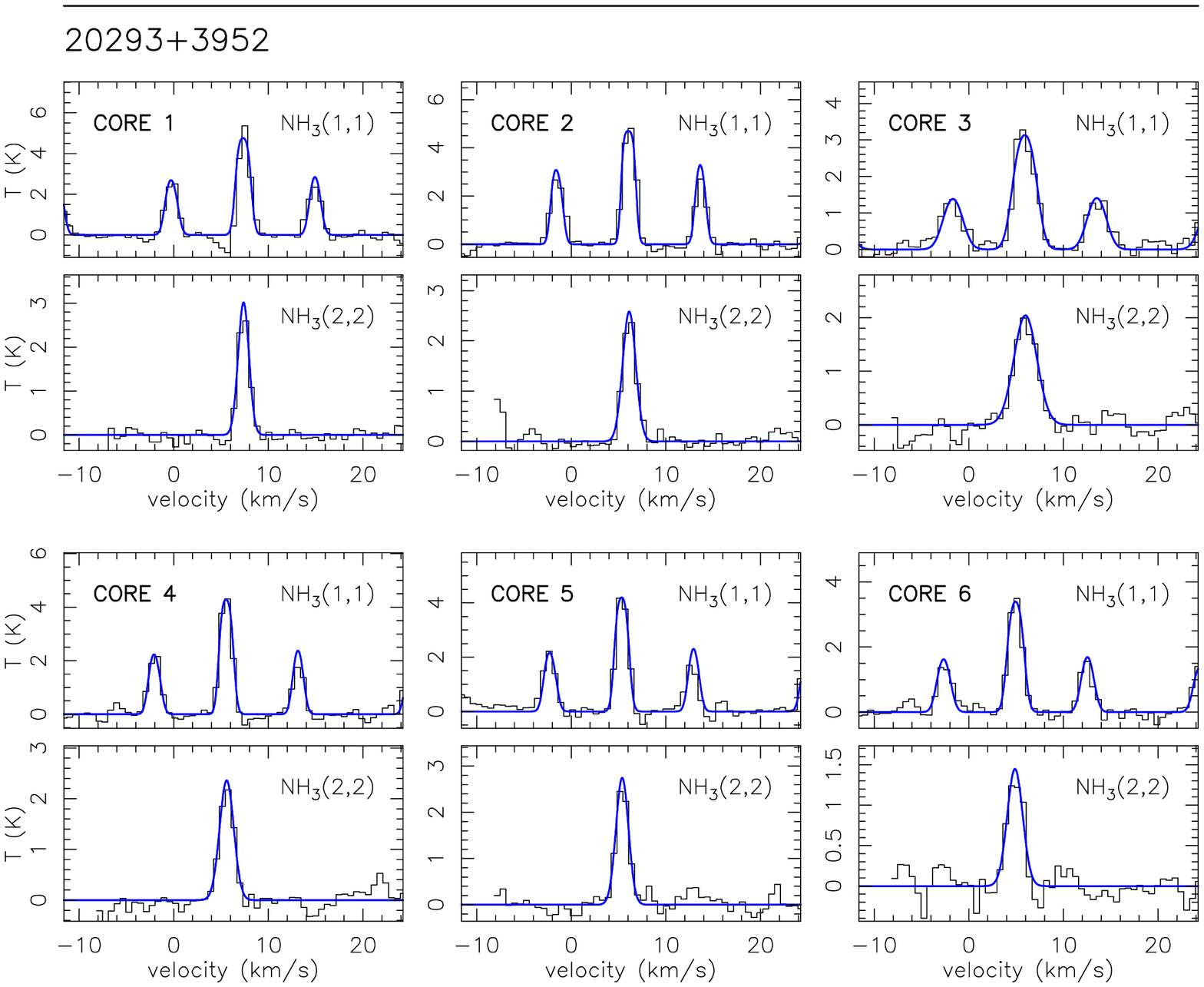, scale=0.85} \\ 
\end{tabular}
\contcaption{}
\end{center}
\end{figure*}
\begin{figure*}
\begin{center}
\begin{tabular}[b]{c}
\vspace{0.5cm}
  \epsfig{file=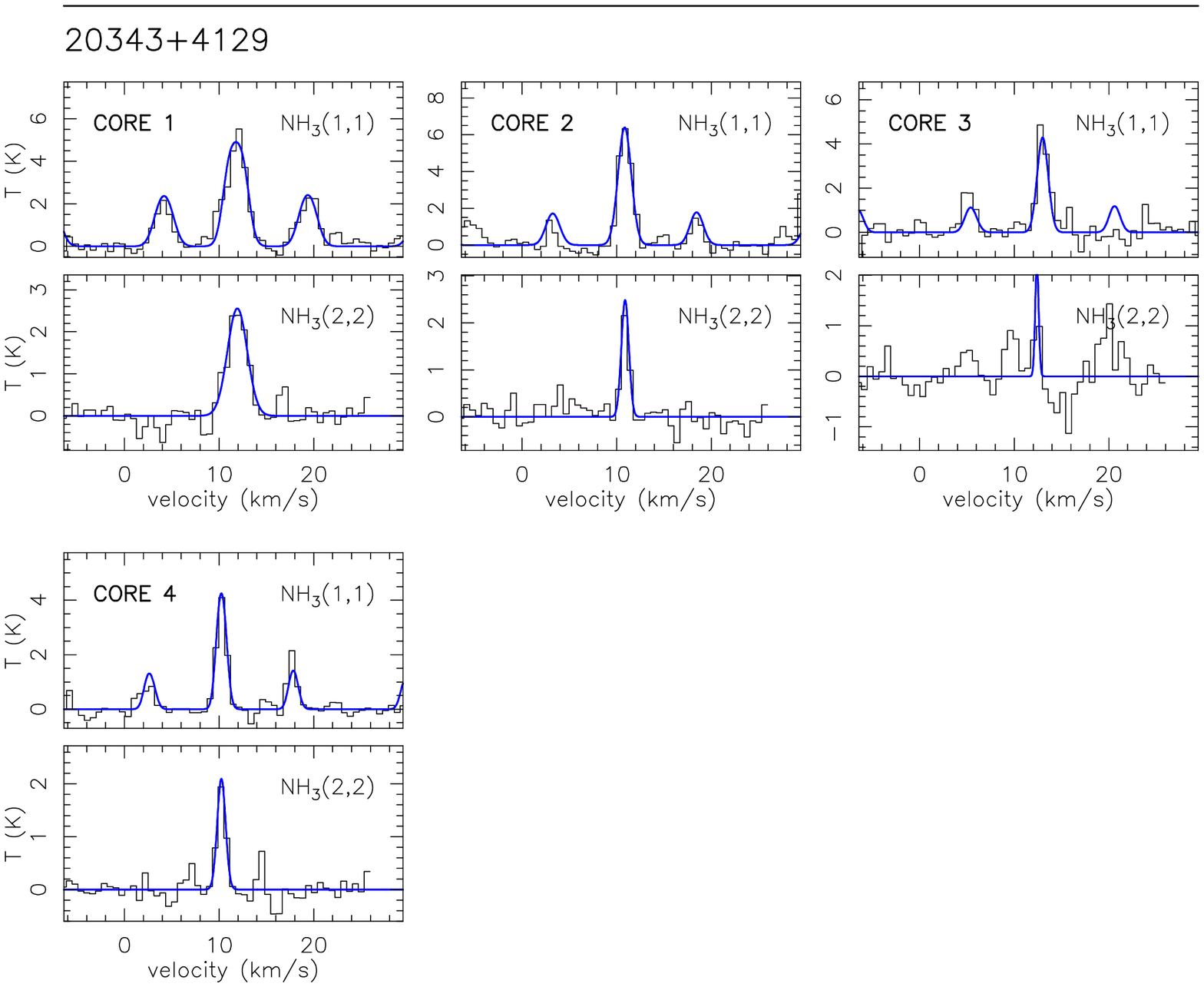, scale=0.85} \\ 
\end{tabular}
\contcaption{}
\end{center}
\end{figure*}
\begin{figure*}
\begin{center}
\begin{tabular}[b]{c}
\vspace{0.5cm}
  \epsfig{file=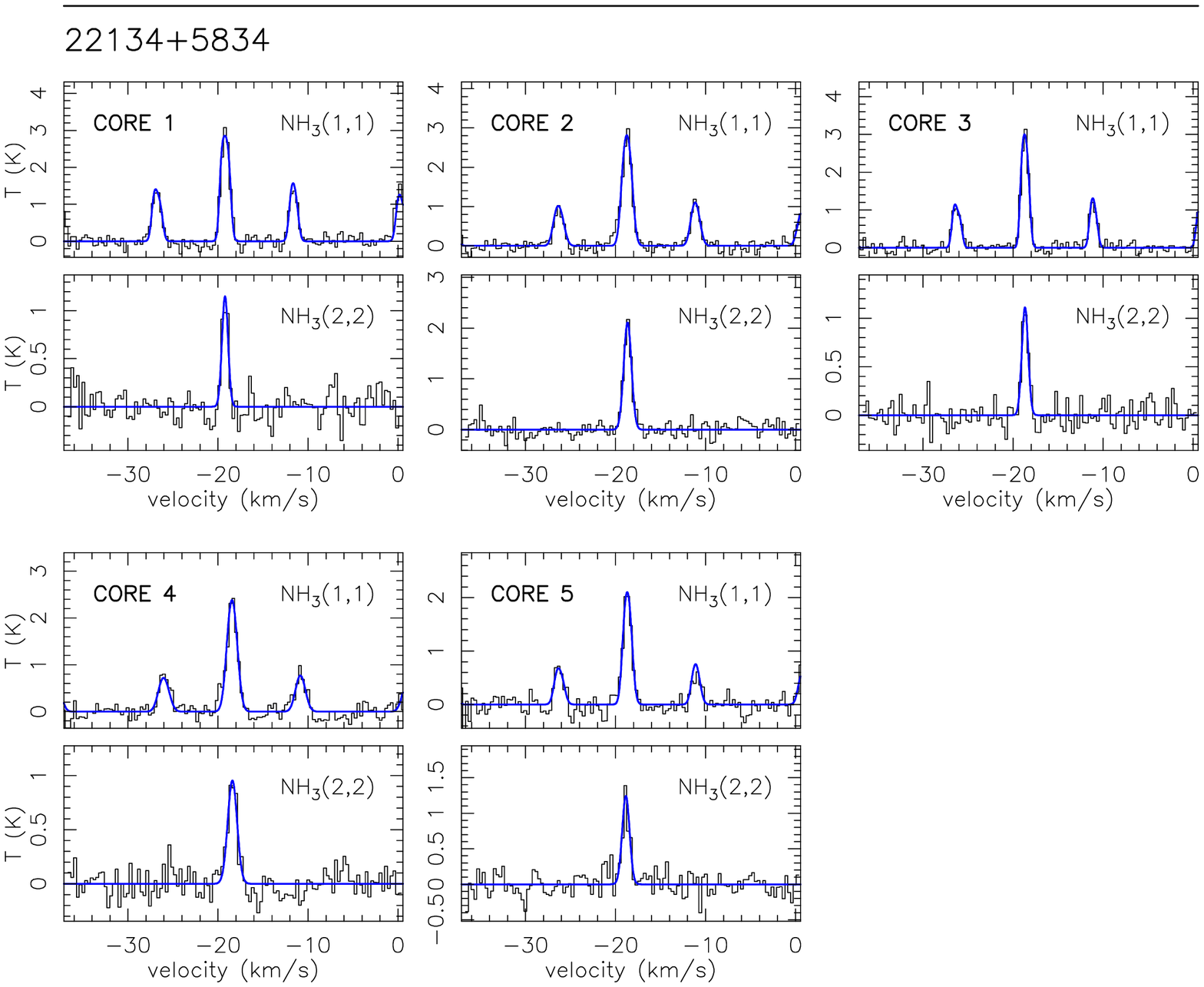, scale=0.85} \\
  \epsfig{file=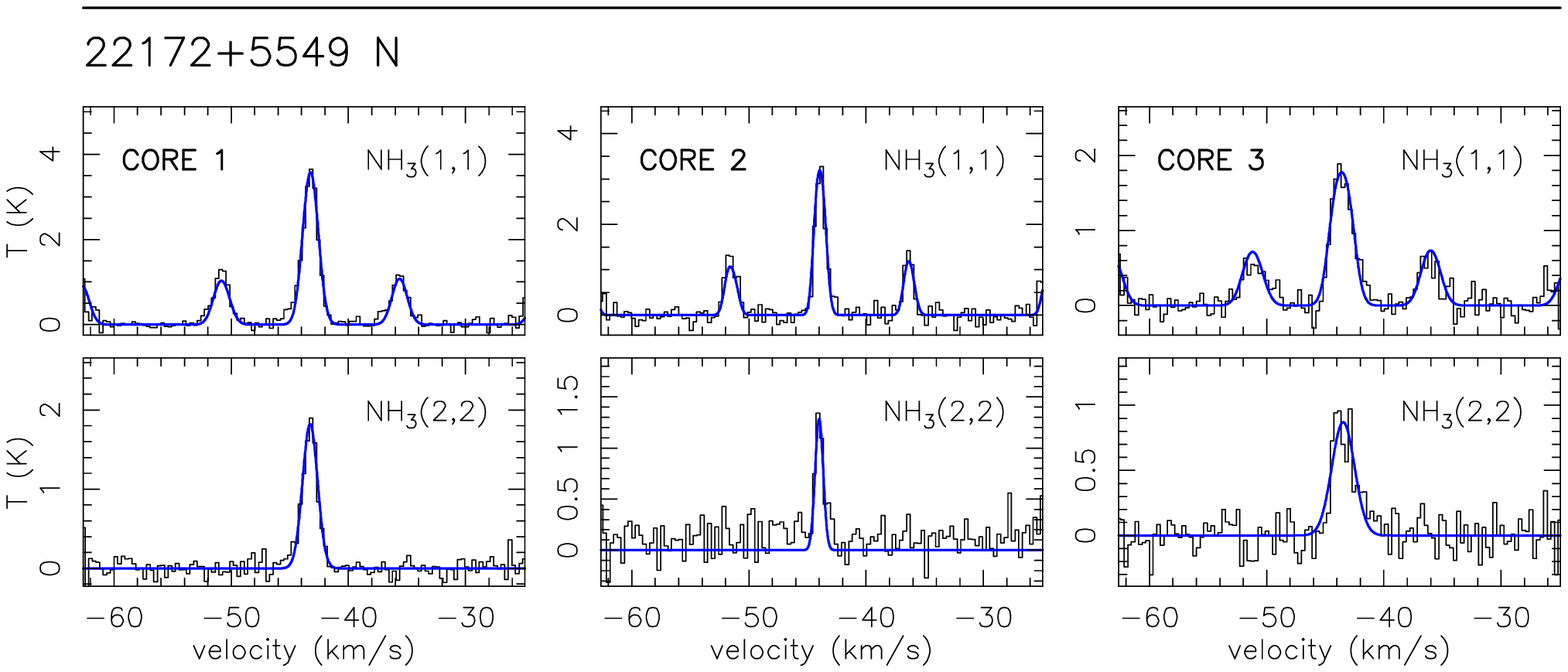, scale=0.85} \\
\end{tabular}
\contcaption{}
\end{center}
\end{figure*}
\begin{figure*}
\begin{center}
\begin{tabular}[b]{c}
\vspace{0.5cm}
  \epsfig{file=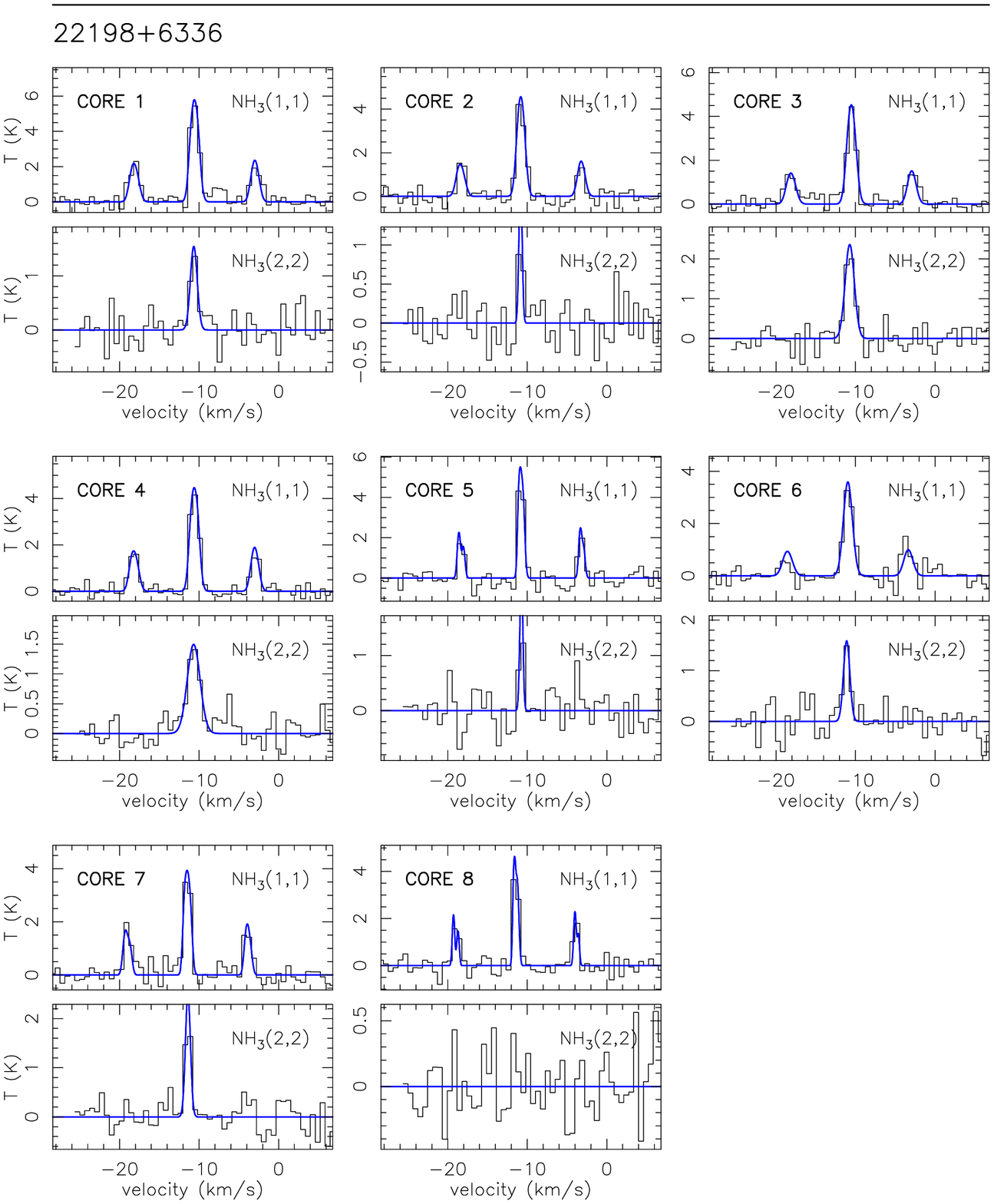, scale=0.85} \\ 
\end{tabular}
\contcaption{}
\end{center}
\end{figure*}
\begin{figure*}
\begin{center}
\begin{tabular}[b]{c}
\vspace{0.5cm}
  \epsfig{file=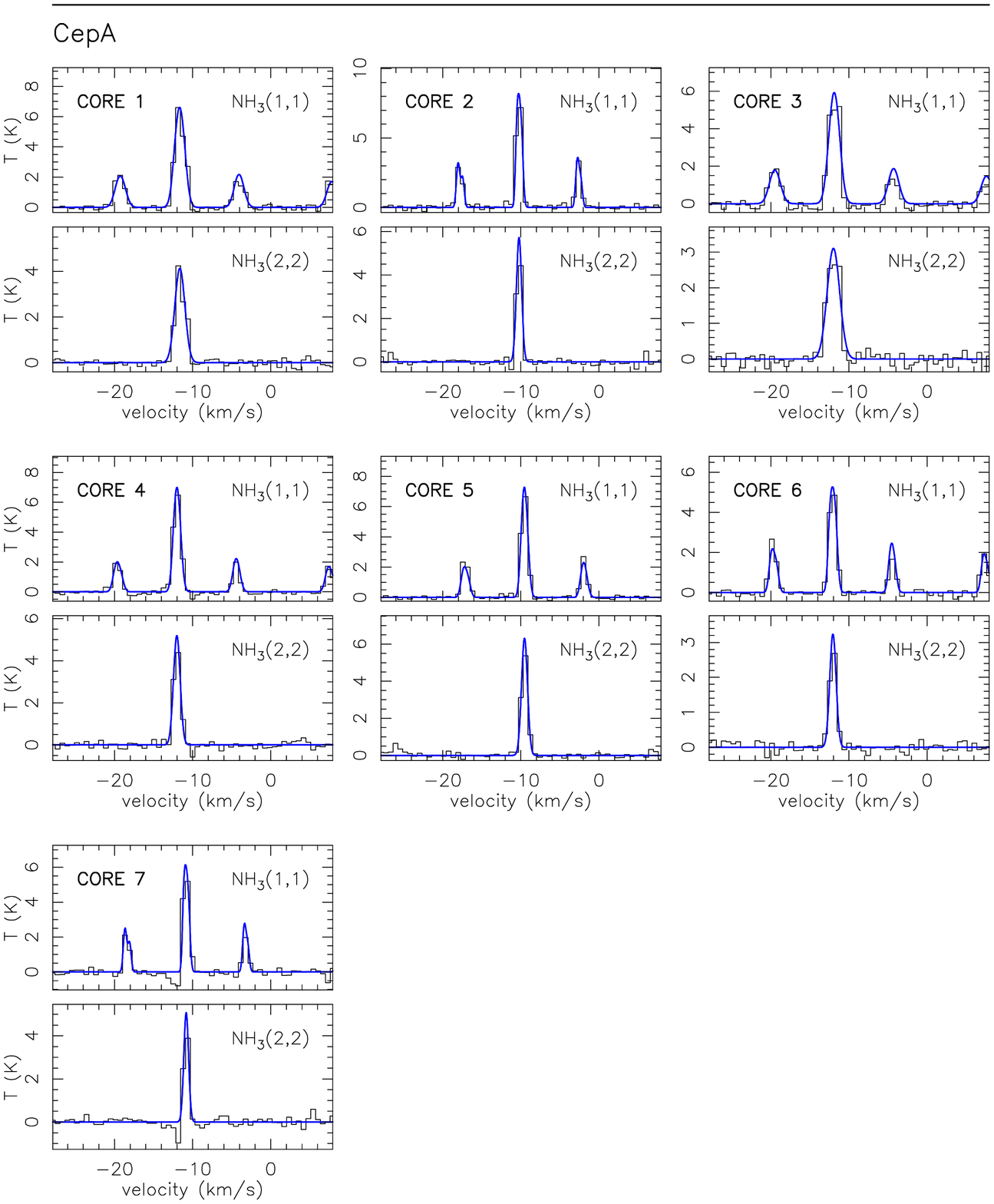, scale=0.85} \\
\end{tabular}
\contcaption{}
\end{center}
\end{figure*}

\end{document}